\documentclass[fleqn,usenatbib]{mnras}

\usepackage{newtxtext,newtxmath}

\usepackage[T1]{fontenc}
\usepackage{ae,aecompl}
\usepackage{array,longtable}

\usepackage{graphicx}	
\usepackage{natbib}
\usepackage[normalem]{ulem}
\usepackage{cancel}

\title[open clusters parameters]
{Updated parameters of 1743 open clusters based on {\it Gaia} DR2}

\author[W. S. Dias et al.]{
W. S. Dias,$^{1}$\thanks{E-mail:wiltonsdias@yahoo.com.br}
H. Monteiro,$^{1}$, A. Moitinho$^{2}$, 
J. R. D. L\'epine$^{3}$, 
G., Carraro,$^{4}$,
E. Paunzen$^{5}$,
\newauthor B. Alessi$^{6}$
and L. Villela$^{1,7}$ 
\\
$^{1}$Instituto de F\'isica e Qu\'imica, Universidade Federal de Itajub\'a, Av. BPS 1303 Pinheirinho, 37500-903 Itajub\'a, MG, Brazil\\
$^{2}$CENTRA, Faculdade de Ci\^encias, Universidade de Lisboa, Ed. C8, Campo Grande, 1749-016 Lisboa, Portugal\\
$^{3}$Universidade de S\~ao Paulo, Instituto de Astronomia, Geof\'isica e Ci\^encias Atmosf\'ericas, S\~ao Paulo,  SP, Brazil\\
$^{4}$Department of Physics and Astronomy, University of Padova, Vicolo dell'Osservatorio 3, I-35122 Padova, Italy 0000-0002-0155-9434\\
$^{5}$Departament of Theoretical Physics and Astrophysics. Marsaryk University. Brno, Czech Republic\\
$^{6}$ Av. Eng. Ant\^onio Heitor Eiras Garcia 1236,  apto 142. Bairro Jardim Esmeralda CEP 05588001 S\~ao Paulo, SP, Brazil \\ 
$^{7}$ Departamento de Engenharia El\'etrica, Universidade Federal do Esp\'irito Santo, Av. Fernando Ferrari, 514, Campus Universit\'ario de Goiabeiras, 29075-710 Vit\'oria, ES, Brazil 
}

\date{Accepted  Received }

\pubyear{2015}

\begin{document}
\label{firstpage}
\pagerange{\pageref{firstpage}--\pageref{lastpage}}
\maketitle

\begin{abstract}
In this study we follow up our recent paper \citep{Monteiro2020} and present a homogeneous sample of fundamental parameters of open clusters in our Galaxy, entirely based on {\it Gaia} DR2 data. 
We used published membership probability of the stars derived from {\it Gaia} DR2 data and applied our isochrone fitting code, updated as in \citet{Monteiro2020},  to $G_{BP}$ and $G_{RP}$ {\it Gaia} DR2 data for member stars.
In doing this we take into account the nominal errors in the data and derive distance, age, and extinction of each cluster.
This work therefore provides parameters for 1743 open clusters and, as a byproduct, a list of likely not physical or dubious open clusters is provided as well for future investigations.
Furthermore, it was possible to estimate the mean radial velocity of 831 clusters (198 of which are new and unpublished so far) using stellar radial velocities from {\it Gaia} DR2 catalog.
By comparing the open cluster distances obtained from isochrone fitting with those obtained from a maximum likelihood estimate of individual member parallaxes, we found a systematic offset of $(-0.05\pm0.04)$mas. 
\end{abstract}

\begin{keywords}
Galaxy: open clusters and associations: general
\end{keywords}


\section{Introduction}

Open clusters (OCs) constitute a privileged class of objects for investigating a range of astronomical topics, from Galactic structure and dynamics, to the formation, structure, and evolution of stars and stellar systems, and are steps in the cosmic distance ladder. Their positions, distances, proper motions and radial velocities, can in general be determined with better precision than those of individual stars, especially for distant objects. Most importantly, with isochrone fitting, their ages can be determined over a broad range with a precision not at reach for most other astronomical objects.

Since the classic works of \citet{1970IAUS...38..205B} and \citet{1982ApJS...49..425J} OCs have played an important role in revealing the structure and evolution of our Galaxy. Some more recent studies include the tracing of spiral arms
 \citep{2006MNRAS.368L..77M, 2008ApJ...672..930V, 2014MNRAS.437.1549B}, the spiral pattern rotation speed and corotation radius \citep{Dias2019MNRAS.486.5726D}, the metallicity gradient \citep{Lepine2011, OCCAMgradient20}, the Galactic warp \citep{2008ApJ...672..930V, Cantat2020a} and the location of the Sun with respect to Galactic plane \citep{Cantat2020a}. 
 A review of pre-{\it Gaia} Galactic structure results with OCs can be found in \citet{2010IAUS..266..106M}. 
 The most widely used catalogs of OCs and their fundamental parameters are the \emph{New catalogue of optically visible open clusters and candidates} \citep[][ hereafter DAML]{Dias2002} and \emph{The Milky Way Star clusters} \citep[][ hereafter MWSC]{Kharchenko2013}.
 Recent, pre-{\it Gaia} examples of OCs used in other topics, such as stellar astrophysics can be found in \citet{2013A&A...558A.103G}, and for constraining the distance scale in \citet{An2007}.

The {\it Gaia} DR2 catalog \citep{GAIA-DR22018} presents more than 1 billion stars with magnitude $G$ $\le$21 with high precision astrometric and photometric data, improving the stellar membership determination and characterisation of thousands of open clusters \citep{Cantat-Gaudin2018A&A...618A..93C, 2018A&A...619A.155S, Monteiro2019MNRAS.487.2385M, Bossini19, Carrera2019A&A...623A..80C, Monteiro2020}. As a consequence, in addition to the direct contribution to Galactic structure studies {\it Gaia} has  allowed the discovery of hundreds of new OCs \citep{ChinesCat, Sim2019JKAS...52..145S, Castro-Ginard2018, Castro-Ginard2019A&A...627A..35C, Castro-Ginard2020A&A...635A..45C, UFMG2020MNRAS.tmp.1833F}, to check the reality of doubtful objects \citep{CantateAnders}, \citet{Monteiro2019MNRAS.487.2385M}.

Despite the exquisite quality of {\it Gaia} DR2, the determination of OC parameters has in many cases remained a challenge. On the one hand, the parameter fitting algorithms have to deal with a number of factors, such as degeneracies, that compromise quality of the fits  over the whole range of parameter space. On the other hand, poor clusters and/or remaining field contamination may lead to ill defined groups. As a result, several clusters observed by Gaia did not have yet their parameters determined with {\it Gaia} DR2 data, and many others have unreliable determinations.

In \citet{Monteiro2020} we have focused on the determination of the parameters of 45 difficult clusters, left-overs of previous large scale {\it Gaia} based studies. This study introduced methodological improvements to isochrone fitting, including an updated extinction polynomial for the {\it Gaia} DR2 photometric band-passes and the Galactic abundance gradient as a prior for metallicity, which led to a successful automatic determination of the fundamental parameters of those clusters. 

This paper is a follow-up to \citet{Monteiro2020}. Here we employ the updated isochrone fitting procedure in a large scale homogeneous, {\it Gaia} DR2 based, determination of parameters for 1743 OCs with previous membership determinations. Both papers are part of an ongoing project to bring DAML into the {\it Gaia} era.

The remainder of this manuscript is organized as follows. In the next section we
describe the sample. Section 3 presents the cluster member stars. In Section 4, we present the determination of the OC astrometric parameters and mean radial velocity. In Section 5, we describe our code of isochrone fitting to determine the distances and ages of the
clusters. In Section 8, we compare the results obtained from {\it Gaia} data with those from UBVRI data for a control sample. Sections 9 and 10 are dedicated to compare our values with published ones.
Section 11 provides general comments on the results. Section 12 is dedicated to statistics of the  parameters determined in previous sections. In Section 13 we compare the distances from isochrone fits and from parallaxes. Finally, in the Section 14 we summarize the results presenting the main conclusions of this work.

\section{The sample of open clusters investigated}

\begin{figure*}
\centering
\includegraphics[scale = 0.55]{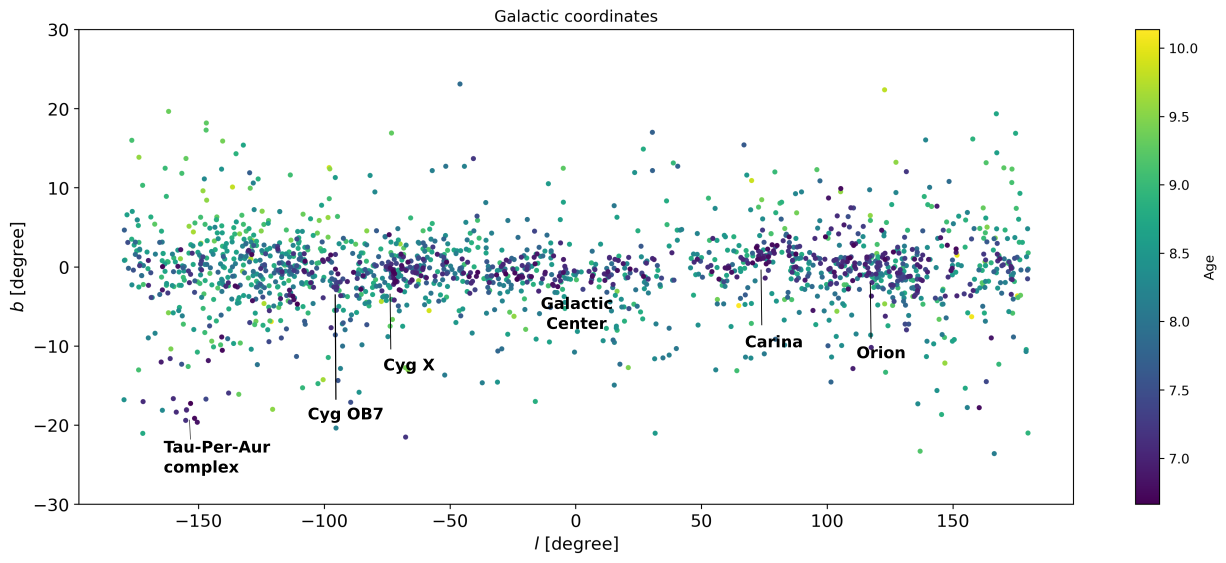}
\caption{Galactic distribution of the 1743 open clusters analysed in this work. The plot present the clusters in Galactic coordinates with the main regions of star formation highlighted. The color is proportional to the age in the sense blue is young, green is intermediate age and yellow is old.}
\label{fig:lxb}
\end{figure*}

\begin{figure}
\centering
\includegraphics[scale = 0.5]{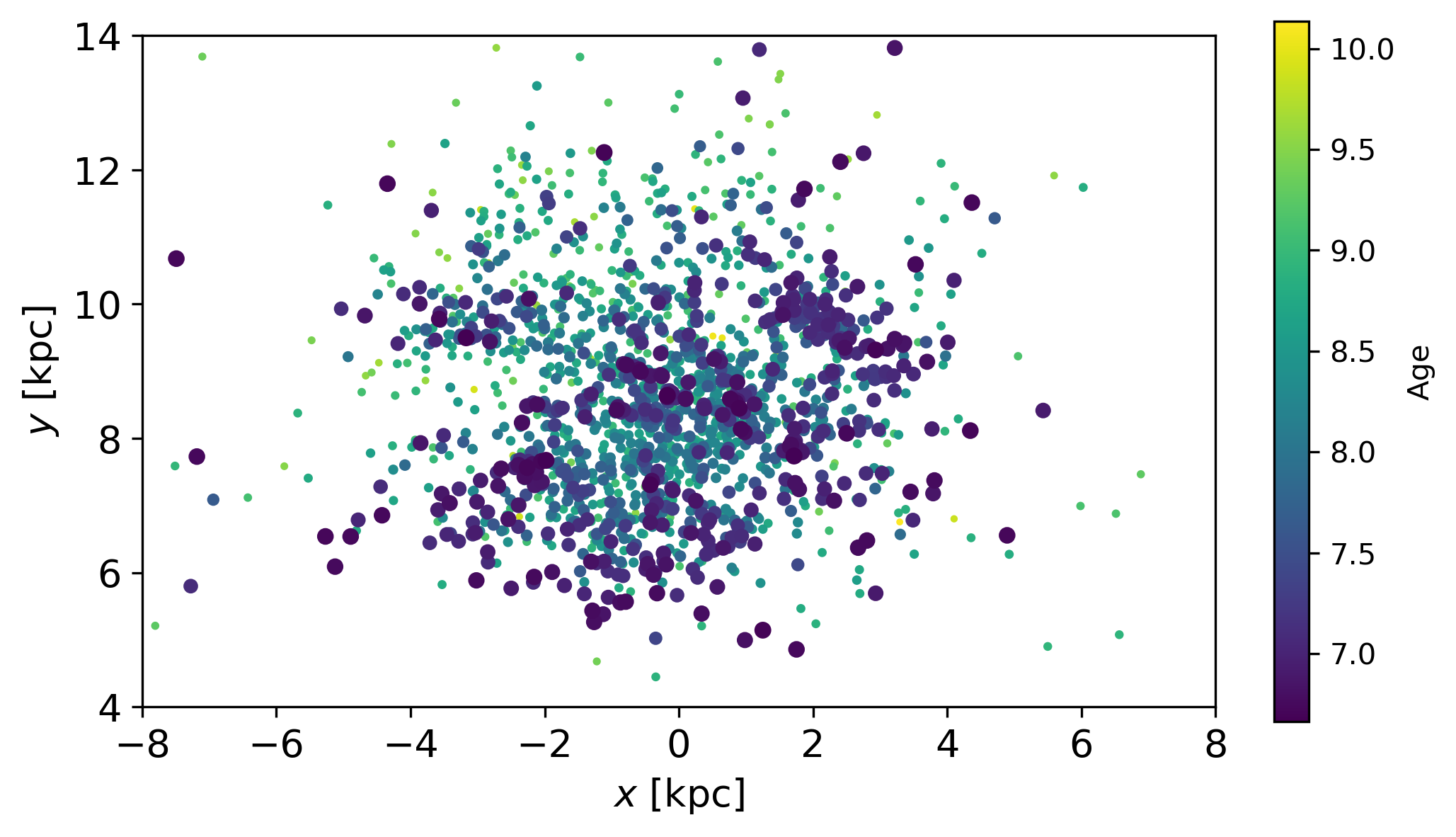}
\caption{Distribution of the 1743 clusters in the plane x-y. The Galactic Centre at (0,0) as reference and the x$-$axis pointing to the Galactic rotation direction. The rotation is clockwise or the vector angular velocity is perpendicular to the x$–$y plane pointing in the direction of the paper. The y$-$axis positive points towards the Galactic anticentre and the Sun is situated at (0,8.3) kpc position.}
\label{fig:plano}
\end{figure}

In this study we investigated 2237 open clusters and present a list of 1743 clusters including the parameters here determined, and their associated uncertainties. The sample is composed of 475 young clusters  (log(age) < 7.5 dex),
1075 intermediate age open clusters (log(age) between 7.5 dex and 9.0 dex) and 193 old clusters (log(age) > 9.0 dex).
365 clusters are located within 1 kpc and 1379 farther than 1 kpc.
Figures \ref{fig:lxb} and \ref{fig:plano} show the distribution of the sample in the Galaxy. The statistics of the parameters are discussed in the Section 12. 

The sample parameters have the main characteristic of being homogeneous since they are determined with the same isochrone fitting method, using the $G_{BP}$ and $G_{RP}$ magnitudes from {\it Gaia} DR2 catalog. 
 
 We also present the mean proper motion and parallax, the number of astrometric member stars, mean radial velocity and the number of stars used to estimate them, the apparent radius $r_{50}$ containing $50\%$ of the identified members, and the maximum radius $r_{max}$, which is the largest angular distance between the member stars and the central coordinates of each cluster considered as the mean positions in RA and DEC of the member stars.
The coordinates are given at epoch J2015.0 (ICRS) as realized by the {\it Gaia} DR2 catalog.

The table with the cluster's parameters and the tables with {\it Gaia} DR2 data of each star in the field with stellar membership probability are given electronically at the {\it Centre de Donn\'ees Stellaires} (CDS).
In this text a portion of the table with the parameters of the clusters is presented in Table \ref{tab:astrometric} and Table \ref{tab:photometric}.

\section{Open cluster memberships}
To determine precise fundamental parameters of open clusters it is necessary to know the member stars. With data from the {\it Gaia} DR2 catalog, astrometric membership determination has become much more precise than using ground-based data \citep{Dias2018}, leading to clearer CMDs and therefore enabling a more precise and reliable parameters' determination.

In this work we selected 2237 clusters with published individual stellar membership determined from {\it Gaia} DR2 astrometric data from the following studies:

\begin{itemize}
\item \citet{CantateAnders}, updated \citet{Cantat-Gaudin2018A&A...618A..93C}, and estimated stellar membership probabilities for 1867 open clusters applying the UPMASK procedure \citep{Krone-Martins2014A&A...561A..57K}; 

\item \citet{Castro-Ginard2020A&A...635A..45C} and \citet{Castro-Ginard2019A&A...627A..35C} found hundreds new open clusters looking for overdensities in
the astrometric space using DBSCAN and applying the machine learning to the CMD to check if the object is real;     

\item \citet{ChinesCat} used the Friend of Friend method to find 2443 open clusters and to select their members;

\item \citet{Sim2019JKAS...52..145S} found 655 cluster candidates (207 new)  by visual inspection of the stellar distributions in proper motion space and spatial distributions in the Galactic coordinates $(l,b)$ space. The members were determined using Gaussian mixture model and mean-shift algorithms;

\item  \citet{Monteiro2020} investigated 45 open clusters applying the maximum likelihood method to estimate stellar membership in the cluster regions;

\item \citet{UFMG2020MNRAS.tmp.1833F} discovered 25 new open clusters and identified the member stars by applying a decontamination procedure to the three-dimensional astrometric space. 

\end{itemize}

In the works published by \citet{ChinesCat} and \citet{Castro-Ginard2020A&A...635A..45C} the memberships were published as having probability equal to one ($P_{i}=1$), for members and zero for non-members. To maintain uniformity with the rest of the sample where individual membership probabilities were available,  we recalculated the individual membership probability of the stars ($P_{i}$) applying a variation of the classic maximum likelihood approach described in \citet{Dias2014} and \citet{Monteiro2020}. In the classic approach, a first step is to determine the bivariate probability density function that represents the distributions of the cluster members and of the field stars in the space formed by proper motions and parallax ($\mu_{\alpha}cos{\delta}$, $\mu_{\delta}$, $\varpi$). The second step is then to use these distributions for assigning cluster/field membership probabilities to the stars under analysis. In the variation here employed, we used the stars reported as members ($P_{i}=1$) in \citet{ChinesCat} and \citet{Castro-Ginard2020A&A...635A..45C} to directly estimate the shapes of the membership probability density functions in proper motion and parallax space, and from there proceed to the membership determinations.

\begin{figure}
\centering
\includegraphics[scale = 0.36]{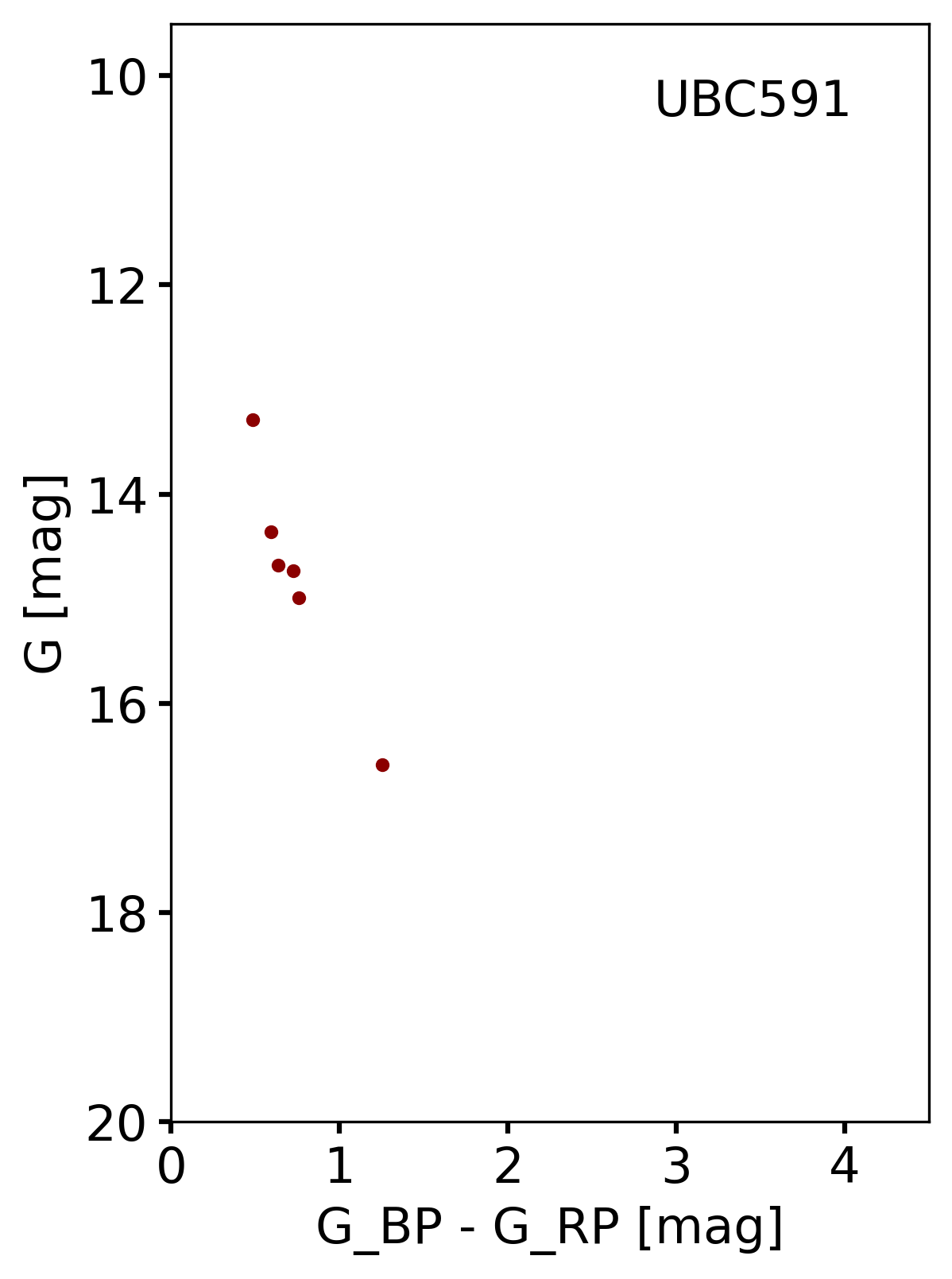}
\includegraphics[scale = 0.36]{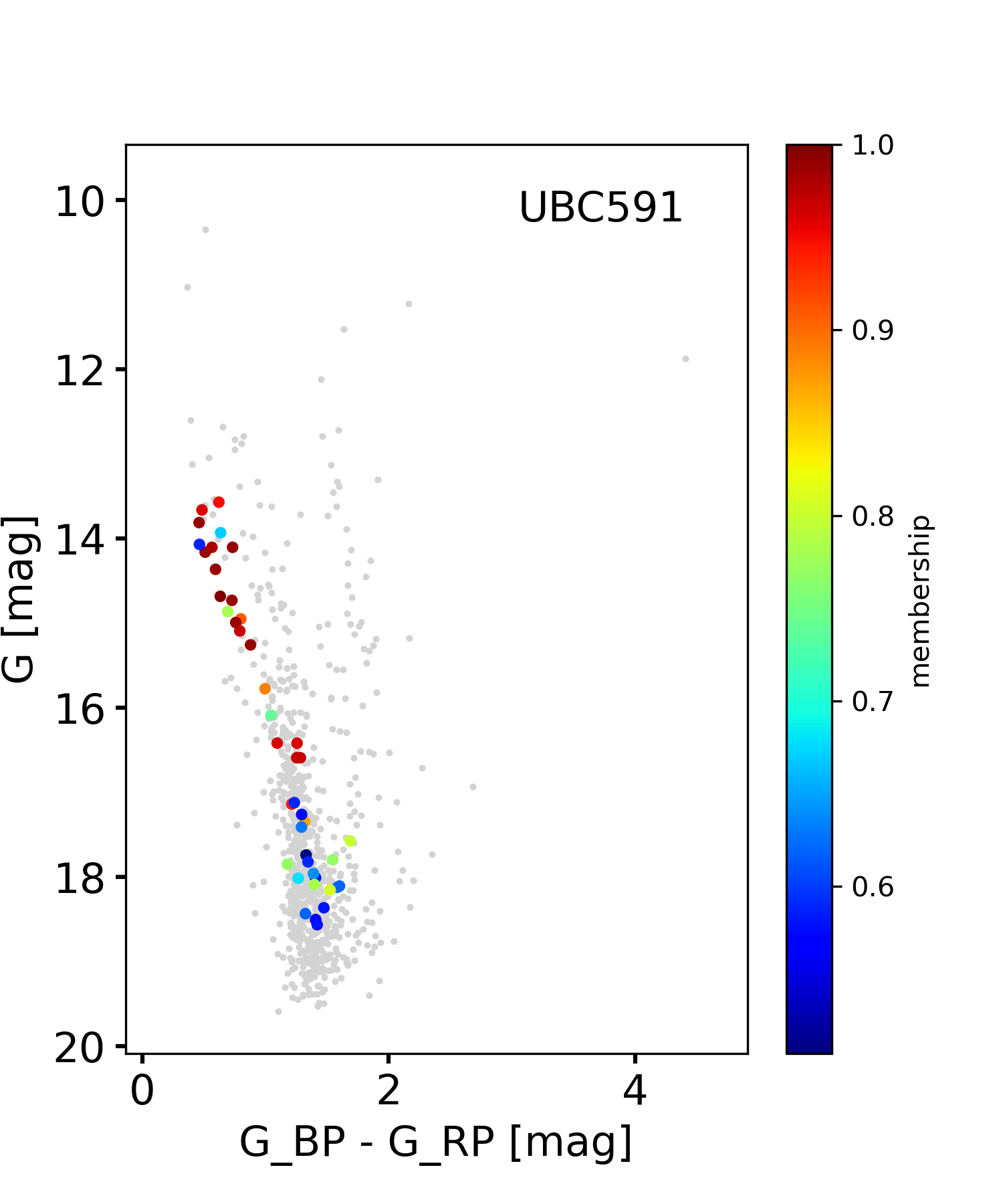}
\caption{Comparison of the members of the open cluster UBC~591 from  \citet{CantateAnders} (left plot) and from the recalculated memberships from this work (right plot).
The stellar membership is proportional to the color, in the sense redder color represent higher membership probability.}
\label{fig:members3D}   
\end{figure}

This procedure provided individual stellar membership probabilities and allowed us to improve the cluster signature in several CMDs, as shown in Figure \ref{fig:members3D} for the case of UBC~591.

Before using the photometric data we apply filters according to \citet{GaiaHRD18}, to guarantee that only stars with good astrometric solutions are used in the isochrone fit. Since we use the G magnitude in the plots we also applied corrections in this to correct for systematic effects as mentioned in \citet{GaiaHRD18} using the recipe in the  {\it Gaia} data archive website (\url{https://www.cosmos.esa.int/web/gaia/dr2-known-issues}).

\section{Parallax, proper motion and radial velocity of the clusters}

We used the stars with membership probability greater than 0.50 to estimate mean proper motion, parallax and radial velocity of the clusters. 
For proper motion and parallax we adopted the simple mean, 
however, to accommodate different numbers of measurements and also measurements with different errors, 
the mean radial velocity of each open cluster was obtained by weighting 
with the number of measurements  and the error of a single measurement, as used in \citet{Dias2014} following the recipe of \citet{Barford1967}. 
Before computing the mean we applied a $3\sigma$ outlier rejection of the radial velocity values.
For mean proper motion and parallax, the standard deviation ($1\sigma$) was adopted to represent the error.

\begin{figure}
\centering
\includegraphics[scale = 0.45]{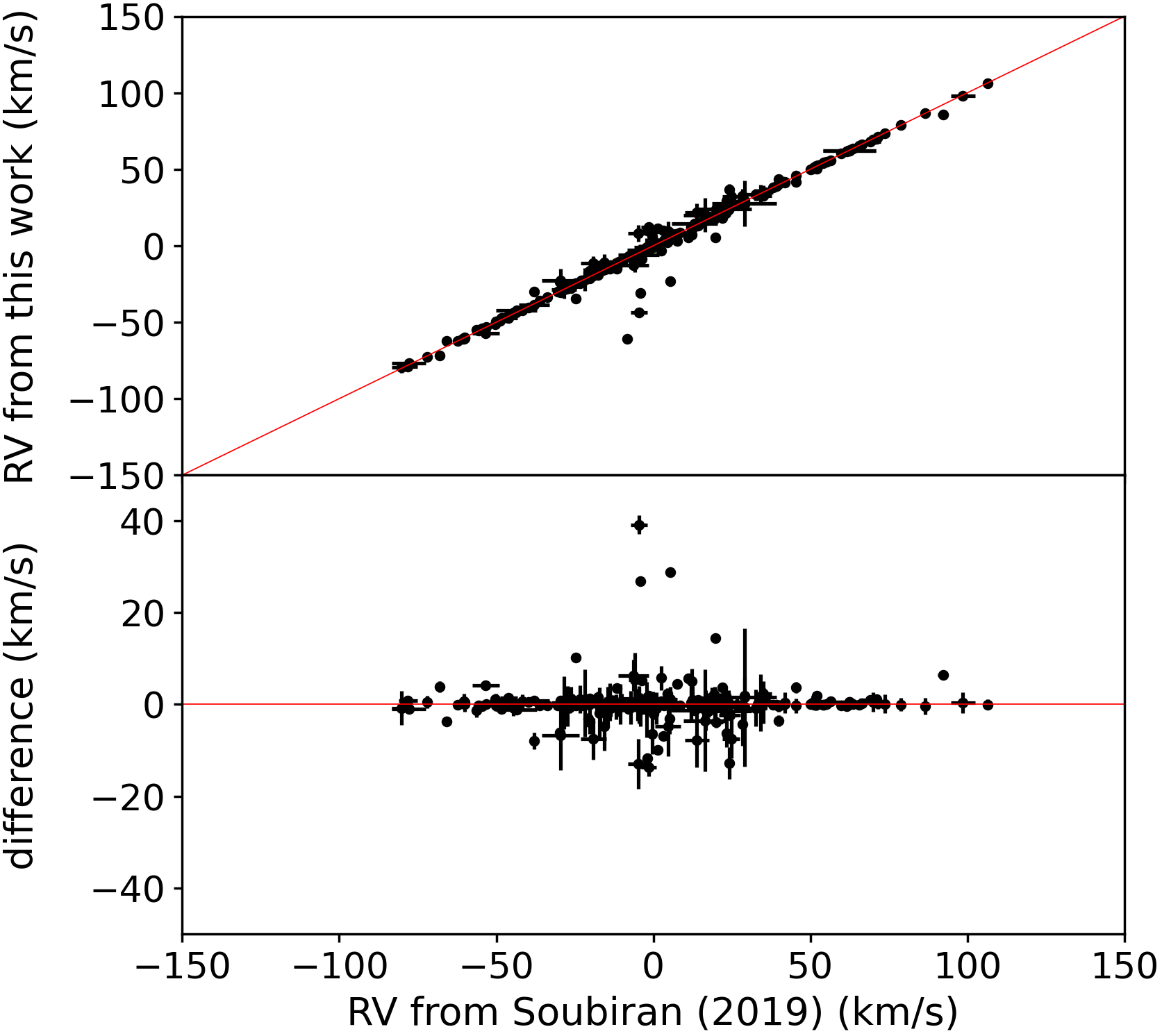}
\caption{Comparison of the mean radial velocity for 229 open clusters in common with \citet{Soubiran2018A&A...619A.155S}. The mean difference is 0.4 km\,s$^{-1}$ with standard deviation of 5.9 km\,s$^{-1}$ in the sense literature minus our value. }
\label{fig:RVSoubiran}
\end{figure}

\begin{figure}
\centering
\includegraphics[scale = 0.45]{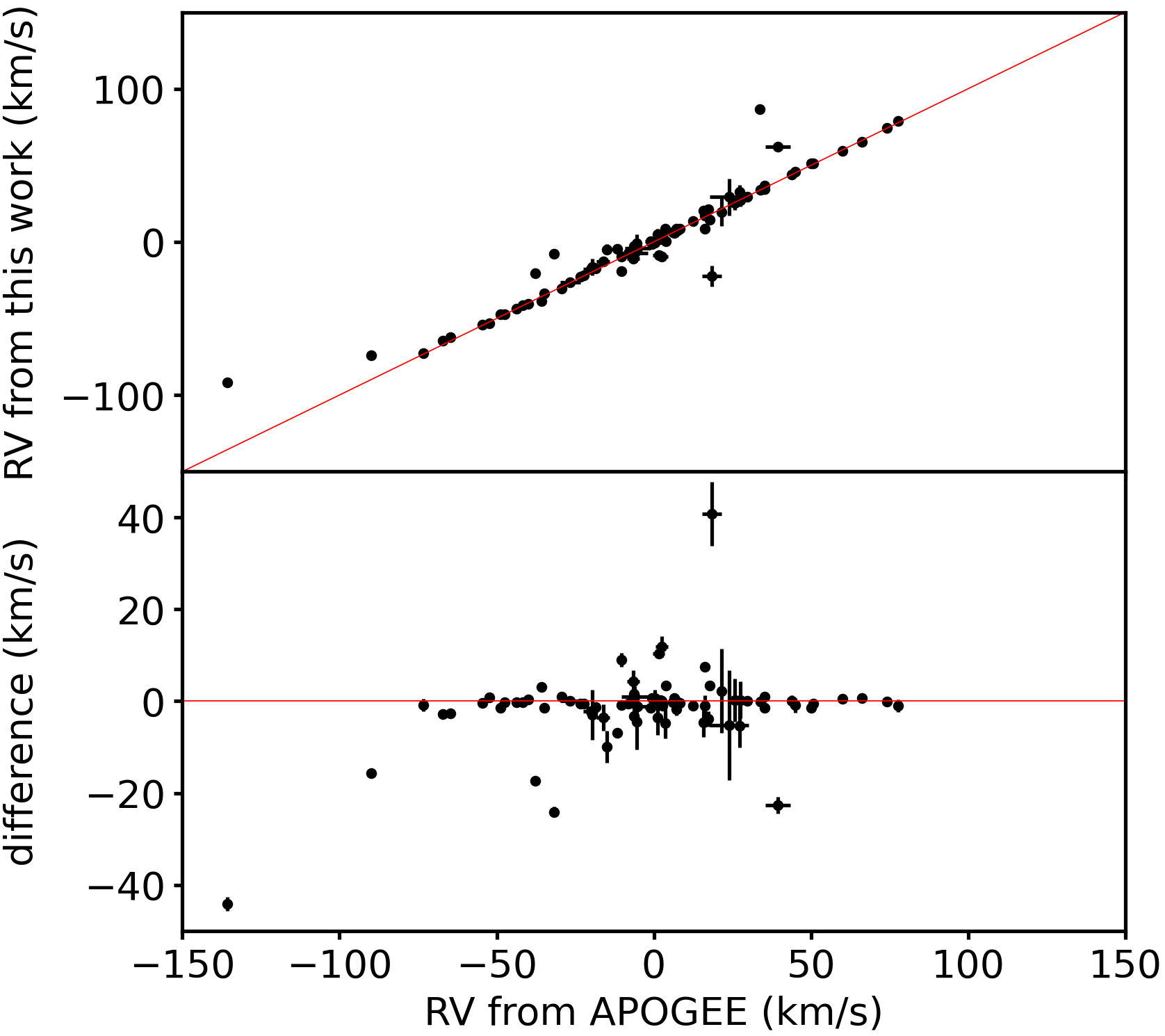}
\caption{Comparison of the mean radial velocity for 79 open clusters in common with \citet{Carrera2019A&A...623A..80C}. The mean difference is -2.0 km\,s$^{-1}$ with standard deviation of 10.4 km\,s$^{-1}$ in the sense literature minus our value. }
\label{fig:RVAPOGEE}
\end{figure}

\begin{figure}
\centering
\includegraphics[scale =0.45]{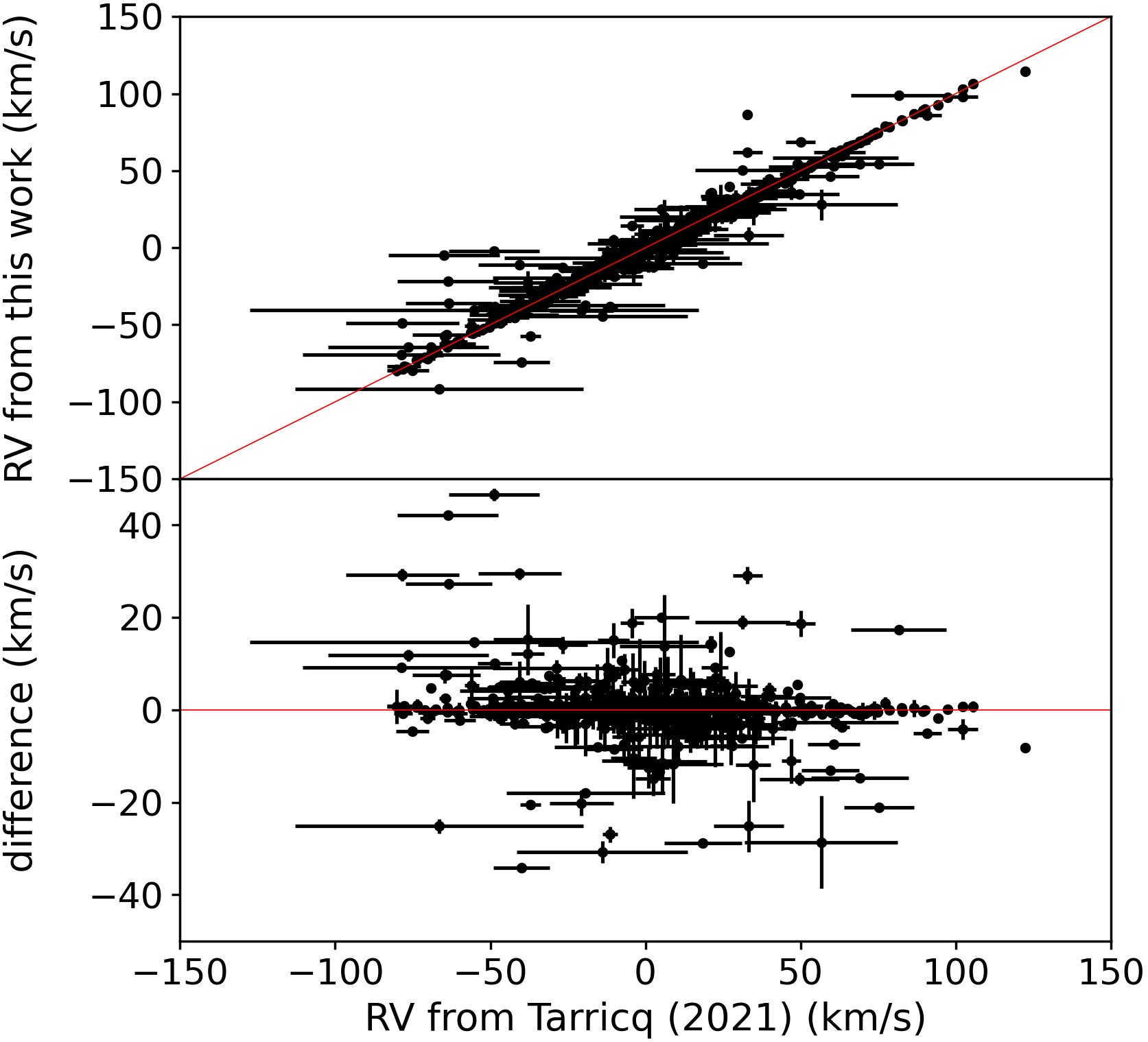}
\caption{Comparison of the mean radial velocity for 631 open clusters in common with \citet{2020arXiv201204017T}. The mean difference is 0.4 km\,s$^{-1}$ with standard deviation of 7.0 km\,s$^{-1}$ in the sense literature minus our value. }
\label{fig:Tarricq}
\end{figure}

As we use only the clusters with member stars derived from {\it Gaia} DR2 data, this was a limiting criterion, since the {\it Gaia} DR2 catalog gives radial velocity for stars with $G$ $\leq$ 12. Even so, mean radial velocity was estimated for 831 open clusters of which 198 had no previous published estimates, of which 106 from \citet{ChinesCat}, 81 from the work of \citet{Castro-Ginard2020A&A...635A..45C}, 1 from \citet{Sim2019JKAS...52..145S}, 7 from \citet{Monteiro2020}, 2 from \citet{Cantat-Gaudin2018A&A...618A..93C} and 1 from \citet{UFMG2020MNRAS.tmp.1833F}. 
In Figures \ref{fig:RVSoubiran}, \ref{fig:RVAPOGEE} and \ref{fig:Tarricq} we check our mean radial velocities against the values derived
by \citet{Soubiran2018A&A...619A.155S}, \citet{Carrera2019A&A...623A..80C} and \citet{2020arXiv201204017T}, all determined using membership from \citet{Cantat-Gaudin2018A&A...618A..93C} and data from {\it Gaia} DR2 and from APOGEE \citep{APOGEE} and GALAH \citep{GALAH} in the case of \citet{Carrera2019A&A...623A..80C}.
The differences of 229 clusters with \citet{Soubiran2018A&A...619A.155S} gives a mean of 0.4 km\,s$^{-1}$ with standard deviation of 5.9 km\,s$^{-1}$  and the differences of 79 clusters with \citet{Carrera2019A&A...623A..80C} gives a mean of -2.0 km\,s$^{-1}$ with standard deviation of 10.4 km\,s$^{-1}$, and considering 631 open clusters in common with \citet{2020arXiv201204017T} we obtain a mean of 0.4 km\,s$^{-1}$ with standard deviation of 7.0 km\,s$^{-1}$. All the values we calculated in the sense literature minus our result. 
 
As for the comparison between our results and those published, there are 5 open clusters with differences which exceed 3$\sigma$. For clusters Alessi 19,  Harvard 16, IC 1311, NGC 1027 our values differ with the \citet{Soubiran2018A&A...619A.155S}. The authors used four member stars for Alessi~19 and one member star for the other clusters.
For Trumpler 3 our value differs from the \citet{Carrera2019A&A...623A..80C}, which used radial velocity from APOGEE DR14 of four members.
The differences can be explained by the fact that different studies  use different ways of calculating mean radial velocity (weighted or not, with rejection or not). Thus, despite the works being based on memberships from \citet{CantateAnders}, either not exactly the same stars were used or different was the way they were analyzed.

In Table \ref{tab:astrometric} we present a portion of the results giving the mean proper motion, mean parallax and mean radial velocity  with the respective errors represented by the one standard deviation.
In the table are also given the number of cluster members, the equatorial coordinates ($\alpha, \delta$) and the radius ($r_{50}$) of each cluster. 

\begin{table*}
\caption[]{A portion of the table of the results of mean astrometric parameters obtained using the {\it Gaia} DR2 stellar proper motion and parallaxes.  
The meaning of the
symbols are as follows:
RA and DE are the central coordinates of the clusters; $r_{50}$ is the radius in which half of
the identified members are located; 
N is the number of cluster stars;
$\mu_{\alpha}cos{\delta}$ and $\mu_{\delta}$ are the  proper motion components in mas yr$^{-1}$;
$\sigma$ is the dispersion of cluster stars' proper motions;
$\varpi$ is the mean parallax of the cluster and $\sigma \varpi$ is the dispersion of the mean parallax.
RV and $\sigma RV$ are the mean and 1$\sigma$ dispersion radial velocity obtained for the cluster using {\it Gaia} DR2 data.}
\label{tab:astrometric}
\begin{center}
\resizebox{17cm}{!}{
\begin{tabular}{lrrccrcrcccrrc}
\hline
   Name             &  
   \multicolumn{1}{c}{$RA$} & 
   \multicolumn{1}{c}{$DE$}  &  
   $r_{50}$    &   
   N  &  
   \multicolumn{1}{c}{$\mu_{\alpha}cos{\delta}$}   &   
   $\sigma_{\mu_{\alpha}cos{\delta}}$  &  
   \multicolumn{1}{c}{$\mu_{\delta}$}  &   
   $\sigma_{\mu_{\delta}}$  &   
   $\varpi$   &   
   $\sigma_{\varpi}$  &       
   \multicolumn{1}{c}{RV} &                
   \multicolumn{1}{c}{$\sigma_{RV}$} & 
   \multicolumn{1}{c}{$N$}\\       
\hline
                       &  [deg]    & [deg]      & [deg]    &        &  [mas\,yr$^{-1}$] & [mas\,yr$^{-1}$] & [mas\,yr$^{-1}$] & [mas\,yr$^{-1}$]& [mas]  & [mas]  & [$km.s^{-1}$] &  [$km.s^{-1}$] & \\   
\hline
Alessi~1               &  151.6211 &    49.5407 &    0.237 &     48 &    6.550 &   0.152  &  -6.252 &    0.293 &   1.396  &  0.063  &       -3.343 &           1.024  &  11\\       
Alessi~10              &   35.3246 &   -10.5442 &    0.272 &     73 &    1.456 &   0.242  &  -7.844 &    0.273 &   2.227  &  0.087  &      -12.671 &           1.819  &   6\\       
Alessi~12              &  308.3153 &    23.9029 &    0.660 &    313 &    4.339 &   0.311  &  -4.650 &    0.222 &   1.825  &  0.098  &       -3.374 &           2.662  &  25\\       
\hline
\end{tabular}}
\end{center}
\end{table*}

\section{Parameters from the isochrone fits} 

The parameters distance, age, and $A_V$ of the clusters were obtained by fitting theoretical isochrones to {\it Gaia} DR2 $G_{BP}$ and $G_{RP}$ photometric data, applying the cross-entropy continuous multi-extremal optimization method (CE), which takes into account the astrometric membership of the star as well the nominal errors of the data. It is exactly the same code used in \citet{Monteiro2020} and to avoid being repetitive, we opted to present a general description in this text.

Basically, the CE method involves an iterative statistical procedure where the following loop is done in each iteration:

\begin{enumerate}
\item random generation of the initial sample of fit parameters, respecting predefined criteria
\item selection of the $10\%$ best candidates based on calculated weighted likelihood values
\item generation of random fit parameter sample derived from a new distribution based on the $10\%$ best candidates calculated in the previous step
\item repeat until convergence or stopping criteria reached.
\end{enumerate}

Our method uses a synthetic cluster obtained from an isochrone, sampling from a pre$-$defined initial mass function (IMF), randomly generating a number of stars in the mass range of the original isochrone.

The code interpolates the Padova (PARSEC version 1.2S) database of stellar evolutionary tracks and isochrones \citep{Bressan2012}, which uses the {\it Gaia} filter passbands of \citet{Maiz18}, scaled to solar metal content with $Z_{\odot} = 0.0152$. The grid used is constructed from isochrones with steps of 0.05 in $log(age)$ and 0.002 in metallicity. A search for the best solutions is performed in the following parameter space:

\begin{itemize}
\item age: from log(age) =6.60 to log(age) =10.15 dex;
\item distance: from 1 to 25000 parsec;
\item $A_V$: from 0.0 to 5.0 mag; 
\item $[Fe/H]$: from -0.90 to +0.70 dex
\end{itemize}

To account for the extinction coefficients dependency on colour and extinction due to the large passbands of {\it Gaia} filters, we used the most updated extinction polynomial for the {\it Gaia} DR2 photometric band-passes, as presented in detail by \citet{Monteiro2020}.

While the isochrones used are provided as a function of $[M/H]$, in this work we assume $[M/H]=[Fe/H]$. The assumption is justified by considering the relation from \citep{SC06} which gives  $[m/H] \sim  [Fe/H] + log(0.694f_{\alpha} + 0.306)$, where $f_{\alpha} = 10^{[\alpha/Fe]}$. From \citep[][ their Fig. 11]{THT09}, we can see that for the range of $[Fe/H]$ in our grid, $[\alpha/Fe]$ is mostly under 0.1. This implies that for most clusters in our sample the assumption is adequate and any difference should be within the uncertainties.  As an example of extreme values, we consider the cluster Messier~11  (also catalogued as NGC~6705) which is known to be $\alpha$-enhanced with $[\alpha/Fe] \sim 0.13 - 0.17$ \citep{M11-study}, which would give $[M/H] \sim [Fe/H] + 0.07$, so there may be some effect for the older clusters. As it will be shown later, apart from a few cases, most of the clusters in this work have $[Fe/H] < 0.5 $.

In this work, as in  \citet{Monteiro2020} we used priors in distance, [Fe/H] and $A_V$, adopting the prior probability for each parameter given by $P({\bf X}) = \prod_{n=0}^{n}P(X_n)$. 
For distance we use $\mathcal{N}(\mu,\,\sigma^{2})$ obtained with Bayesian inference from the parallax ($\varpi$) and its uncertainty ($\sigma_{\varpi}$) and the variance ($\sigma^2$) is obtained from the distance interval calculated from the inference using the uncertainty as $1\sigma_{\varpi}$. The prior in $A_V$ is also adopted as a normal distribution with $\mu$ and variance ($\sigma^{2}$) for each cluster taken from the 3D extinction map produced by \citet{3Debv}\footnote{The 3D extinction map is available online at \url{https://stilism.obspm.fr/}}. The prior for [Fe/H] was estimated from the Galactic metallicty gradient published by \citet{OCCAMgradient20}. 
For age there is no prior so we adopt $P(X_n) = 1$.

In our code we adopt a likelihood function given in the usual manner for the
maximum likelihood problem as
\begin{equation}
\mathcal L (D_N|{\bf X}) = \prod_{i=1}^{N}\Phi(I({\bf X}),D_N) \times  P_{i},
\label{eq:likelihood}
\end{equation}
where ${\bf X}$ is the vector of parameters ($Av$, distance $d$, age $log(age)$ and $[Fe/H]$)
that define the maximum $\mathcal L$ and $N$ the number of randomly generated, independent sets of model parameters. $P_{i}$ is the membership probability of the star.

The likelihood function above is used to define the objective function
  ($S({\bf X}) = -log( P({\bf X}) \times \mathcal L (D_N|{\bf X}) )$) of the optimization algorithm that define a given
isochrone ${\bf I_N}$. The optimization is done with respect to N.

Finally, our method does not take into account the differential extinction in clusters.
The grid of PARSEC isochrones does not include ages younger than $\sim$ 4 Myr ($log(age) = 6.60$ dex) and therefore is not suited for pre-main sequence evolutionary phases where this effect is expected to be large. 

In any case, when we look at the results obtained for clusters with known differential extinction, we get good agreement with the literature values. We have 18 clusters from DAML catalog for which we know there is differential extinction. When comparing the results for these objects we find the following
mean and standard deviation differences: distance = ($ -0.056 \pm 0.428$) kpc, $log(age)$ =  ($0.21 \pm 0.42$) dex, $Av$ =  ($0.03 \pm 0.29$) mag and $[Fe/H]$ = ($ 0.14 \pm 0.17$) dex.

Considering the comparison between our results and those from DAML there is no indication of any systematic trend in $Av$ assuring that both sets agree within the uncertainties.
The biggest discrepancies are in age which is expected since all of the most discrepant objects are young clusters with the typical problems of poorly defined turn-off, poor membership estimation, and small number of members.

In Table \ref{tab:photometric} we present a portion of the results obtained by the CE.

\begin{table*}
\caption[]{A portion of the fundamental parameters obtained from the isochrone fits is given.  
 }
\label{tab:photometric}
\begin{center}
\begin{tabular}{lccccrccc}
\hline
   Name                       &    $dist$     & $\sigma_{dist}$  &     $log(age)$          & $\sigma_{log(age)}$ &                \multicolumn{1}{c}{$[Fe/H]$}        &               $\sigma_{[Fe/H]}$          &                  $A_{V}$         &                $\sigma_{A_{V}}$      \\
\hline     
                              &  [pc]          &  [pc]          &        [dex]         &       [dex]     &       [dex]        &  [dex]       &       [mag]       & [mag]    \\  
\hline
Alessi~1                      &     681        &      15         &       8.993        &     0.045      &        -0.109      & 0.132        &        0.416      &       0.150       \\   
Alessi~10                     &     445        &       9         &       8.547        &     0.131      &         0.019      & 0.076        &        0.496      &       0.066       \\   
Alessi~12                     &     539        &      10         &       8.195        &     0.127      &        -0.023      & 0.079        &        0.248      &       0.087       \\   
\hline
\end{tabular}
\end{center}
\end{table*}

\section{Internal uncertainties}

\begin{figure}
\centering
\includegraphics[scale = 0.23]{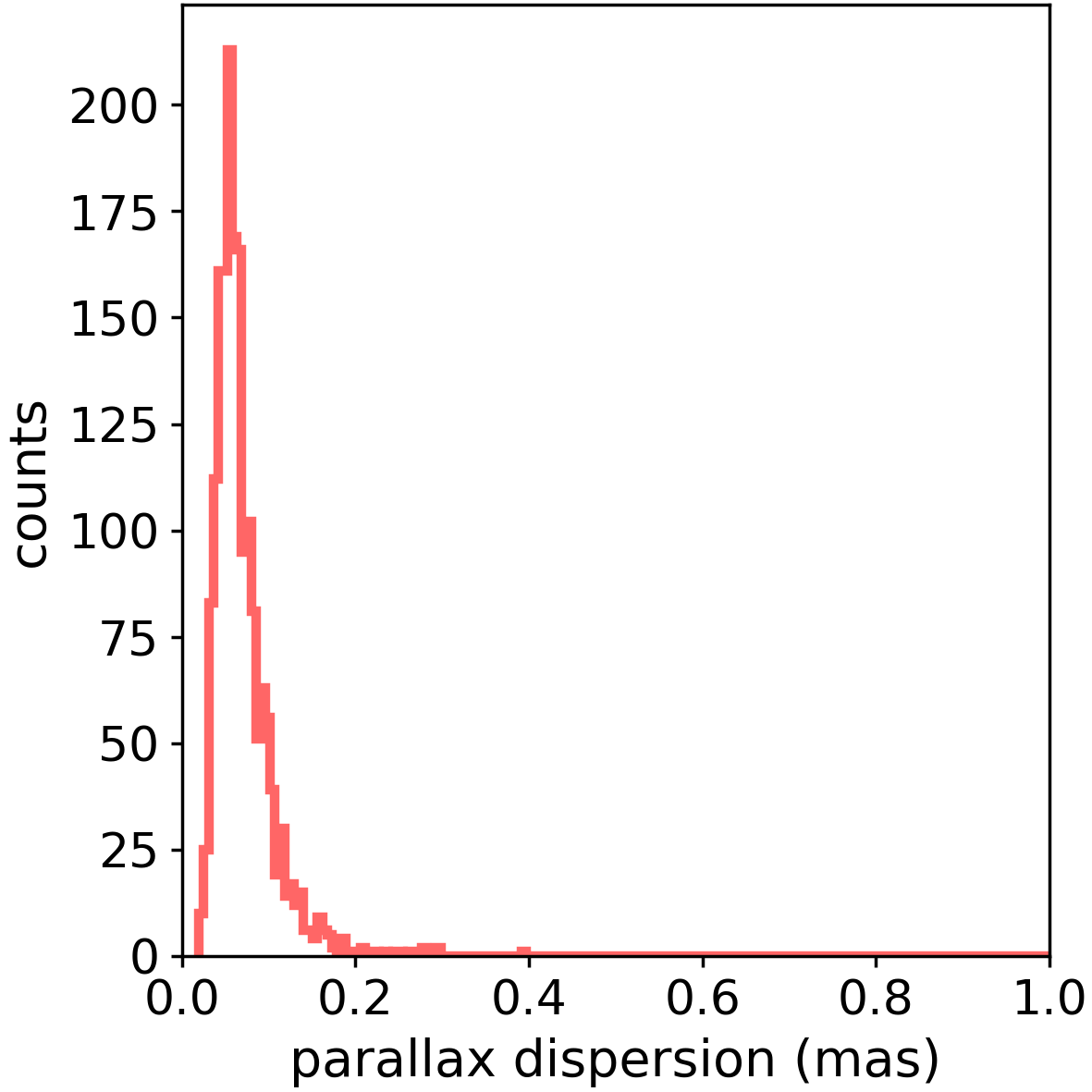}
\includegraphics[scale = 0.23]{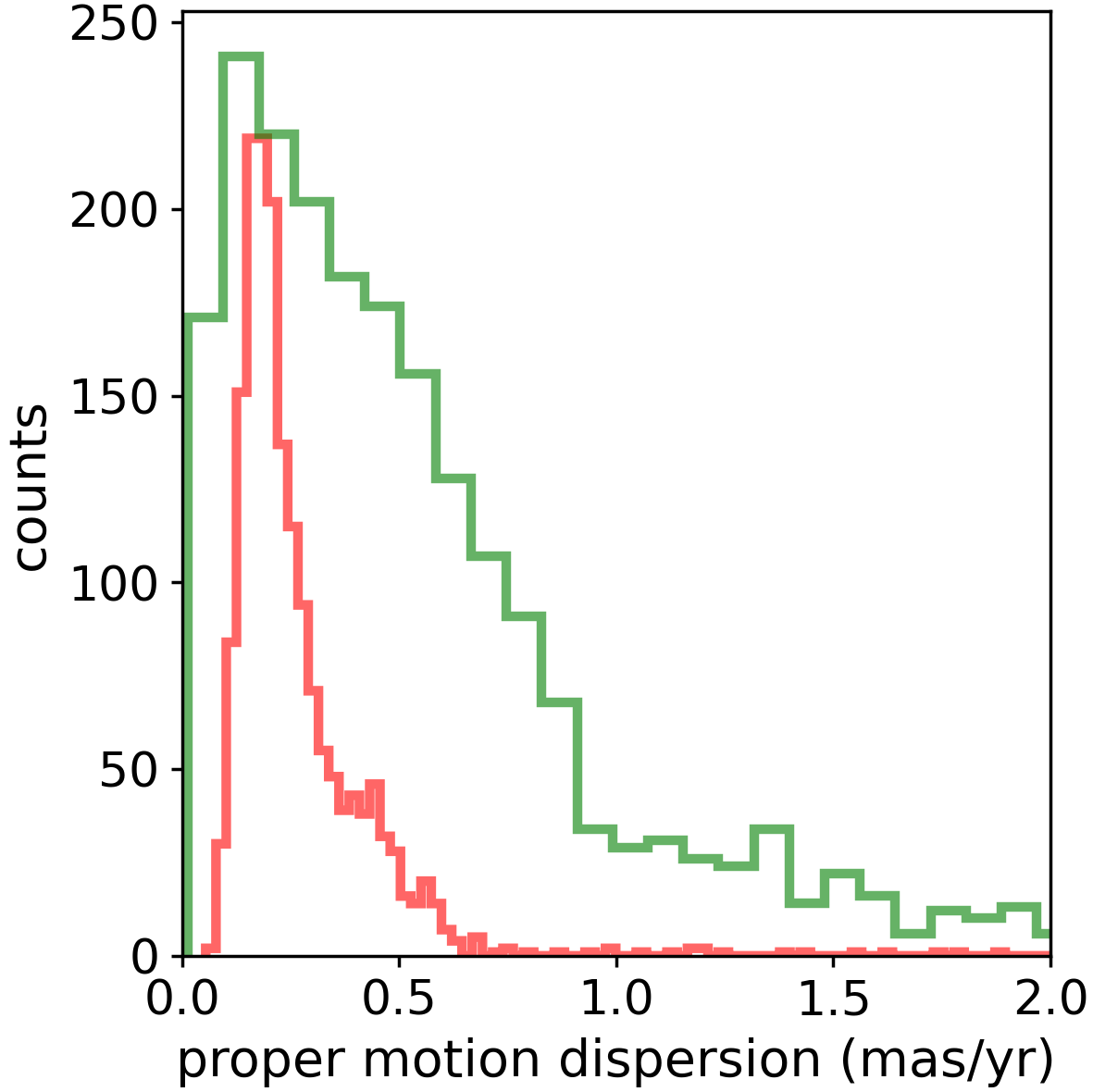}
\includegraphics[scale = 0.23]{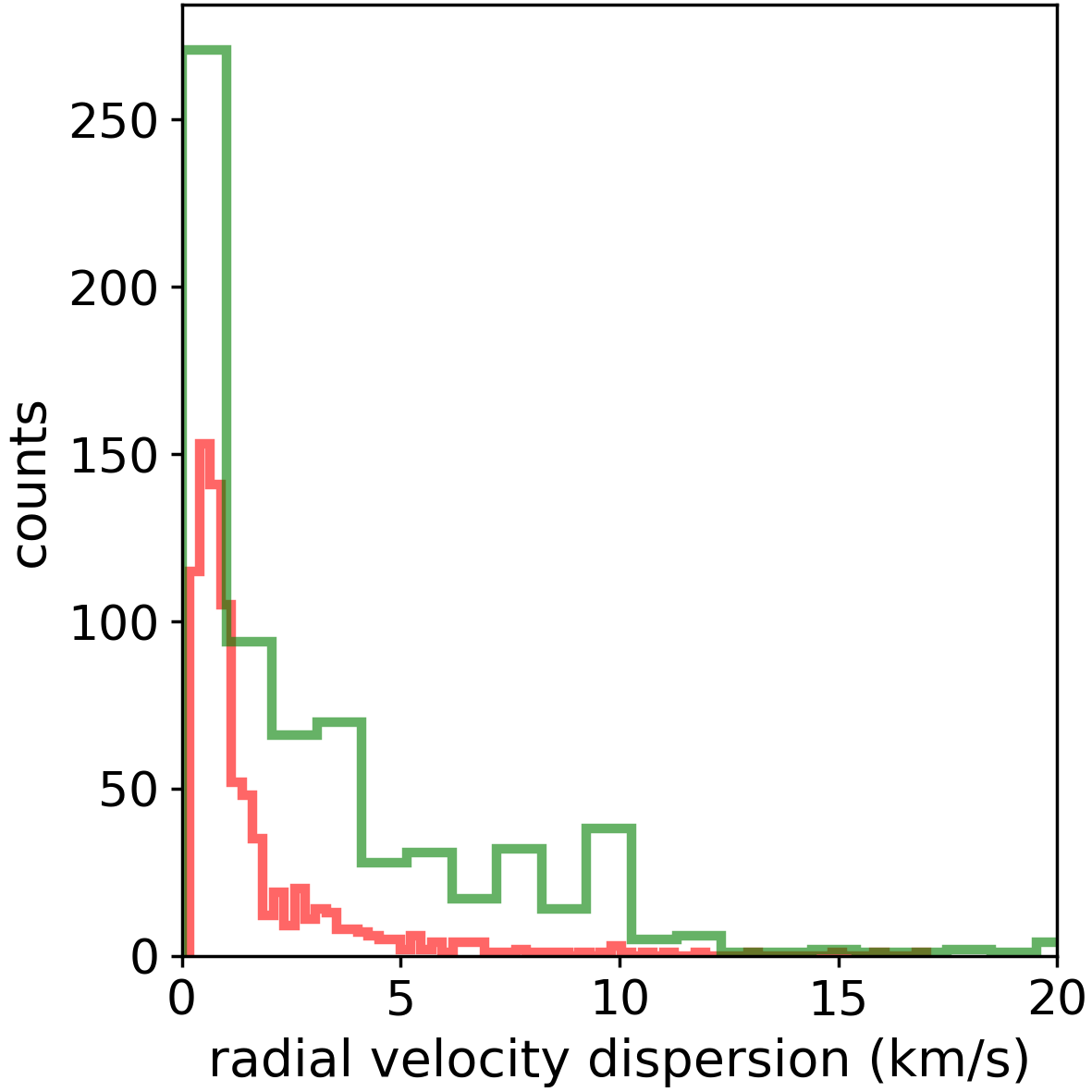}
\caption{Distribution of the uncertainty of the parallax, proper motion and radial velocity determined in this study (red color). The uncertainty of proper motion and and radial velocity distributions are compared with those from DAML which are presented in green.}
\label{fig:astrometric-uncertainty}
\end{figure}

Figure \ref{fig:astrometric-uncertainty} shows the distribution of the uncertainties in mean parallax, mean proper motion and mean radial velocity, determined in this study. The average uncertainties for the sample are about 0.28 mas\,yr$^{-1}$ in proper motion, 0.07 mas in parallax and 1.60 km\,s$^{-1}$ in radial velocity. The figure also displays the distributions of the corresponding uncertainties listed in DAML, showing that this work brings a clear improvement.

The uncertainties of the parameters determined from the isochrone fitting, were estimated applying a Monte-Carlo re-sampling of the member stars with replacement (bootstrapping). The synthetic clusters were also re-generated in each run. The final parameters and their uncertainties were estimated as the mean and one standard deviation of ten runs. 

\begin{figure}
\centering
\includegraphics[scale = 0.23]{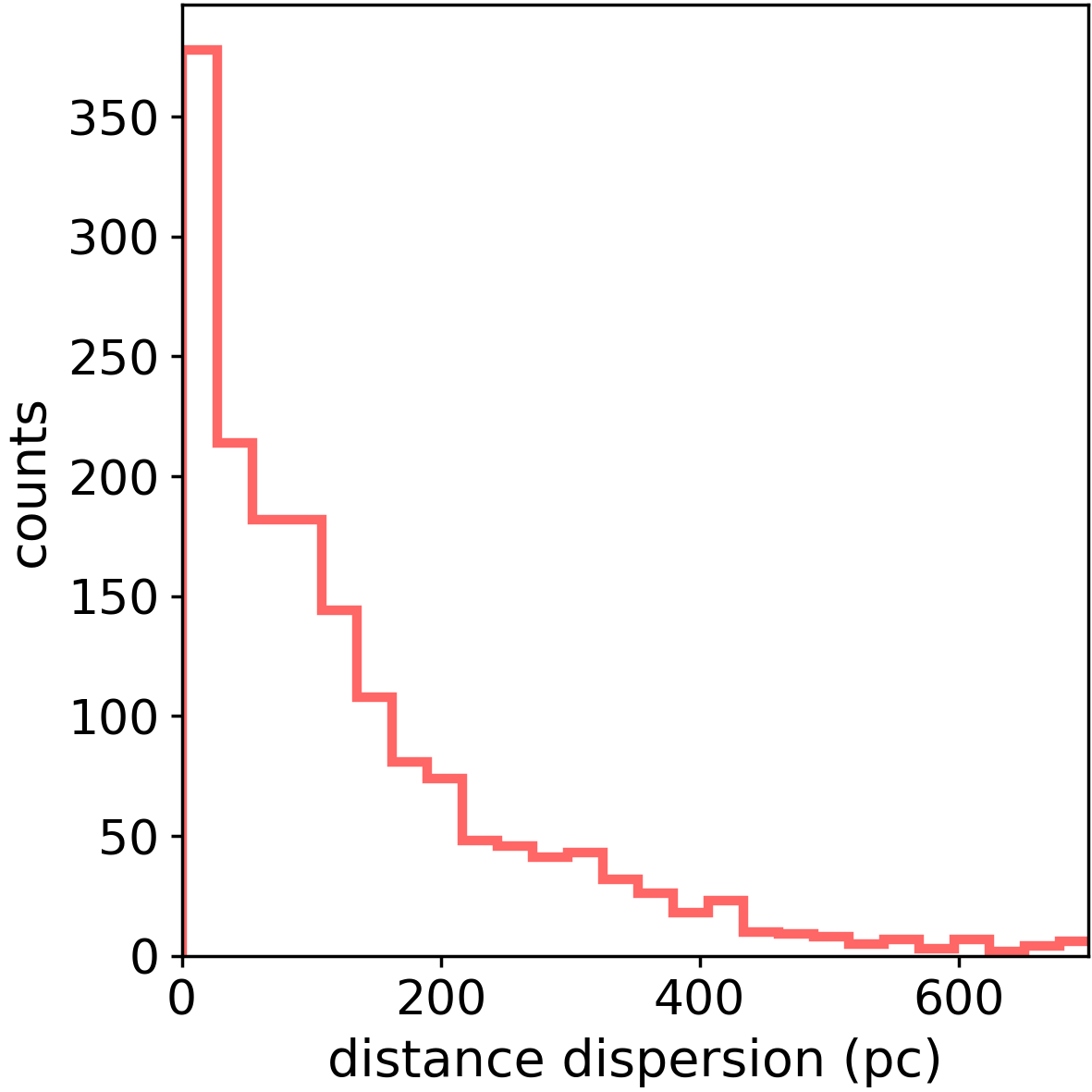}
\includegraphics[scale = 0.23]{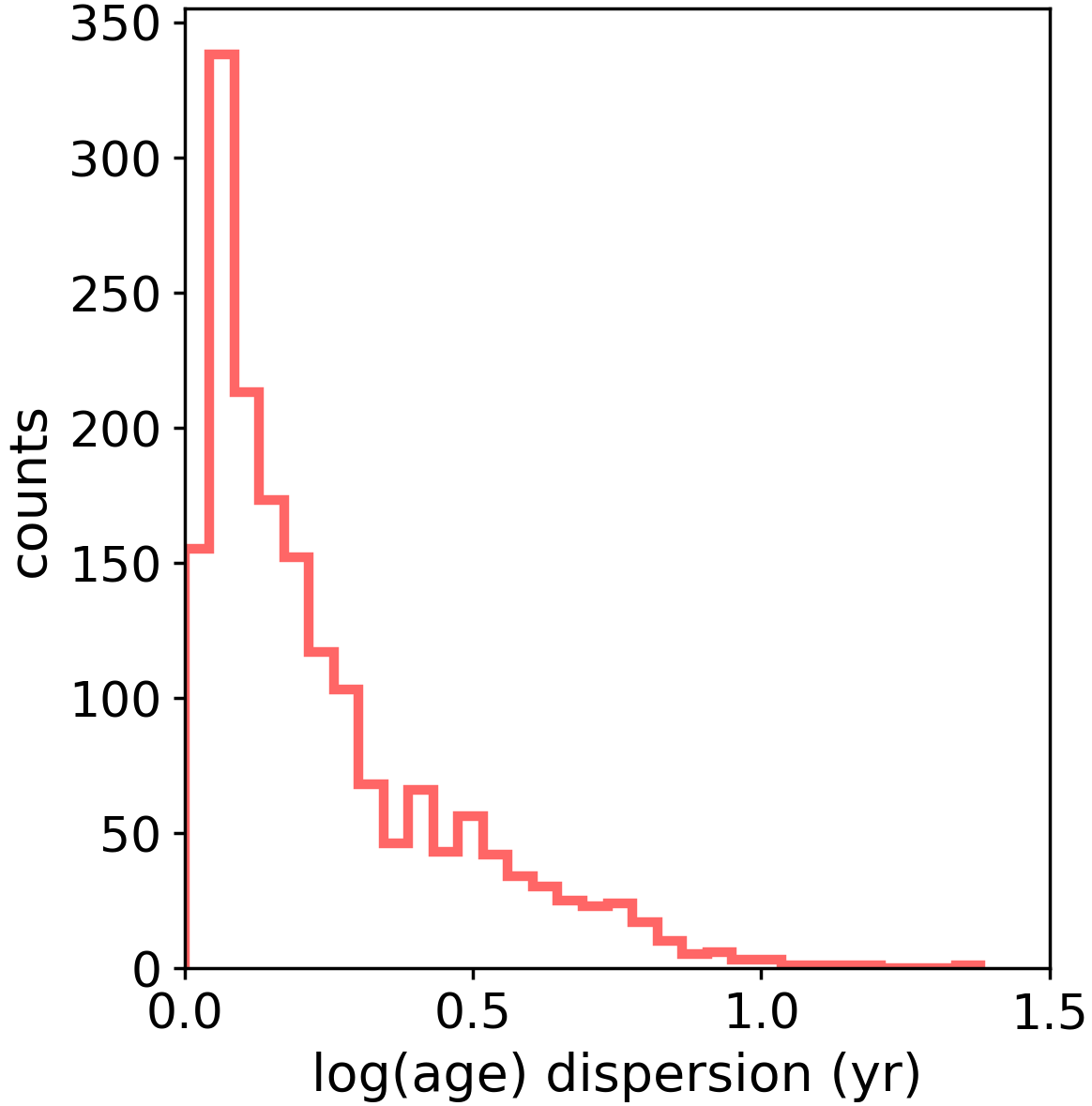}
\includegraphics[scale = 0.23]{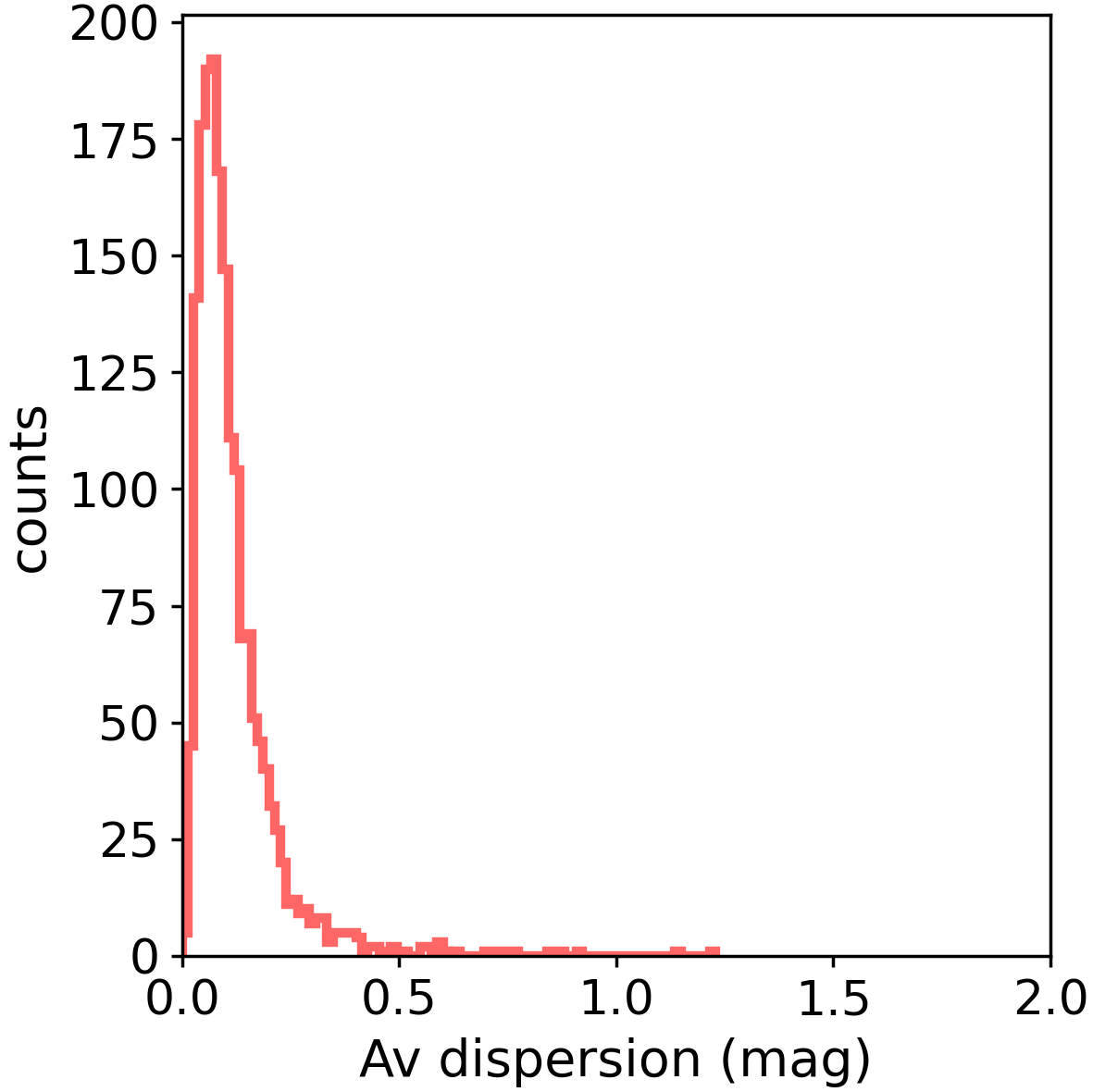}
\caption{Distribution of the uncertainty of the distance, log(age) and $A_V$ determined in this study.}
\label{fig:phtometric-uncertainty}
\end{figure}

\begin{figure*}
\centering
\includegraphics[scale = 0.32]{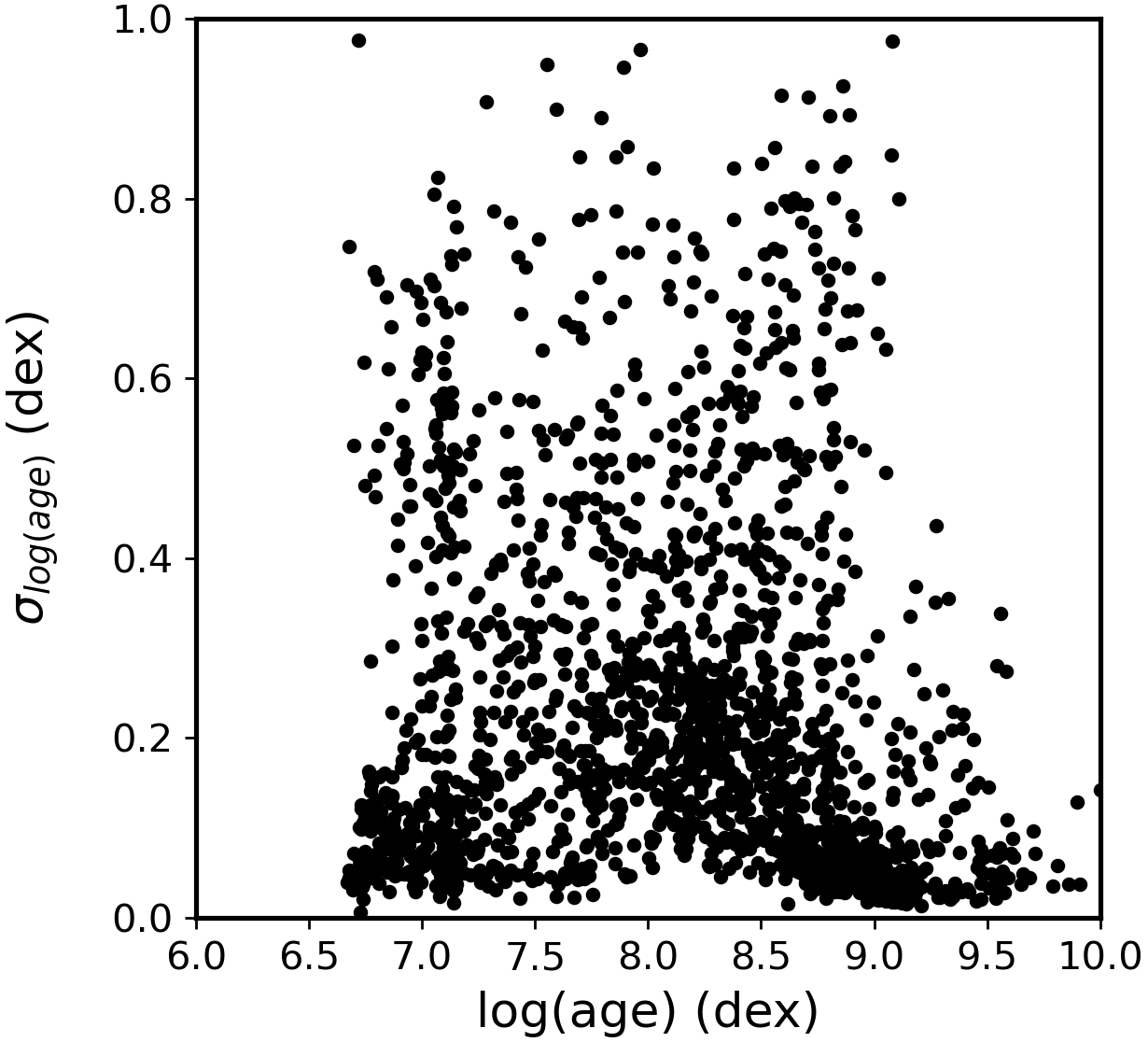}
\includegraphics[scale = 0.32]{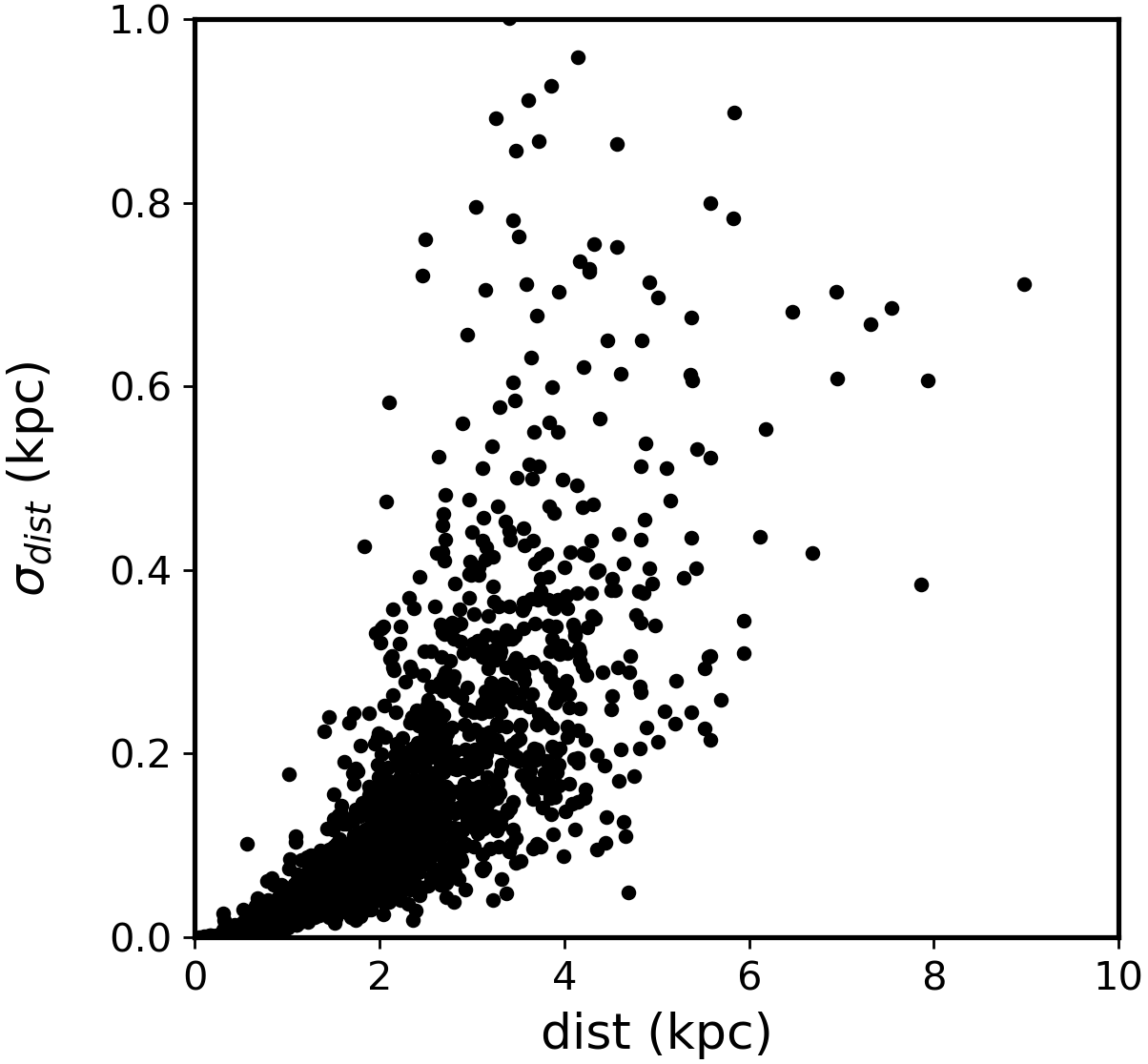}
\includegraphics[scale = 0.32]{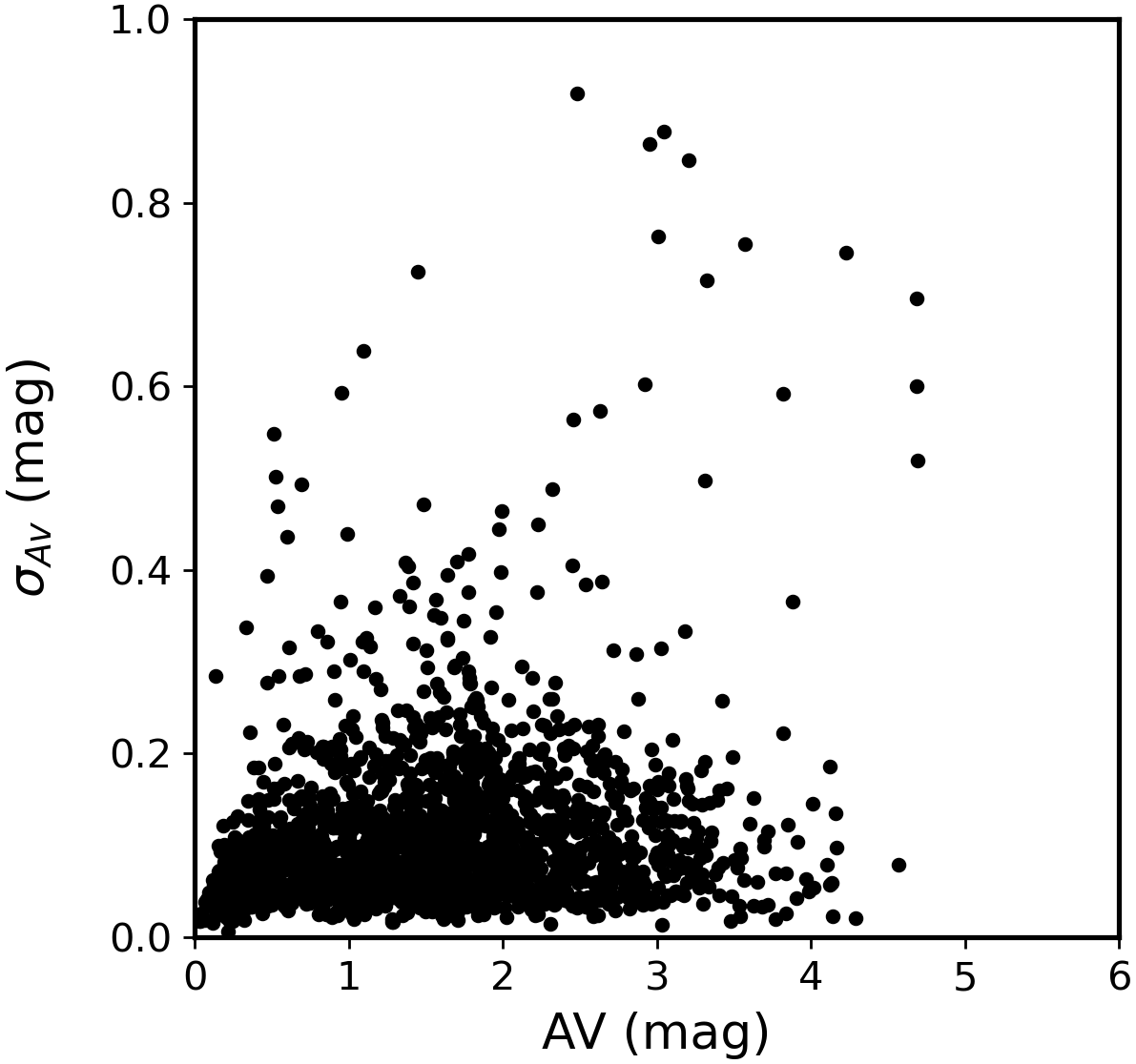}
\includegraphics[scale = 0.32]{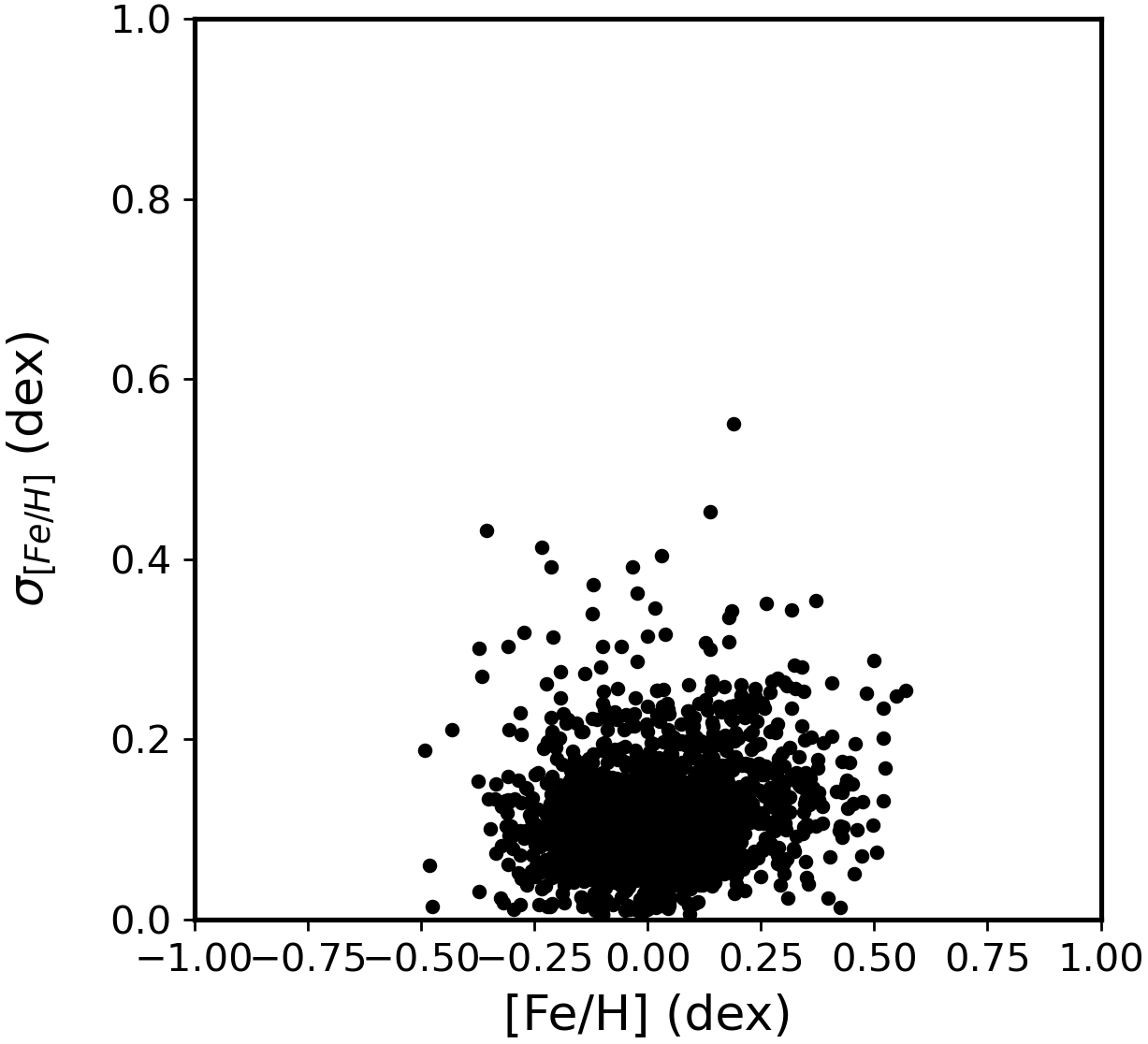}
\caption{Distribution of the internal uncertainties for the parameters from isochrone fit determined by the cross-entropy method.}
\label{fig:errors}
\end{figure*}

The distribution of the uncertainties of the derived parameters are given in Figure \ref{fig:phtometric-uncertainty}. Figure \ref{fig:errors} presents the internal uncertainties as a function of the parameters. 
The derived parameters and their uncertainties are given in Table~\ref{tab:photometric}, available electronically.

\section{The effect of the priors} \label{sec:priors}

\begin{figure*}
\centering
\includegraphics[scale = 0.42]{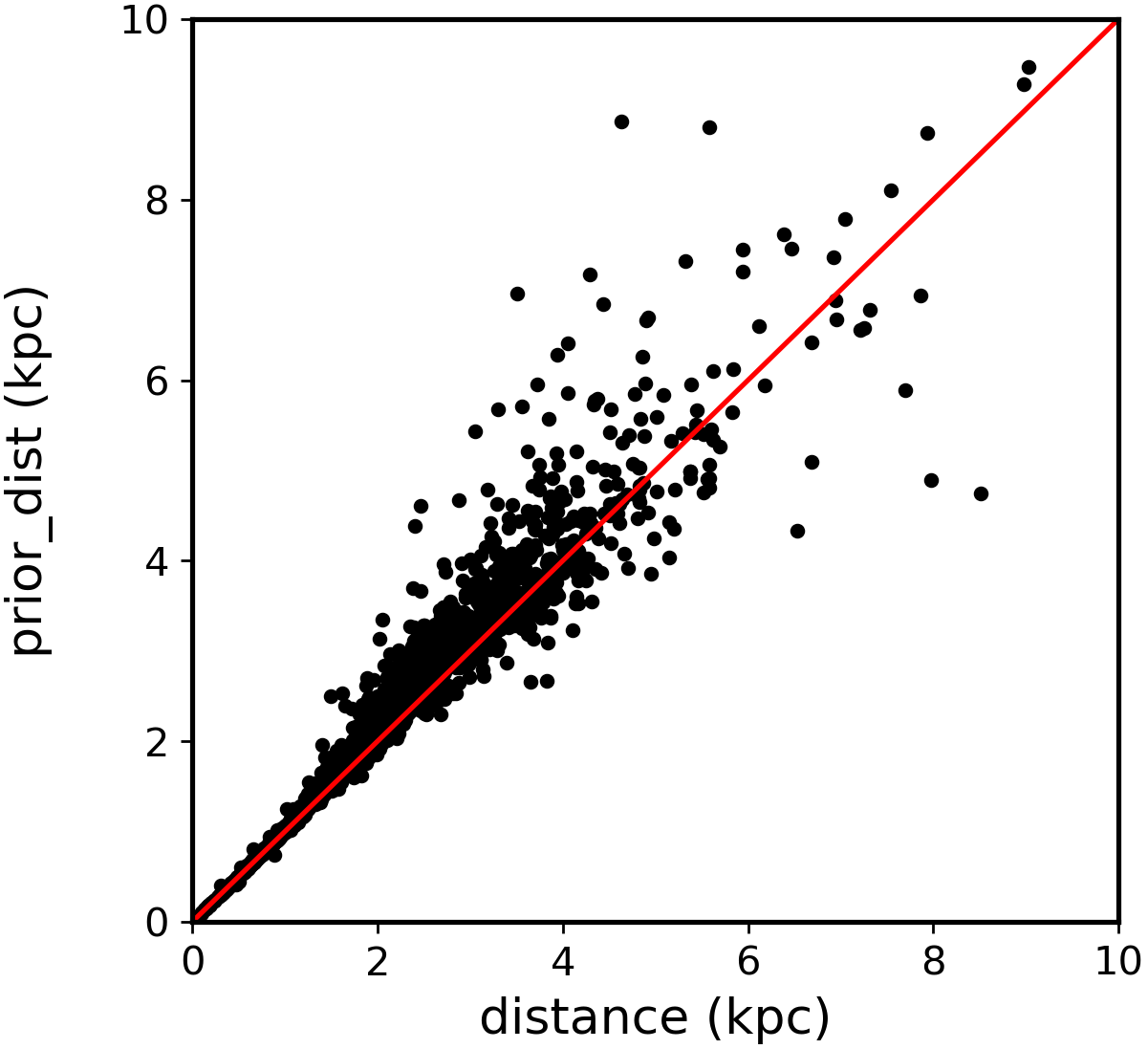}    
\includegraphics[scale = 0.42]{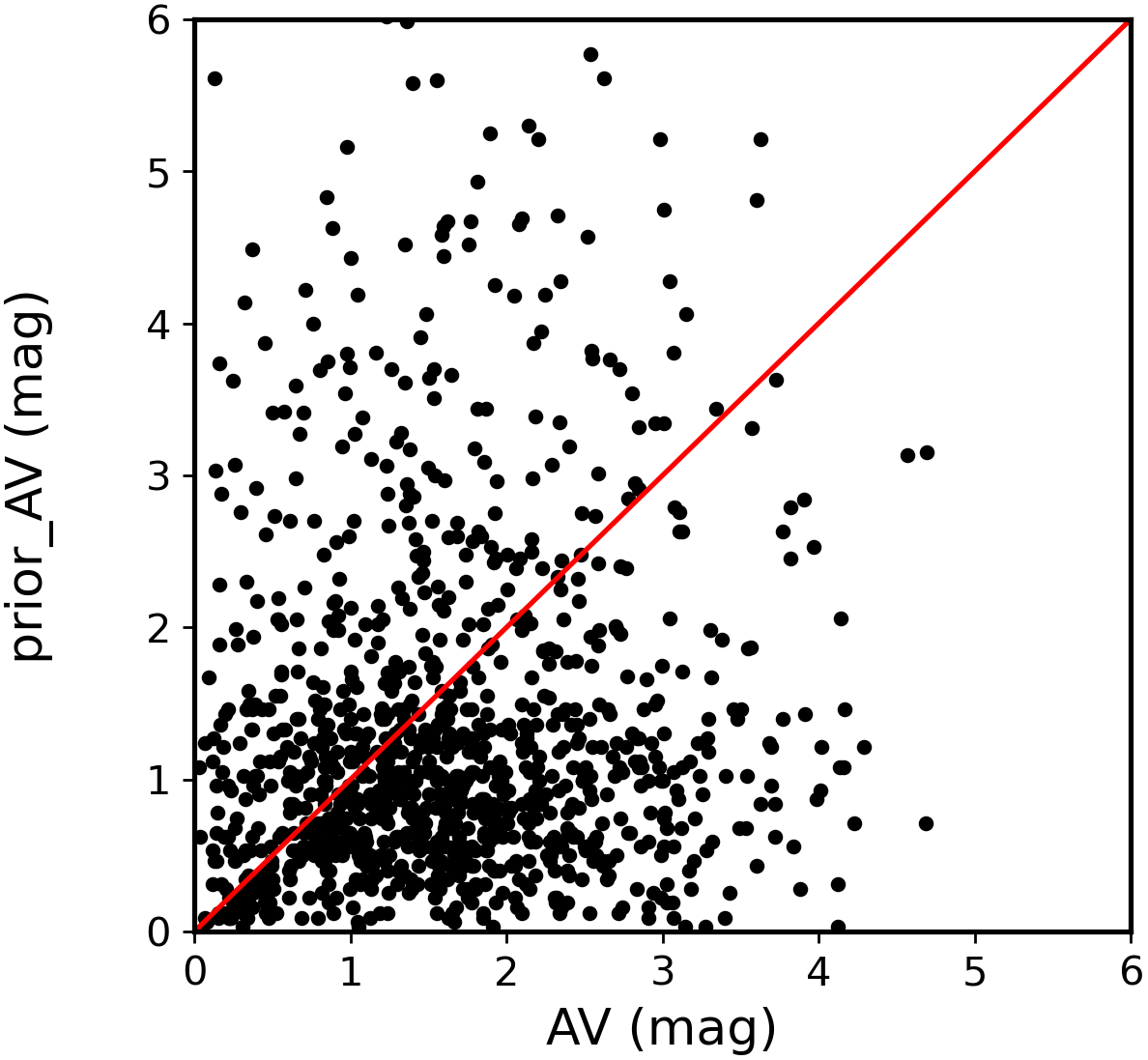}    
\includegraphics[scale = 0.42]{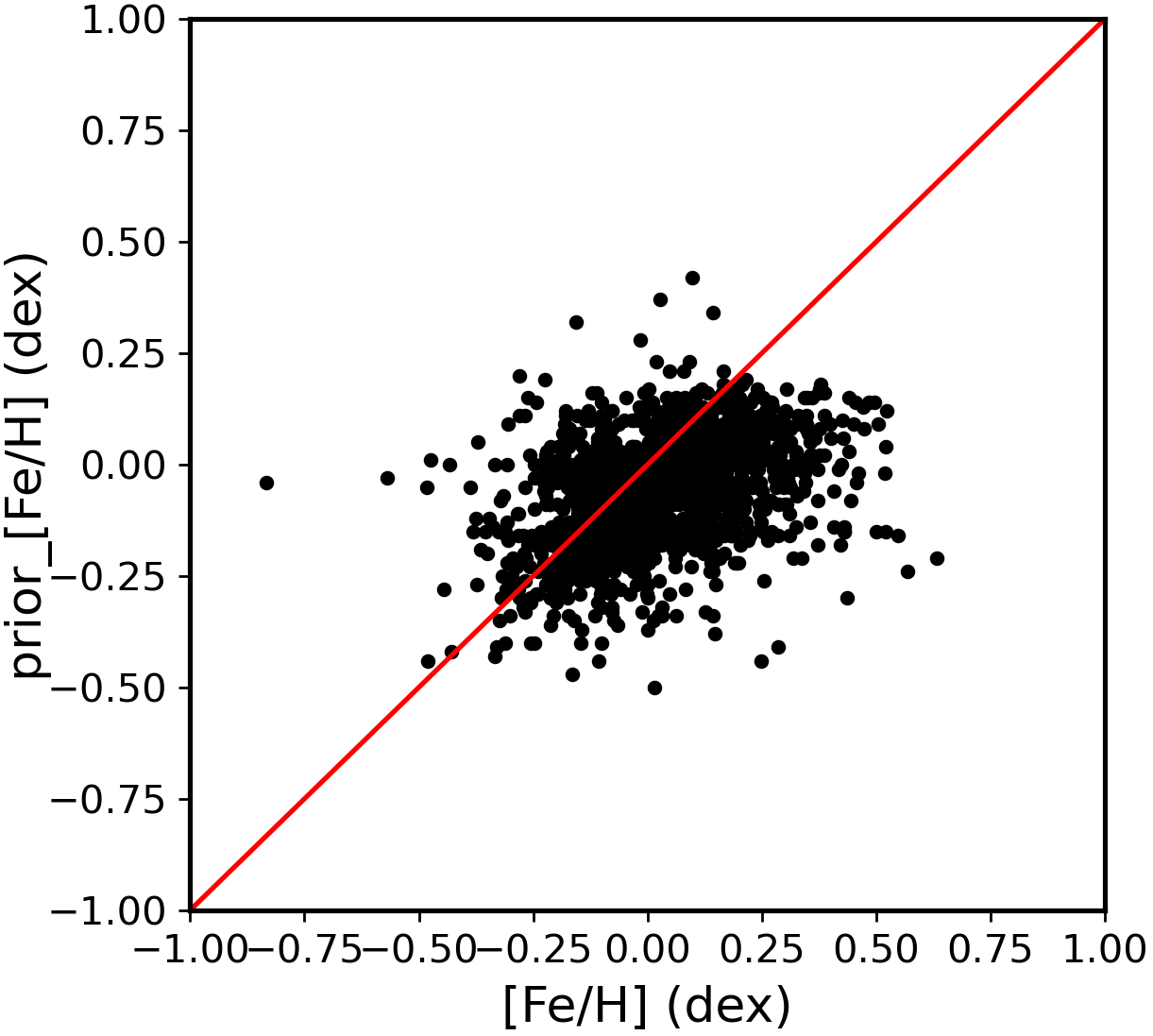}
\caption{Comparison of the values of the parameters obtained from the isochrone fit with the values of the priors used.}
\label{fig:priorsxvalues}
\end{figure*}

To assess how much the priors constrain the results we took the sample of 67 clusters with metallicity determinations for high resolution spectroscopy compiled by \citet[][ table A.1]{Monteiro2020}.
The sample covers practically the full range in distance, age and $A_V$ as well in metallicity.

We start by comparing the values of the parameters obtained from the isochrone fit with the values of the priors. 
Figure \ref{fig:priorsxvalues} shows how the prior in the distance is very restrictive, which reflects our trust on the quality of Gaia parallaxes. The prior in [Fe/H] is moderately restrictive. Because the fits are less sensitive to metallicity, this avoids solutions with extreme values of metallicity while allowing for reasonable uncertainty in the metallicity gradient. Finally the prior in the $A_V$ is shown to be clearly non restrictive, which reflects our lower trust in the extinction model as well as our high trust that the data is sensitive enough to this parameter as verified in  \citet{Monteiro2020}. These considerations are also expressed in the correlation coefficients between the parameters obtained from the isochrone fits and the values of the priors, which are 0.95, 0.68 and 0.30 for distance, [Fe/H] and $A_V$, respectively.

\begin{figure*}
\centering
\includegraphics[scale=0.25]{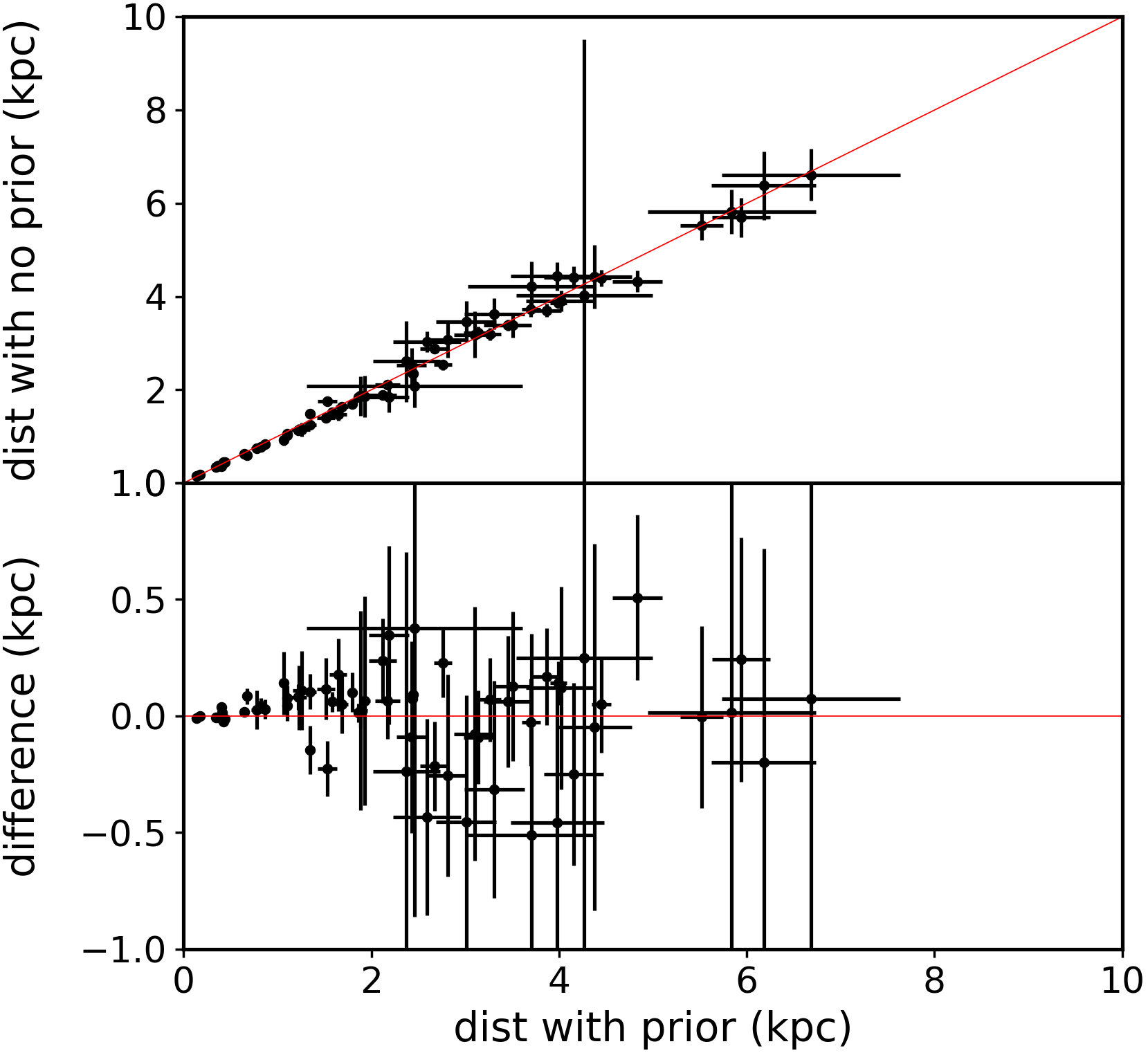}  
\includegraphics[scale = 0.25]{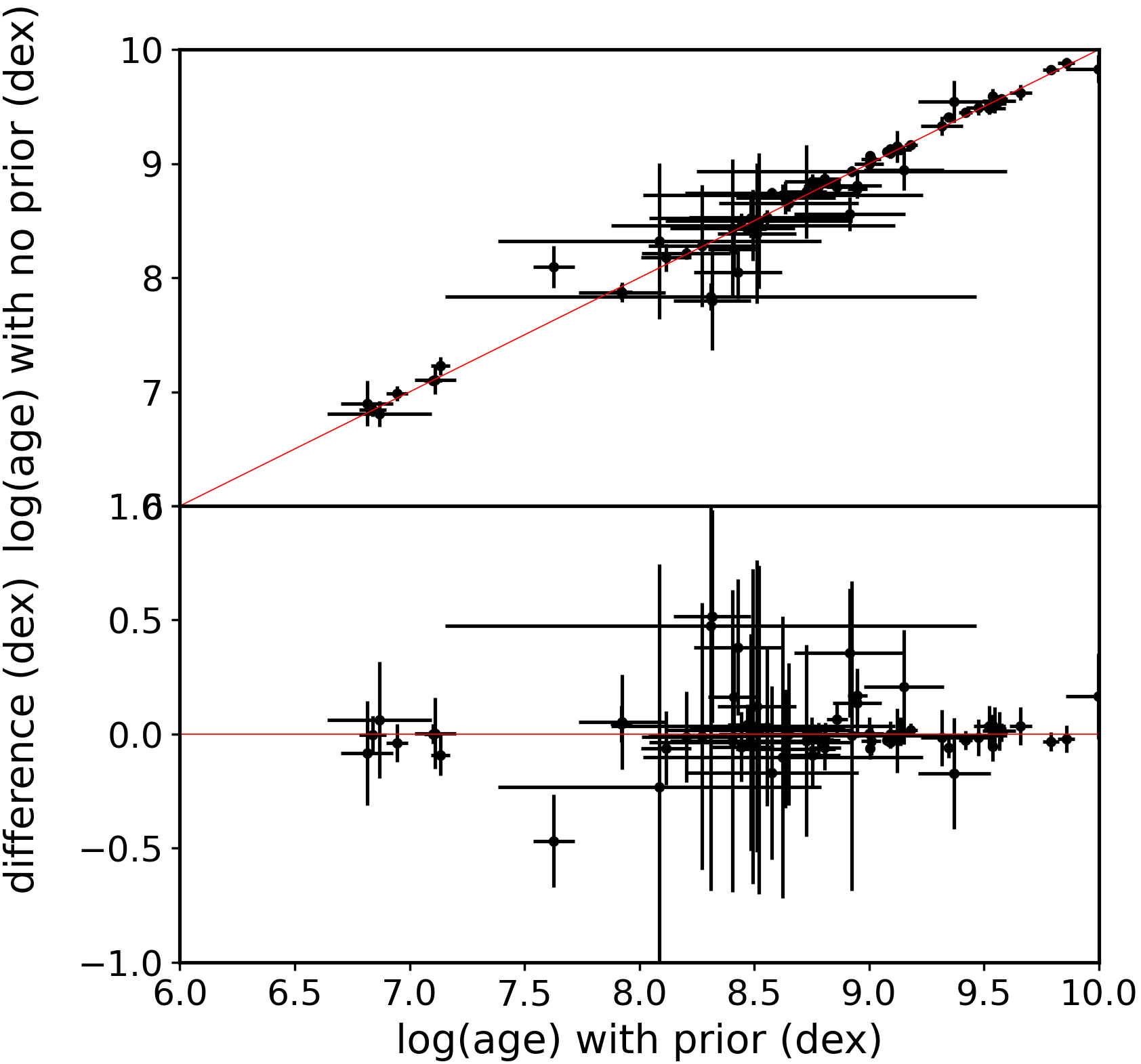} 
\includegraphics[scale = 0.25]{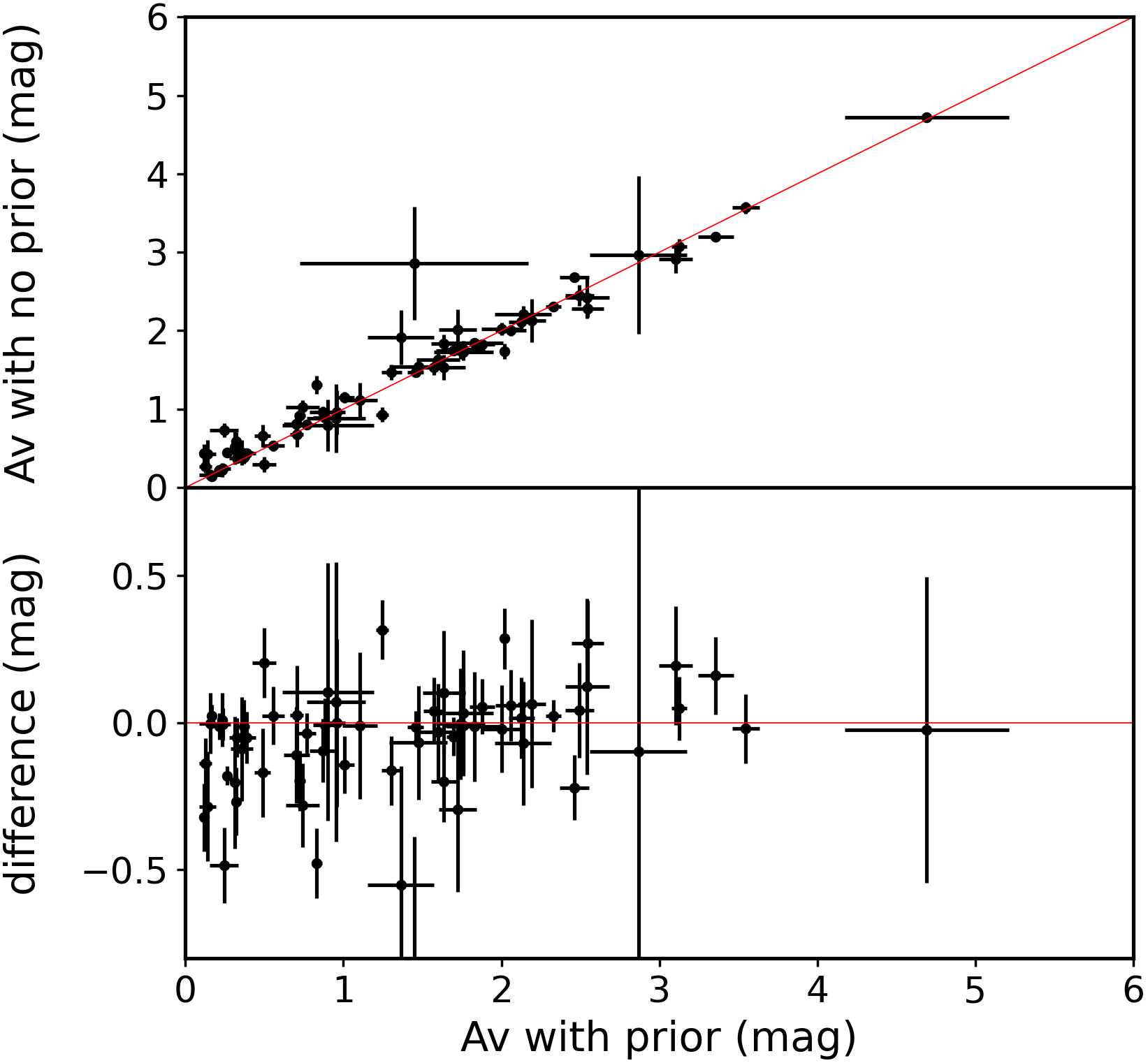}
\includegraphics[scale = 0.25]{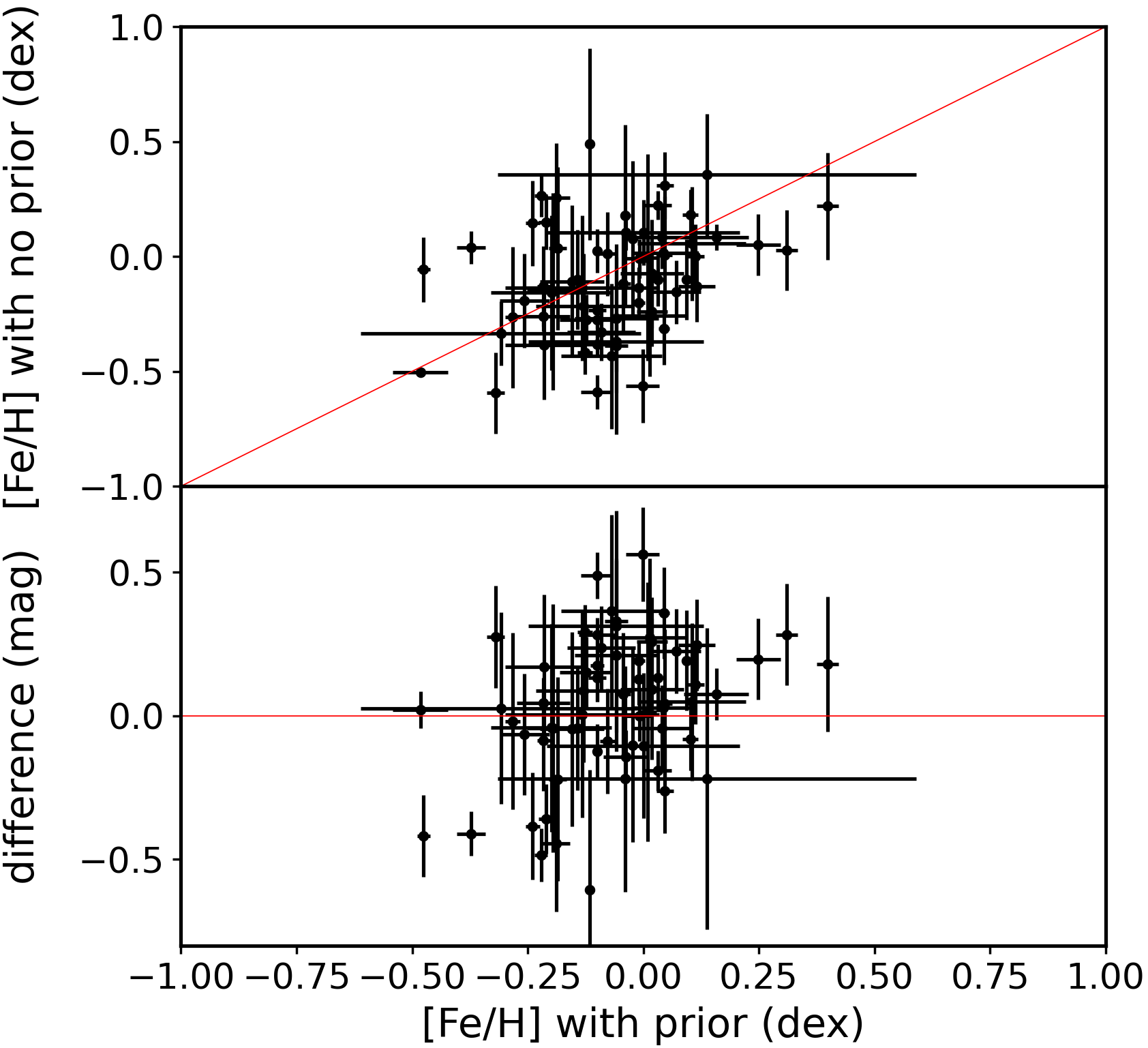}
\caption{Comparison of the values of the parameters obtained from the isochrone fit with and without priors.}
\label{fig:priorsxnoprios}
\end{figure*}

We now present in Figure \ref{fig:priorsxnoprios} the comparison of the parameters from the isochrone fits with and without the adopted priors.
The correlation of the parameters obtained with and without priors are 0.99, 0.98, 0.97 and 0.33 for distance, age, $A_V$ and [Fe/H], respectively.

The differences between the parameters determined with priors and with no priors are in agreement considering the
errors as pointed by the mean differences 0.010 kpc in distance, 0.014 dex in log(age),  -0.069 mag in $A_V$ and 0.032 dex in [Fe/H] with standard deviation of 0.187 kpc, 0.142 dex, 0.234 mag, 0.237 dex.

\begin{figure*}
\centering
\includegraphics[scale=0.36]{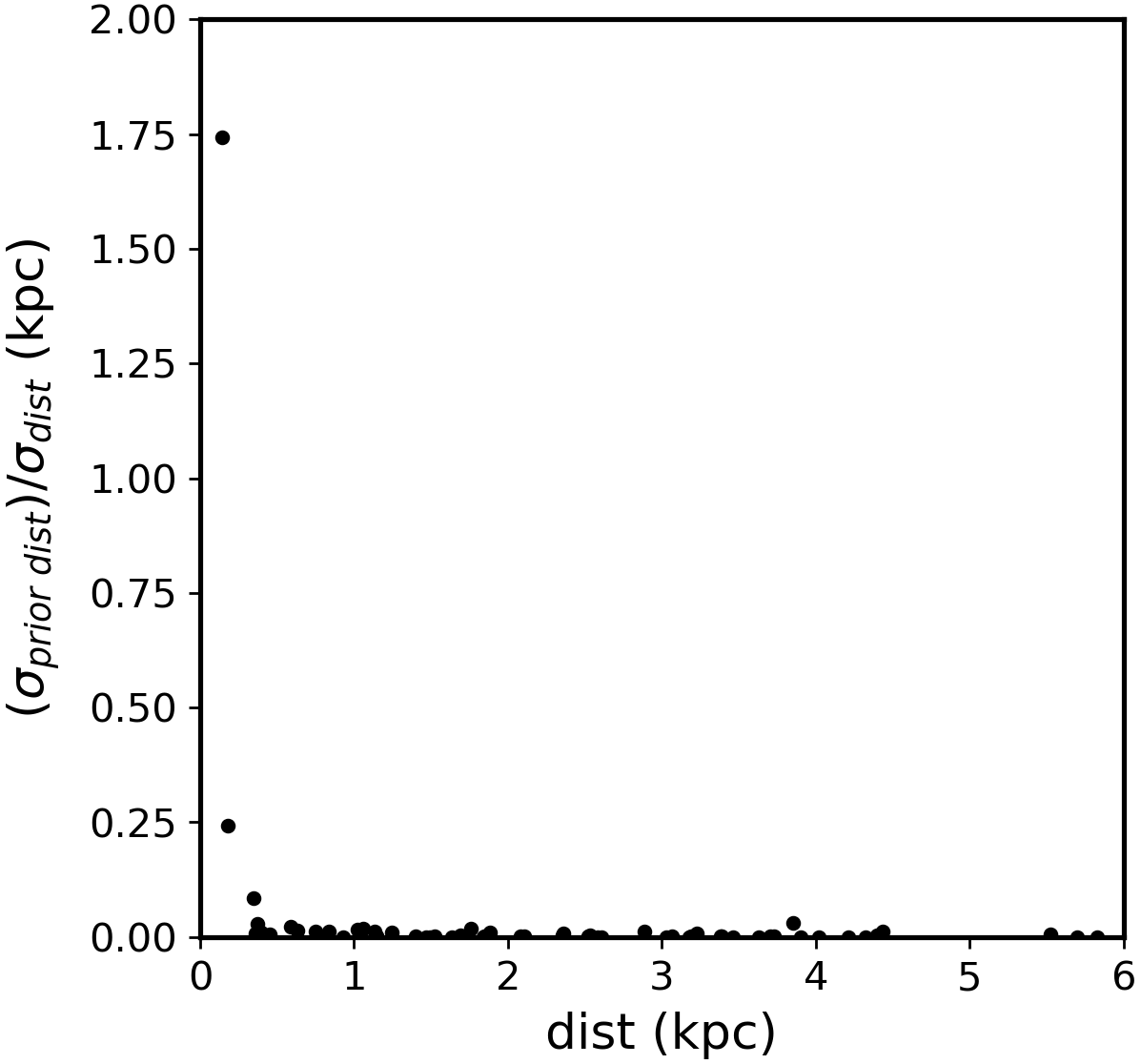}  
\includegraphics[scale = 0.36]{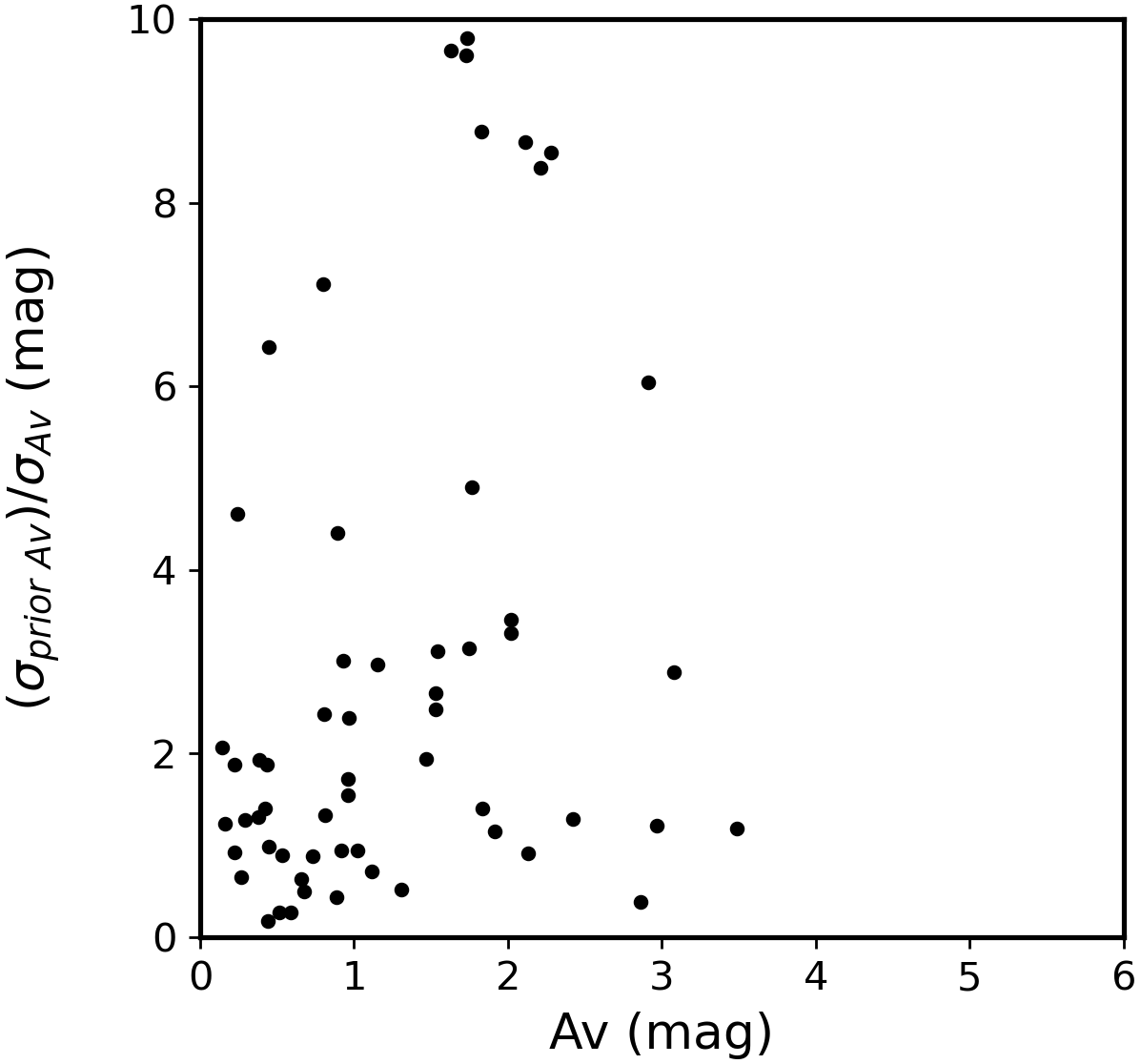} 
\includegraphics[scale = 0.36]{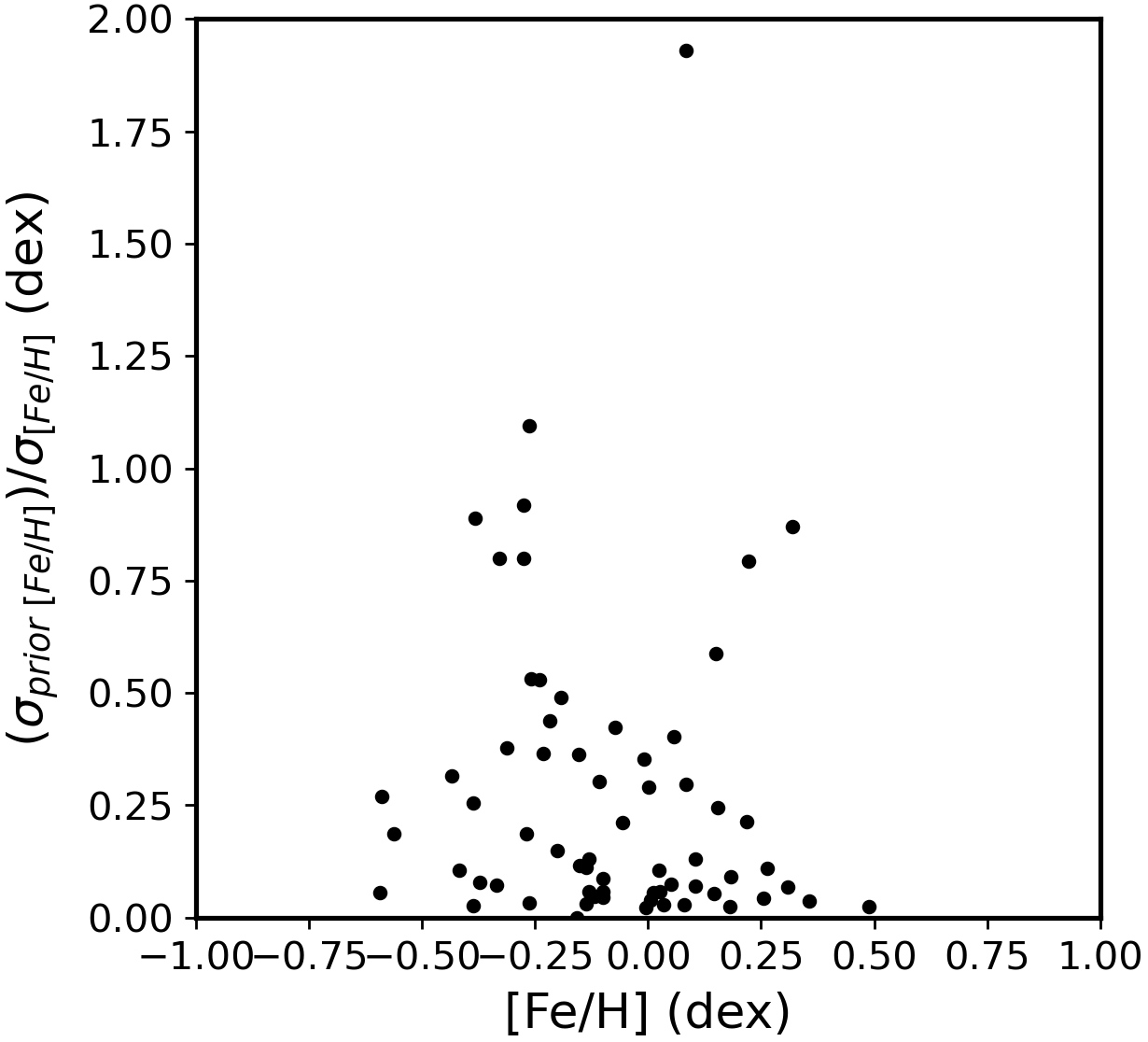}
\caption{Ratio of the width of the prior distribution, $\sigma_{prior}$, to the error of the respective parameter: $\sigma_{dist}$ (left), $\sigma_{Av}$ (middle), $\sigma_{[Fe/H]}$ (right).}
\label{fig:razao}
\end{figure*}

Another indication of how much the priors constrain the results is given by the ratio of the width of the prior distribution, $\sigma_{prior}$, to the error of the respective parameter ($\sigma_{dist}$, $\sigma_{Av}$, $\sigma_{[Fe/H]}$) estimated from the isochrone fit with no priors. These ratios are presented in Figure \ref{fig:razao}.
Note that the errors in the parameters estimated without prior are essentially a measure of the dispersion of the likelihood distribution. In this sense we can say that priors restrict results if the ratio is much less than the likelihood dispersion. In other words, a ratio $\sigma_{prior-parameters}/\sigma_{parameter}$  greater than 1 means that a prior is not very restrictive and smaller than 1 that it is more restrictive.

\section{The control sample}

In the analysis of UBVRI data the U band clearly improves the reliability of the parameters estimated for the open clusters, since it removes the spectral-type/reddening degeneracy in the color$-$color diagrams.
In this way the agreement of the results derived from {\it Gaia} DR2, in this study using $G_{BP}$ and $G_{RP}$ magnitudes, with those from the literature obtained with $U$ filter is a quality indicator of the efficiency in the determination of cluster's parameters.

To perform this check we use high quality homogeneous UBVRI photometry of 20 open clusters published by \citet{Moitinho2001}.

Each star observed by \citet{Moitinho2001} was cross-matched with those from \citet{CantateAnders} for assigning {\it Gaia} based memberships to the UBVRI data. 
In the procedure we used the SKY algorithm from TOPCAT \citep{topcat} to cross-match the data with coordinates in the J2000 equinox and same epoch.
Note that the different number of stars in each sample is due to the smaller field of view of the ground-based observations.

To fit the isochrones to the UBVRI data we used our code and the same procedure adopted for the {\it Gaia} isochrone fitting described in Section~5, including the prior in distance and [Fe/H] and adopting the reddening law from \citet{CCM89} as used in \citet{Dias2018}.
So, the method applied to both data samples is as similar as possible.

\begin{figure*}
\centering
\includegraphics[scale = 0.45]{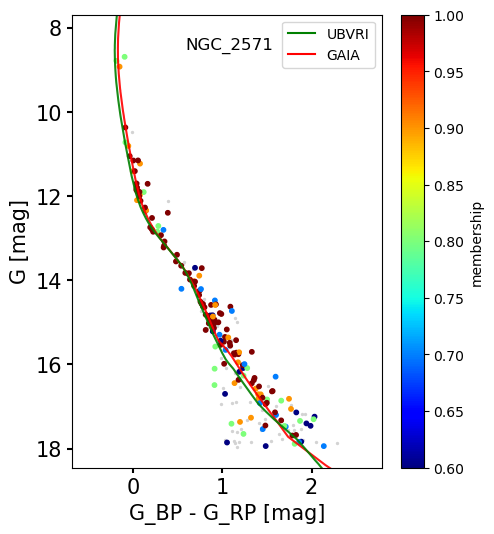}
\includegraphics[scale = 0.45]{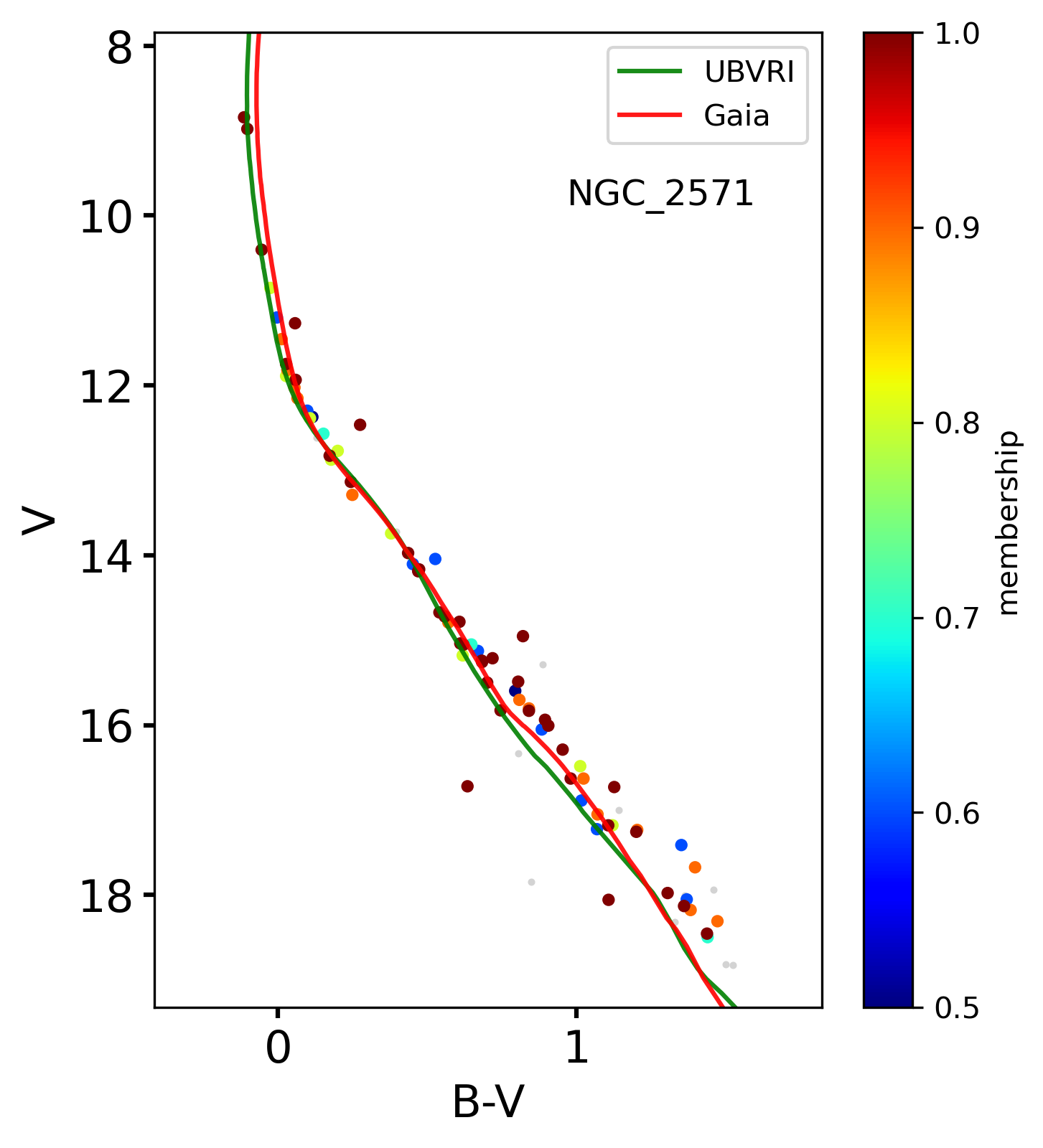}
\includegraphics[scale = 0.45]{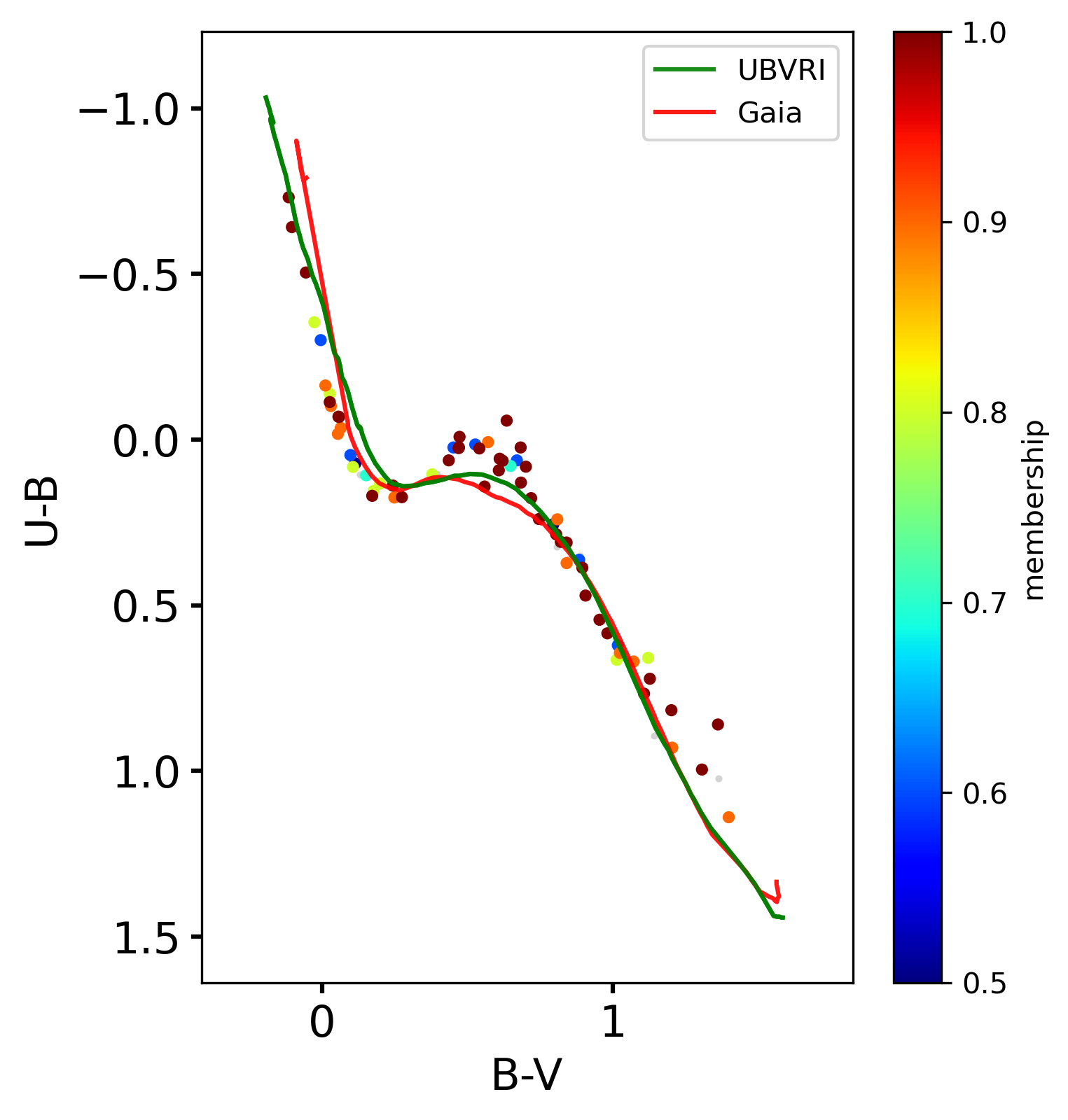}
\caption{Comparison of the parameters obtained from {\it Gaia} data (red isochrones) and from UBVRI data (green isochrones) for NGC 2571. The memberships from \citet{CantateAnders} are proportional to the color in the sense of redder indicating higher membership probability.}
\label{fig:NGC2571}   
\end{figure*}

\begin{figure*}
\centering
\includegraphics[scale = 0.45]{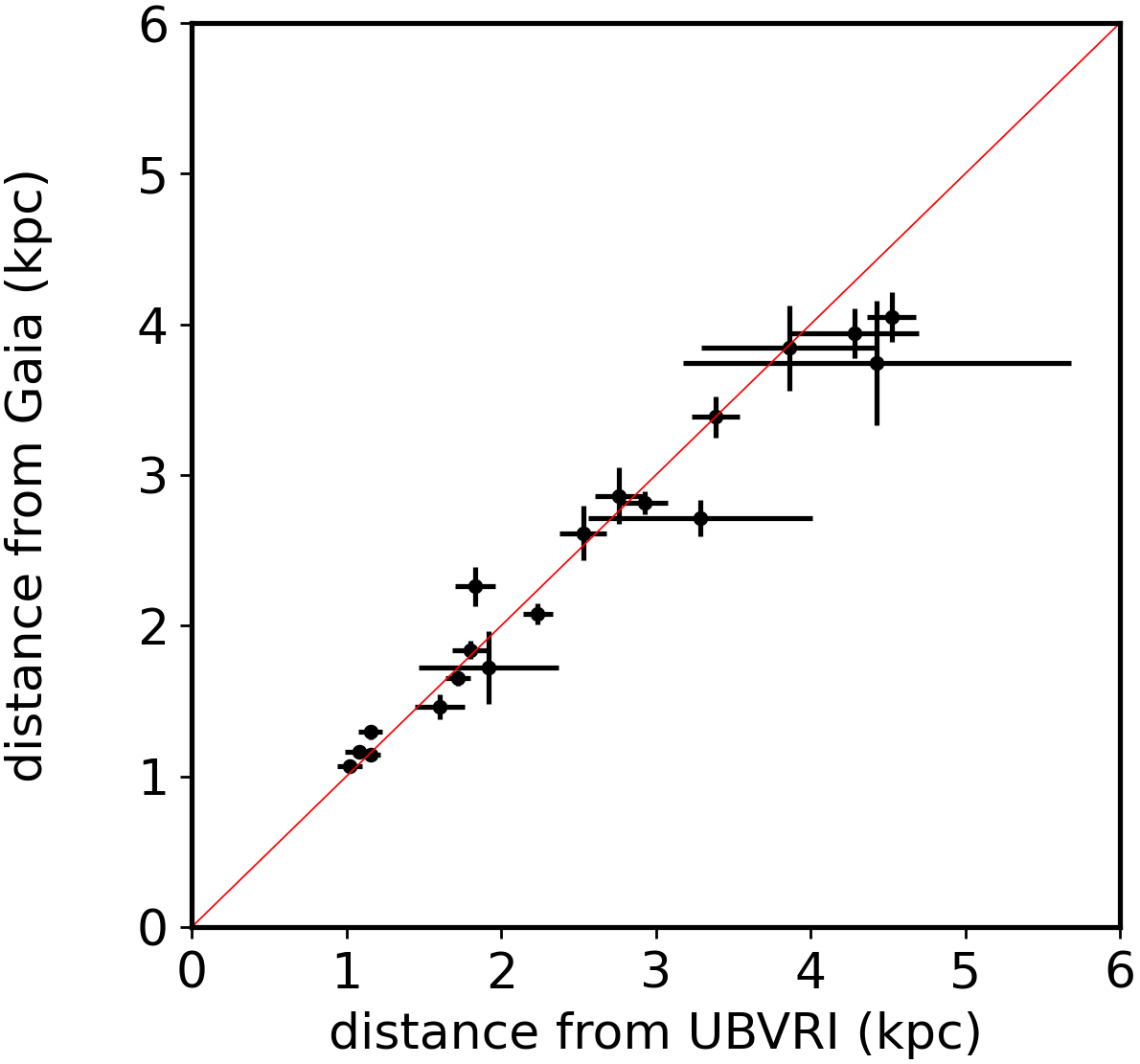}
\includegraphics[scale = 0.45]{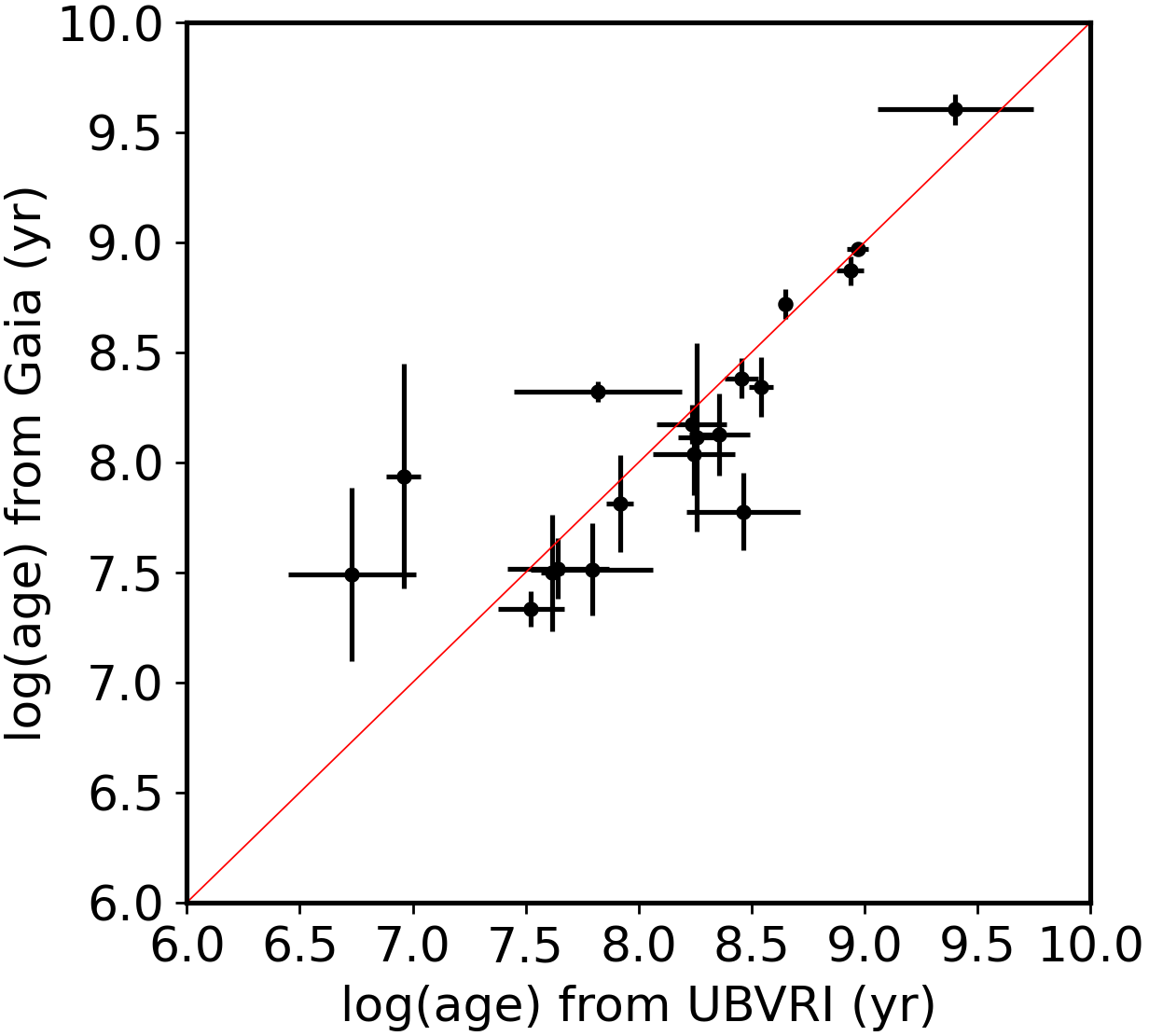}
\includegraphics[scale = 0.45]{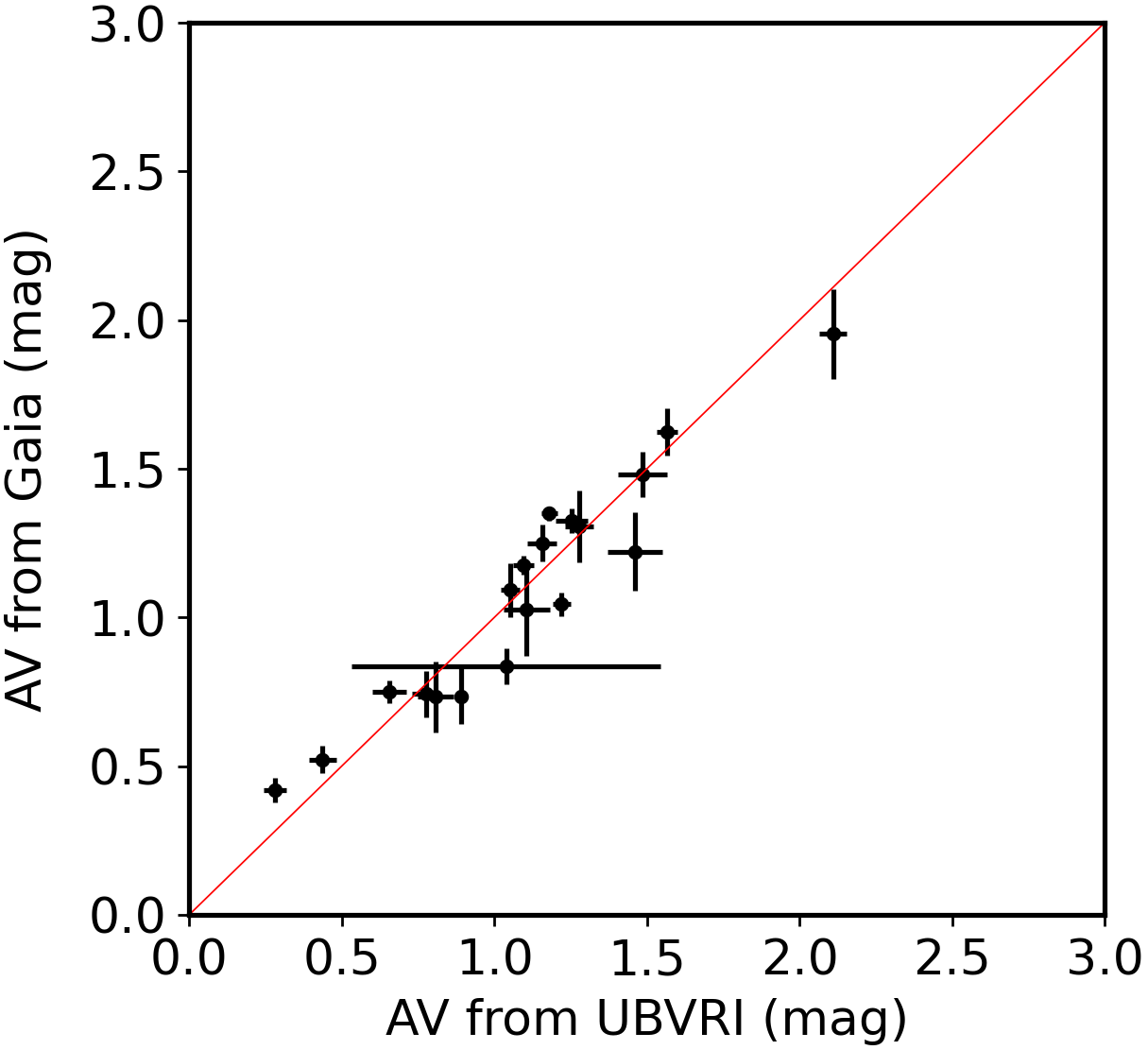}
\caption{Comparison of the parameters obtained from {\it Gaia} $G_{BP}$, and $G_{RP}$ and UBVRI data for a sample of 20 open clusters from \citet{Moitinho2001}.}
\label{fig:control}
\end{figure*}

In Figure \ref{fig:NGC2571} we present the CMDs and color$-$color diagram for NGC~2571 as examples of the results obtained both using {\it Gaia} $G_{BP}$ and $G_{RP}$ and UBVRI data. 
In Figure \ref{fig:control} and Table \ref{tab:results} we show the comparison of the values of the parameters obtained with our code applied to the {\it Gaia} and UBVRI data.

The differences between the parameters determined from UBVRI data and from {\it Gaia} data are in agreement at a 3$\sigma$ level and considering the errors indicated by the mean differences (0.005 dex in log(age), -59 pc in distance, -0.007 mag in $A_{V}$ and 0.11 dex in $[Fe/H]$)  with root mean square difference (0.302 dex in log(age), 218 pc in distance, 0.099 mag in $A_{V}$ and 0.16 dex in $[Fe/H]$), in the sense of results from {\it Gaia} data minus UBVRI data. 

However, we can notice some differences at the 1$\sigma$ level. The most discrepant cases are clusters with poorer fits in {\it Gaia} DR2 data.
The difference in distance and/or age of OCs Haffner~19, NGC~2311, NGC~2367, NGC~2401 occurs due to the poorly defined main sequence (MS) and/or turn-off (TO) in {\it Gaia} DR2 data. The differences in distance for NGC~2425 and NGC~2635 can be explained by the differences in $A_{V}$ and metallicity.  For the distant clusters NGC~2401 and Haffner~19 the poorer parameters are in line with what is expected from the internal uncertainties shown in Figure~\ref{fig:errors}. NGC~2367 is a very young cluster with sparse and poorly defined sequence. 

The $[Fe/H]$ values are presented for comparison purposes and should be considered with caution. As noted in \citet{Monteiro2020} and in Section~\ref{sec:priors}, the  isochrone fits employ a moderately restrictive metallicity gradient prior. Thus, while individual metallicities are reasonable estimates, as a group they may be biased towards the Galactic metallicity gradient used as prior. For $A_{V}$, the results are hardly affected by the prior which is mostly non-restrictive.

The results for the clusters in this control sample from \citet{Moitinho2001}, together with the extensive validation performed in \citet{Monteiro2020} indicate that the parameters here obtained with {\it Gaia} $G_{BP}$ and $G_{RP}$ data are reliable. Still, it should be kept in mind that the analysis has been performed with a small number of clusters, with different numbers of stars used in each data set due to the different sizes of the fields. We note that the results of the isochrone fit depend on the number of stars used and clearly in the {\it Gaia} data there are many more stars than in the UBVRI data considered in this comparison. On the other hand, UBVRI photometry can better constrain the results since it is not so affected by degeneracies in the parameter determinations.
We also tested the isochrone fitting with UBVRI data using no prior in distance and $[Fe/H]$. The results of the parameters agree within the estimated errors showing no statistical distinction.

In the next Sections we present more evidence supporting that the results obtained in this work are reliable and improve upon published ones that used {\it Gaia} DR2 data .

\begin{table*} 
  \caption{Parameters obtained for 20 clusters with UBVRI data from \citet{Moitinho2001}. Columns two to six list the parameters determined using {\it Gaia} $G_{BP}$, and $G_{RP}$ data.
Columns seven, eight and nine list the values obtained from UBVRI data. }
\label{tab:results} 
  \begin{tabular}{lcccccccc}
    \hline\hline
              & \multicolumn{4}{c}{Results from {\it Gaia} data}                      & &\multicolumn{3}{c}{Results from UBVRI data} \\
    {Cluster}          & {Distance}     & {log(Age)}      & {$A_{v}$}      &   {$[Fe/H]$}      &  {Distance}        & {log(Age)}        & {$A_{v}$}          & {$[Fe/H]$}            \\
    \hline
             &    [pc]        & [dex]             &[mag]             & [dex]               &  [pc] &              [dex]                  &   [mag]             &   [dex]  \\    
\hline
Haffner~16   & $2862\pm186 $  & $7.513\pm0.210$  & $0.750\pm0.037$  & $-0.160\pm0.127$    &  $2759\pm154    $   &    $7.794\pm0.271$  & $0.656\pm0.055$     &   $-0.258\pm0.113$       \\
Haffner~19   & $3745\pm413 $  & $7.938\pm0.510$  & $1.221\pm0.131$  & $-0.210\pm0.159$    &  $4428\pm1254   $   &    $6.959\pm0.076$  & $1.462\pm0.090$     &   $-0.473\pm0.492$       \\
NGC~2302     & $1162\pm 27 $  & $8.038\pm0.186$  & $0.733\pm0.119$  & $-0.025\pm0.043$    &  $1084\pm096    $   &    $8.245\pm0.183$  & $0.806\pm0.059$     &   $-0.338\pm0.127$       \\
NGC~2309     & $2076\pm 71 $  & $8.343\pm0.136$  & $1.326\pm0.041$  & $-0.111\pm0.073$    &  $2235\pm095    $   &    $8.542\pm0.053$  & $1.254\pm0.053$     &   $-0.143\pm0.069$       \\
NGC~2311     & $1839\pm 62 $  & $7.777\pm0.175$  & $1.092\pm0.091$  & $-0.079\pm0.057$    &  $1801\pm120    $   &    $8.465\pm0.252$  & $1.053\pm0.031$     &   $-0.295\pm0.106$       \\
NGC~2335     & $1462\pm 82 $  & $7.812\pm0.220$  & $1.480\pm0.076$  & $-0.071\pm0.101$    &  $1603\pm163    $   &    $7.917\pm0.061$  & $1.486\pm0.080$     &   $ 0.118\pm0.239$       \\
NGC~2343     & $1068\pm 36 $  & $8.174\pm0.088$  & $0.743\pm0.078$  & $ 0.063\pm0.110$    &  $1018\pm080    $   &    $8.236\pm0.155$  & $0.776\pm0.045$     &   $-0.181\pm0.082$       \\
NGC~2353     & $1140\pm 16 $  & $8.127\pm0.185$  & $0.522\pm0.046$  & $-0.055\pm0.049$    &  $1158\pm058    $   &    $8.357\pm0.134$  & $0.437\pm0.046$     &   $-0.104\pm0.068$       \\
NGC~2367     & $1722\pm244 $  & $7.490\pm0.394$  & $1.025\pm0.156$  & $-0.065\pm0.125$    &  $1918\pm452    $   &    $6.731\pm0.282$  & $1.106\pm0.075$     &   $-0.383\pm0.327$       \\
NGC~2383     & $2815\pm 79 $  & $8.382\pm0.091$  & $1.175\pm0.031$  & $-0.174\pm0.043$    &  $2925\pm154    $   &    $8.454\pm0.074$  & $1.095\pm0.033$     &   $-0.171\pm0.100$       \\
NGC~2401     & $3940\pm165 $  & $8.322\pm0.046$  & $1.044\pm0.040$  & $-0.168\pm0.081$    &  $4281\pm417    $   &    $7.820\pm0.373$  & $1.220\pm0.029$     &   $-0.346\pm0.227$       \\
NGC~2425     & $2714\pm123 $  & $9.604\pm0.071$  & $0.836\pm0.060$  & $ 0.047\pm0.089$    &  $3288\pm726    $   &    $9.403\pm0.344$  & $1.038\pm0.507$     &   $-0.314\pm0.219$       \\
NGC~2432     & $1651\pm 51 $  & $8.971\pm0.031$  & $0.735\pm0.095$  & $ 0.014\pm0.105$    &  $1722\pm080    $   &    $8.970\pm0.047$  & $0.891\pm0.022$     &   $-0.183\pm0.074$       \\
NGC~2439     & $3386\pm136 $  & $7.334\pm0.081$  & $1.350\pm0.024$  & $-0.117\pm0.074$    &  $3385\pm155    $   &    $7.523\pm0.147$  & $1.180\pm0.026$     &   $-0.205\pm0.094$       \\
NGC~2453     & $3842\pm283 $  & $7.498\pm0.265$  & $1.624\pm0.079$  & $-0.171\pm0.063$    &  $3865\pm571    $   &    $7.617\pm0.051$  & $1.565\pm0.034$     &   $-0.217\pm0.318$       \\
NGC~2533     & $2613\pm180 $  & $8.871\pm0.066$  & $1.307\pm0.121$  & $-0.103\pm0.080$    &  $2529\pm151    $   &    $8.937\pm0.059$  & $1.277\pm0.047$     &   $-0.261\pm0.102$       \\
NGC~2571     & $1293\pm 46 $  & $7.517\pm0.138$  & $0.419\pm0.041$  & $ 0.058\pm0.106$    &  $1154\pm076    $   &    $7.644\pm0.225$  & $0.280\pm0.037$     &   $-0.118\pm0.117$       \\
NGC~2635     & $4049\pm167 $  & $8.720\pm0.066$  & $1.250\pm0.061$  & $-0.218\pm0.117$    &  $4523\pm160    $   &    $8.649\pm0.030$  & $1.157\pm0.048$     &   $-0.177\pm0.060$       \\
Ruprecht~18  & $2259\pm130 $  & $8.114\pm0.427$  & $1.953\pm0.152$  & $ 0.123\pm0.155$    &  $1830\pm130    $   &    $8.256\pm0.083$  & $2.110\pm0.045$     &   $-0.500\pm0.168$       \\
Trumpler~7   & $1546\pm 60 $  & $8.185\pm0.278$  & $1.027\pm0.141$  & $-0.100\pm0.083$    &  $1439\pm162    $   &    $8.052\pm0.285$  & $0.968\pm0.198$     &   $-0.286\pm0.198$       \\
\hline      
  \end{tabular} 
\end{table*}

\section{Comparison with the literature after {\it Gaia}}

The large scale results of cluster parameters, mainly distance and age, published after the {\it Gaia} DR2 catalog can briefly summarized as follows:

\begin{itemize}
\item \citet{Cantat2020a} estimate the distance, age, and interstellar reddening for 1867 clusters using artificial neural networks applied to $G$, $G_{BP}$ and $G_{RP}$ magnitudes of member stars brighter than $G = 18$ 

\item \citet{Bossini19} estimated age, distance modulus and extinction for a sample of 269 open clusters applying BASE-9 \citep{BASE92006ApJ...645.1436V}, to fit stellar isochrones on the observed $G$, $G_{BP}$ and $G_{RP}$ magnitudes of the high probability member stars from \citet{Cantat-Gaudin2018A&A...618A..93C}

\item \citet{ChinesCat} used its own algorithm to automatic isochrone fitting applied to CMD of 2443 cluster candidates, considering stars with $G\le17$

\item \citet{Sim2019JKAS...52..145S} used least-squares fitting of isochrones to CMDs, considering the mode of distribution of the parallaxes of the member stars as cluster distance. The authors used for interstellar reddening E($G_{BP}-G_{RP}$) and the extinction in the $G$ band, $A_{G}$, using the extinction values from {\it Gaia} DR2.
 
\end{itemize}

\begin{figure*}
\centering
\includegraphics[scale = 0.36]{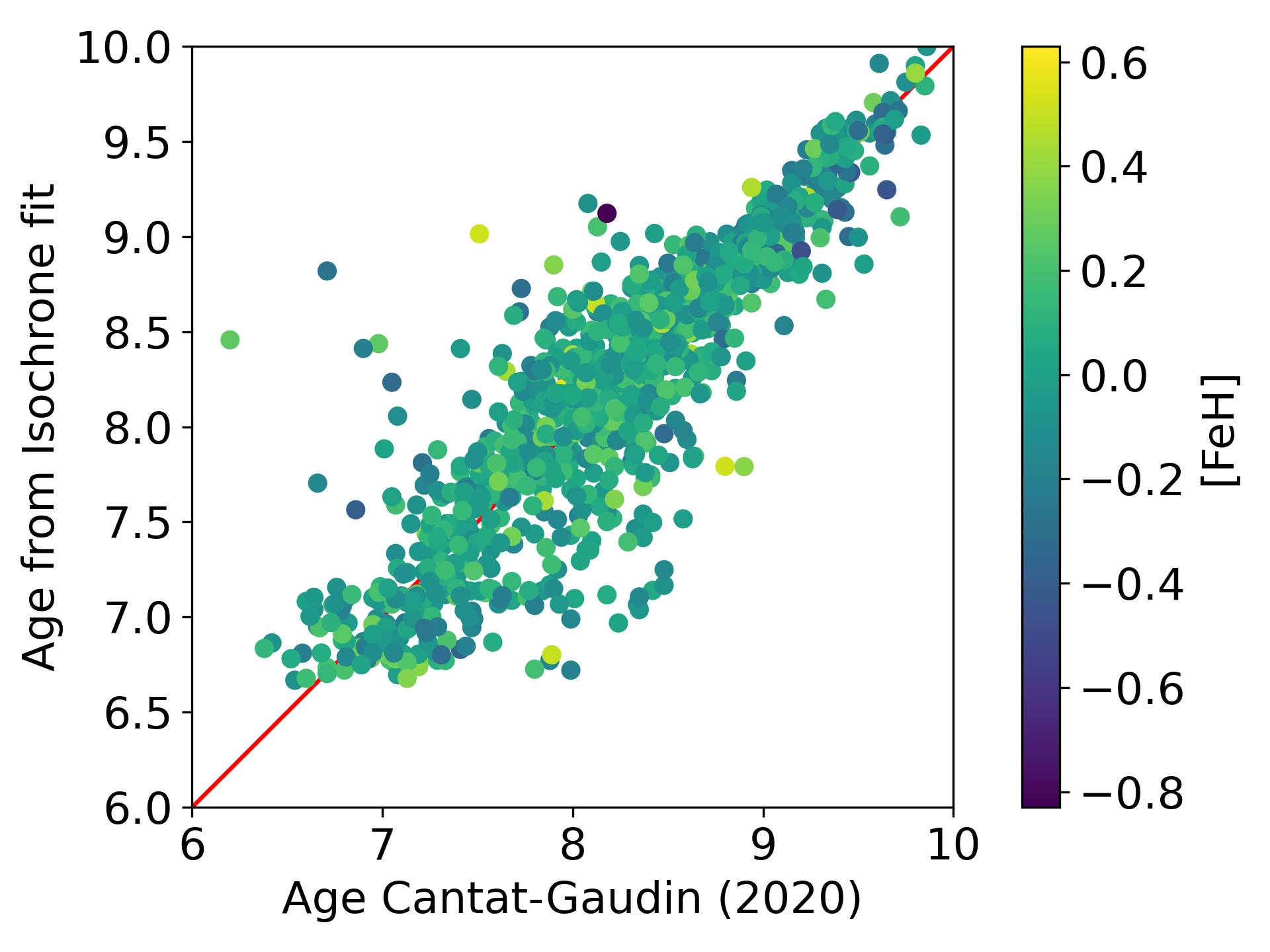}
\includegraphics[scale = 0.36]{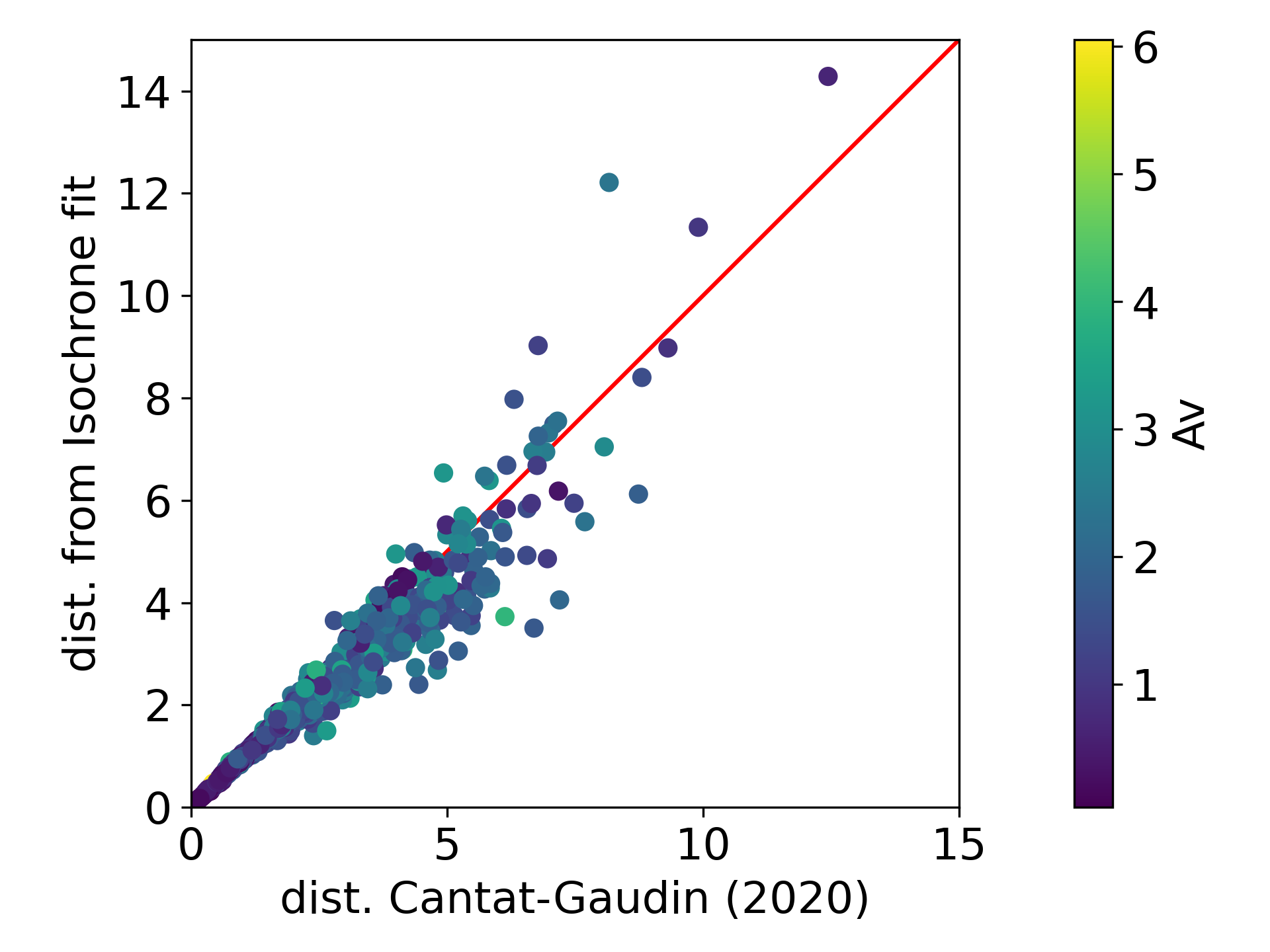}
\includegraphics[scale = 0.36]{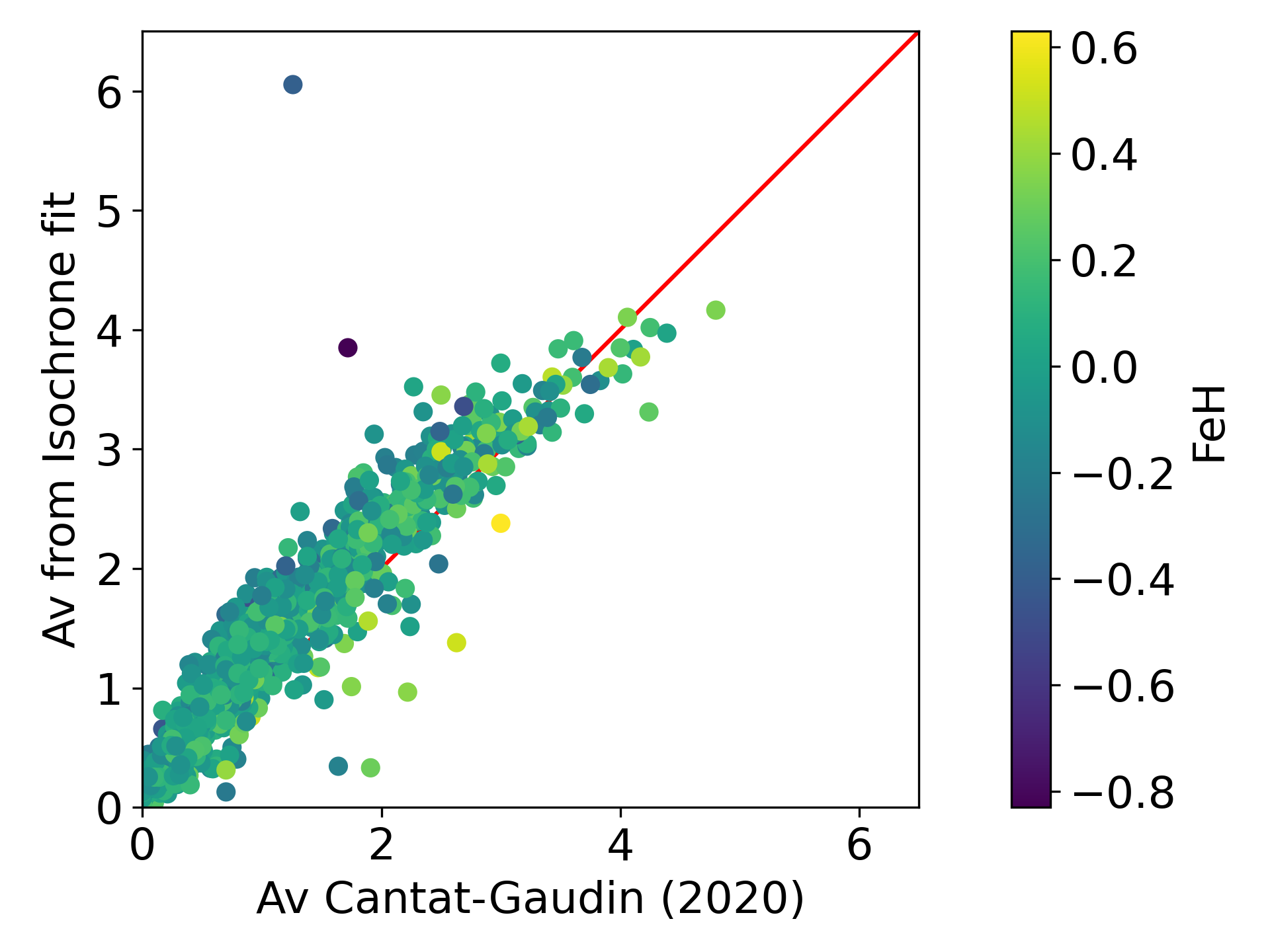} \\
\includegraphics[scale = 0.36]{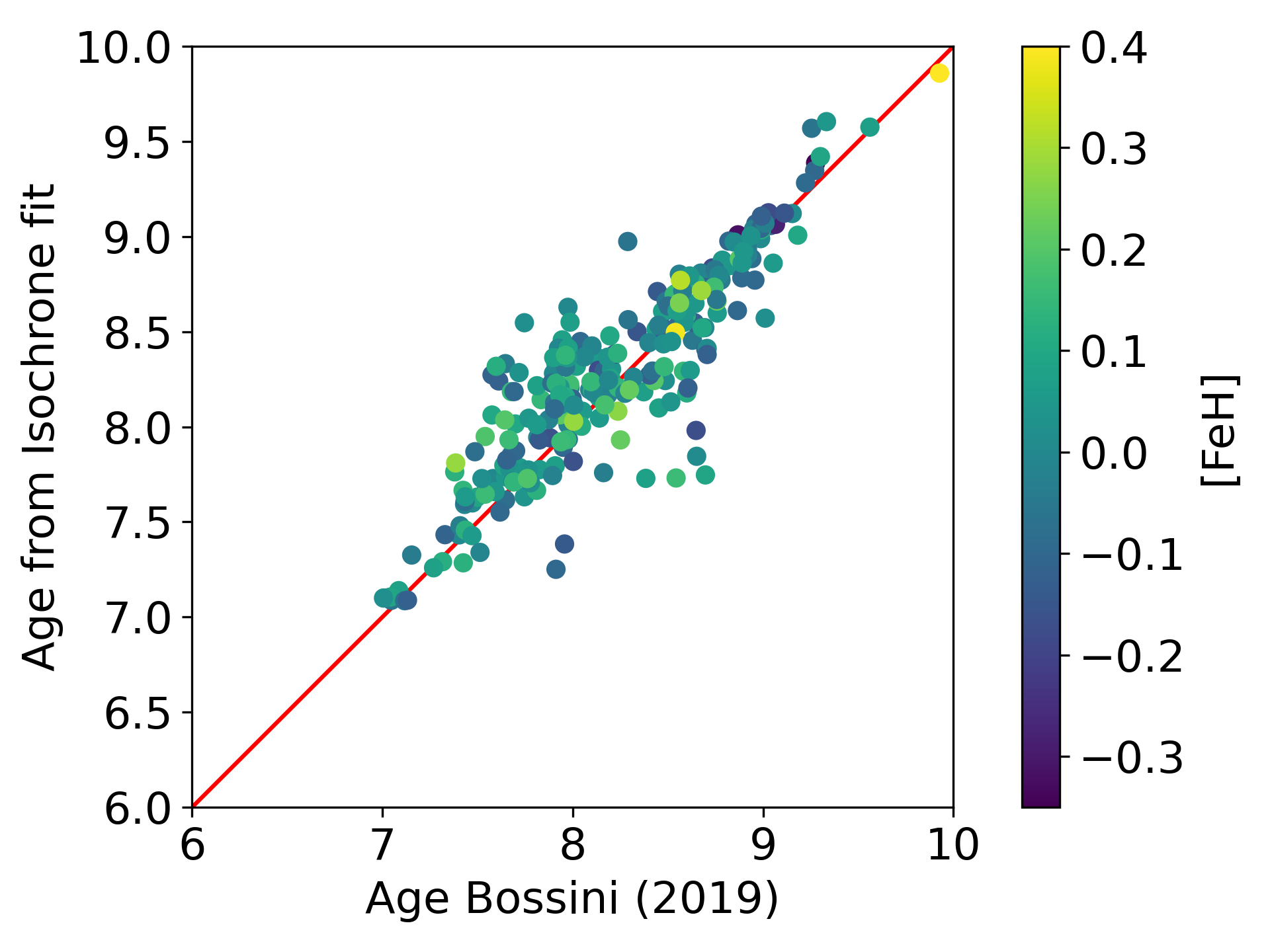}
\includegraphics[scale = 0.36]{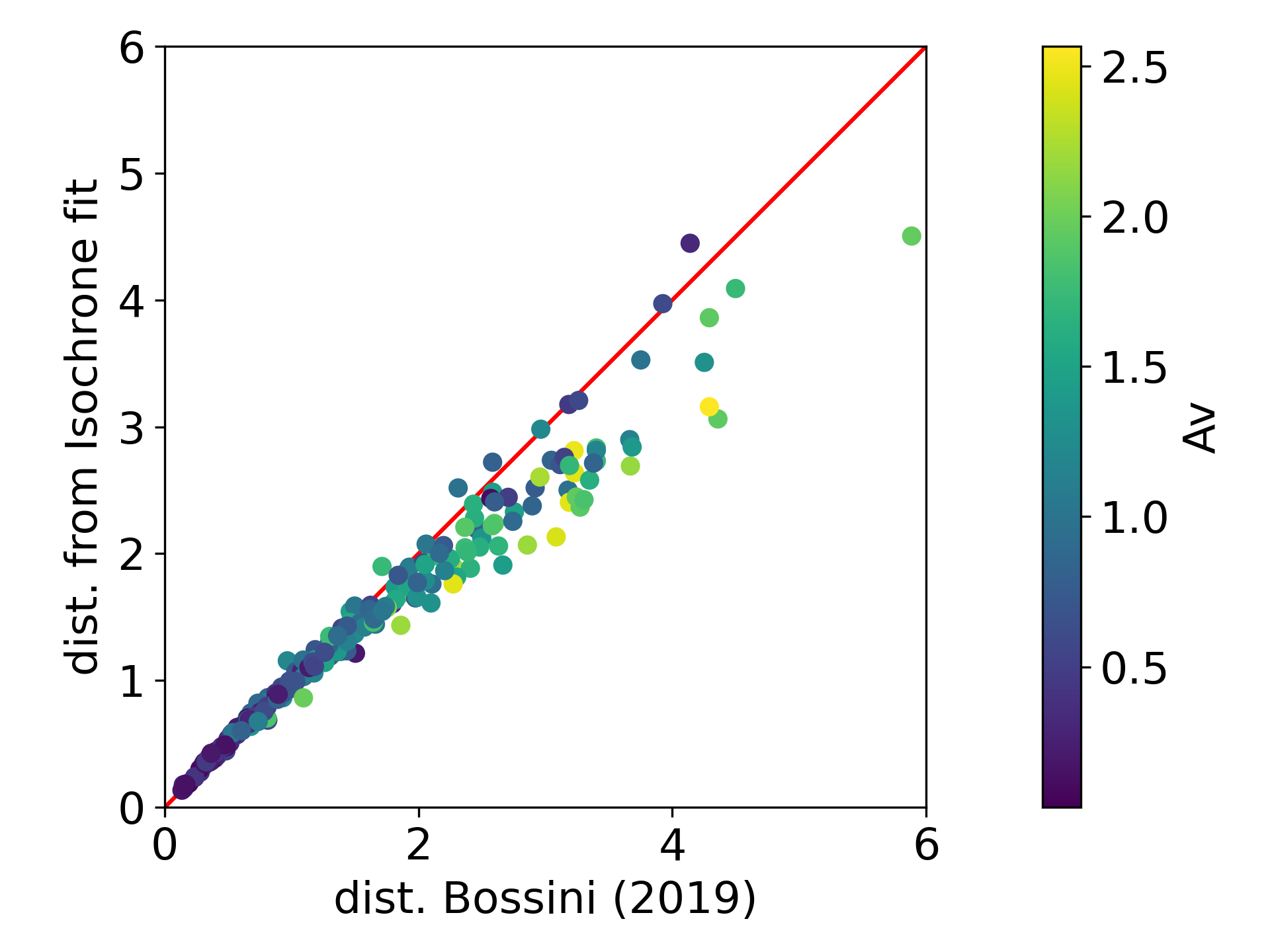}
\includegraphics[scale = 0.36]{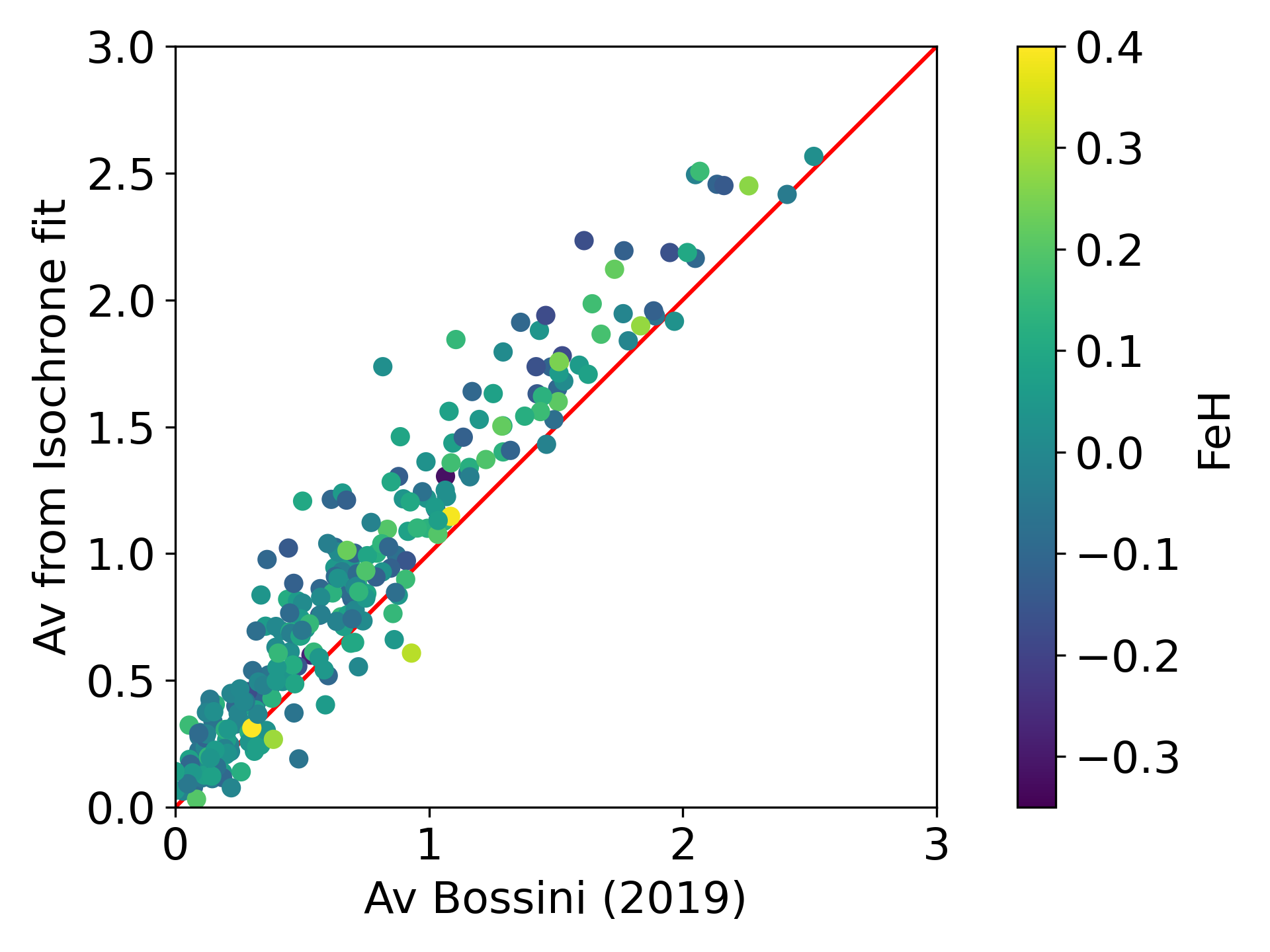} \\
\includegraphics[scale = 0.36]{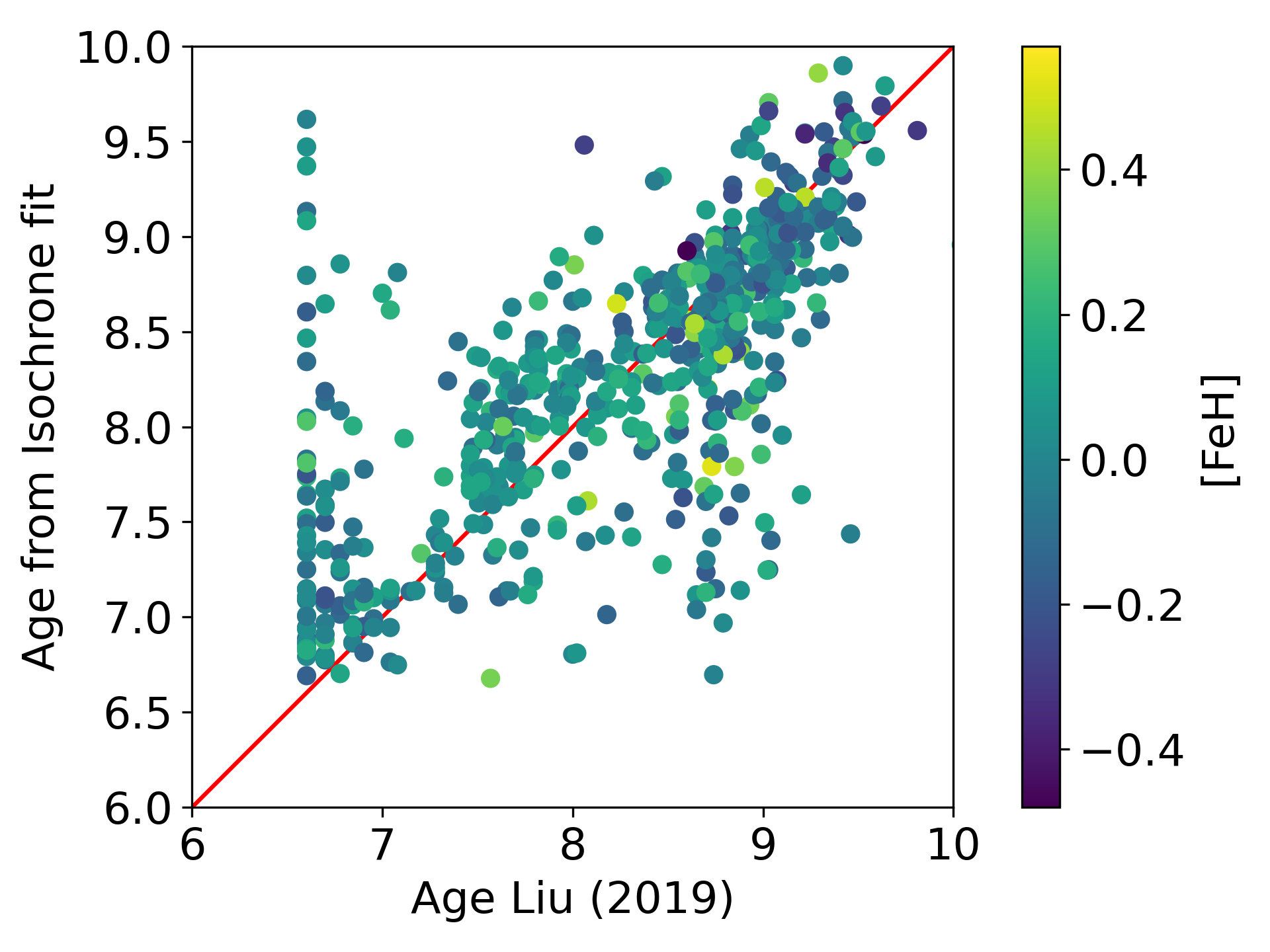}
\includegraphics[scale = 0.36]{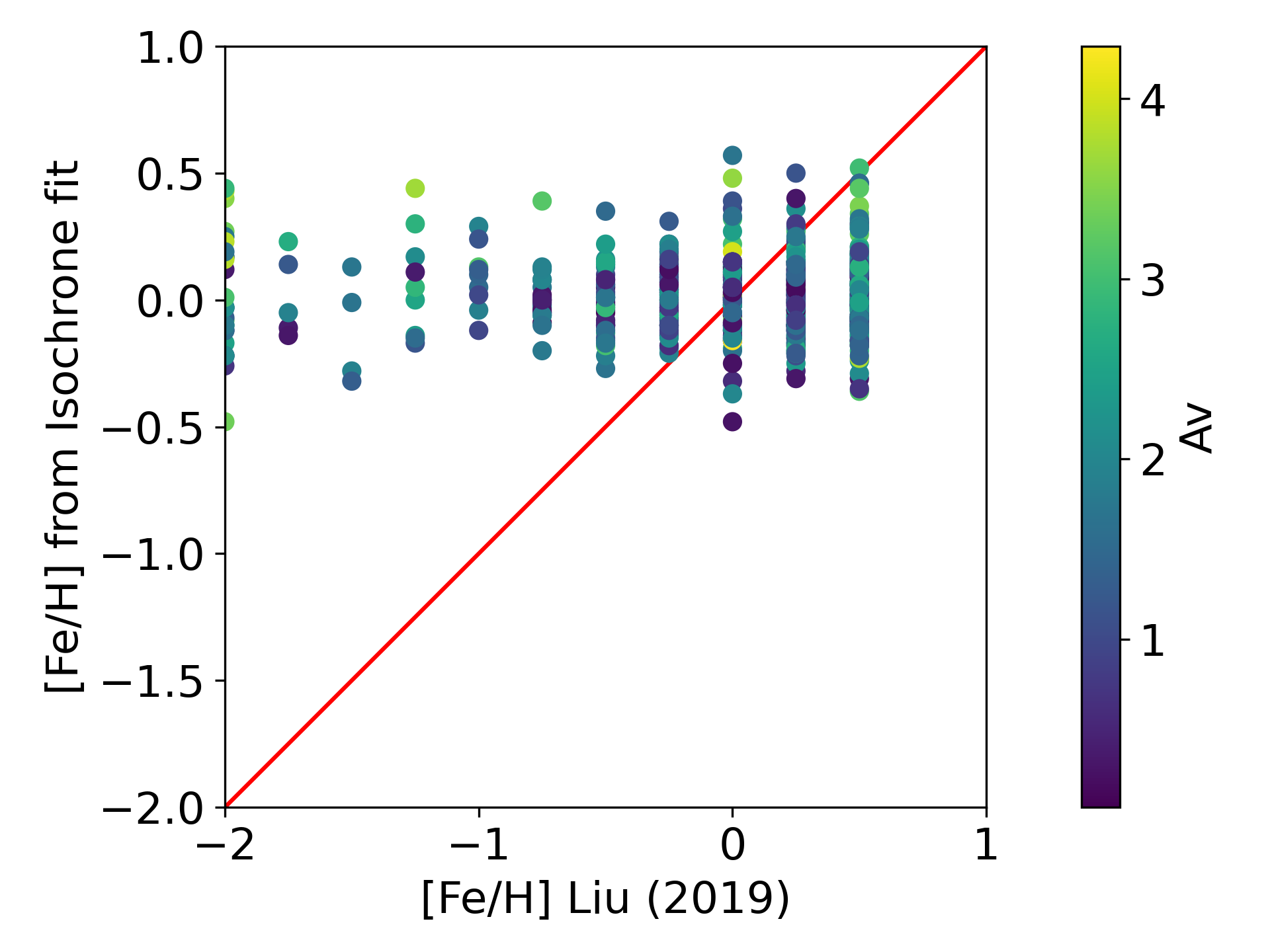}\\
\includegraphics[scale = 0.36]{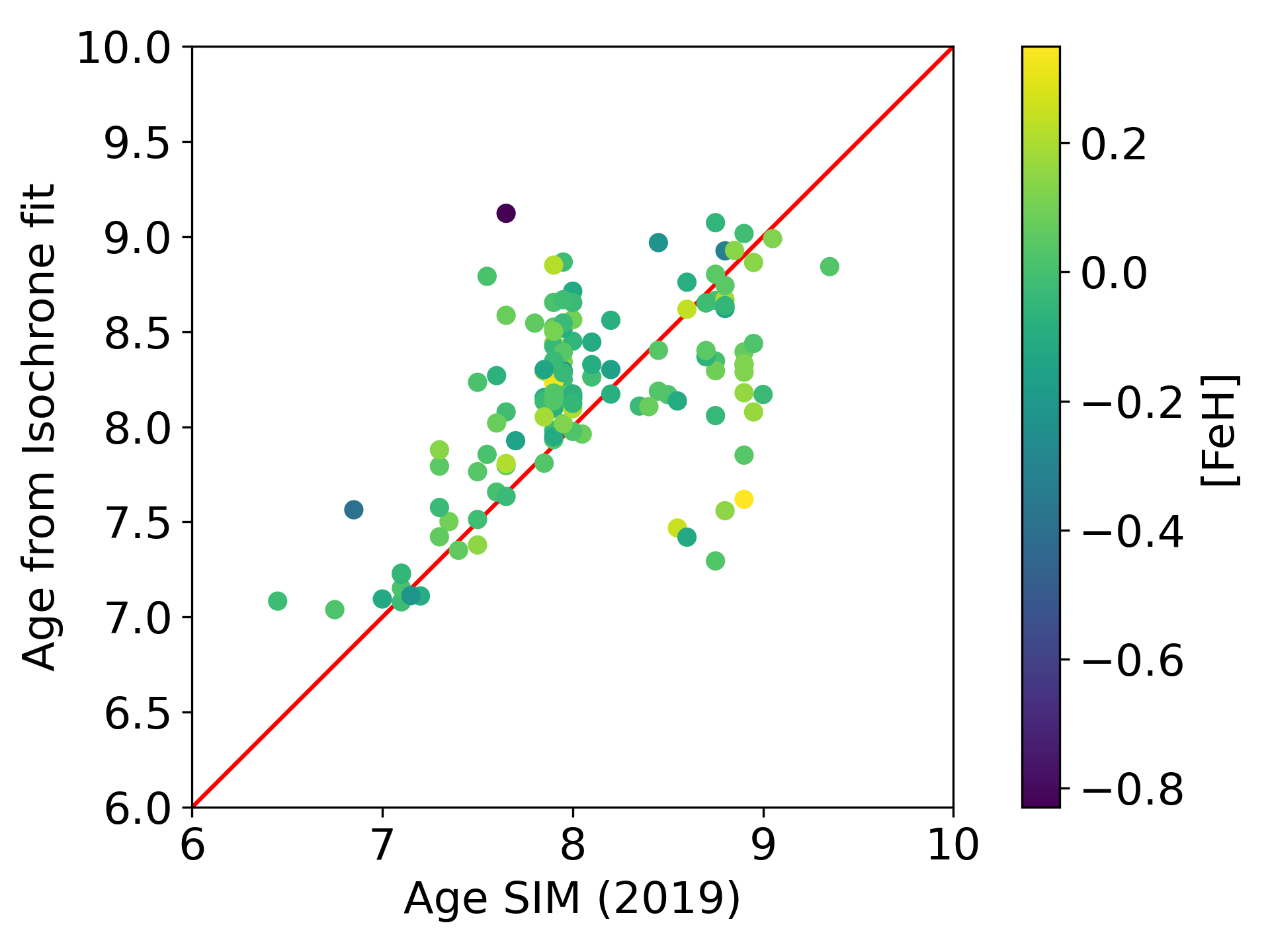}
\includegraphics[scale = 0.36]{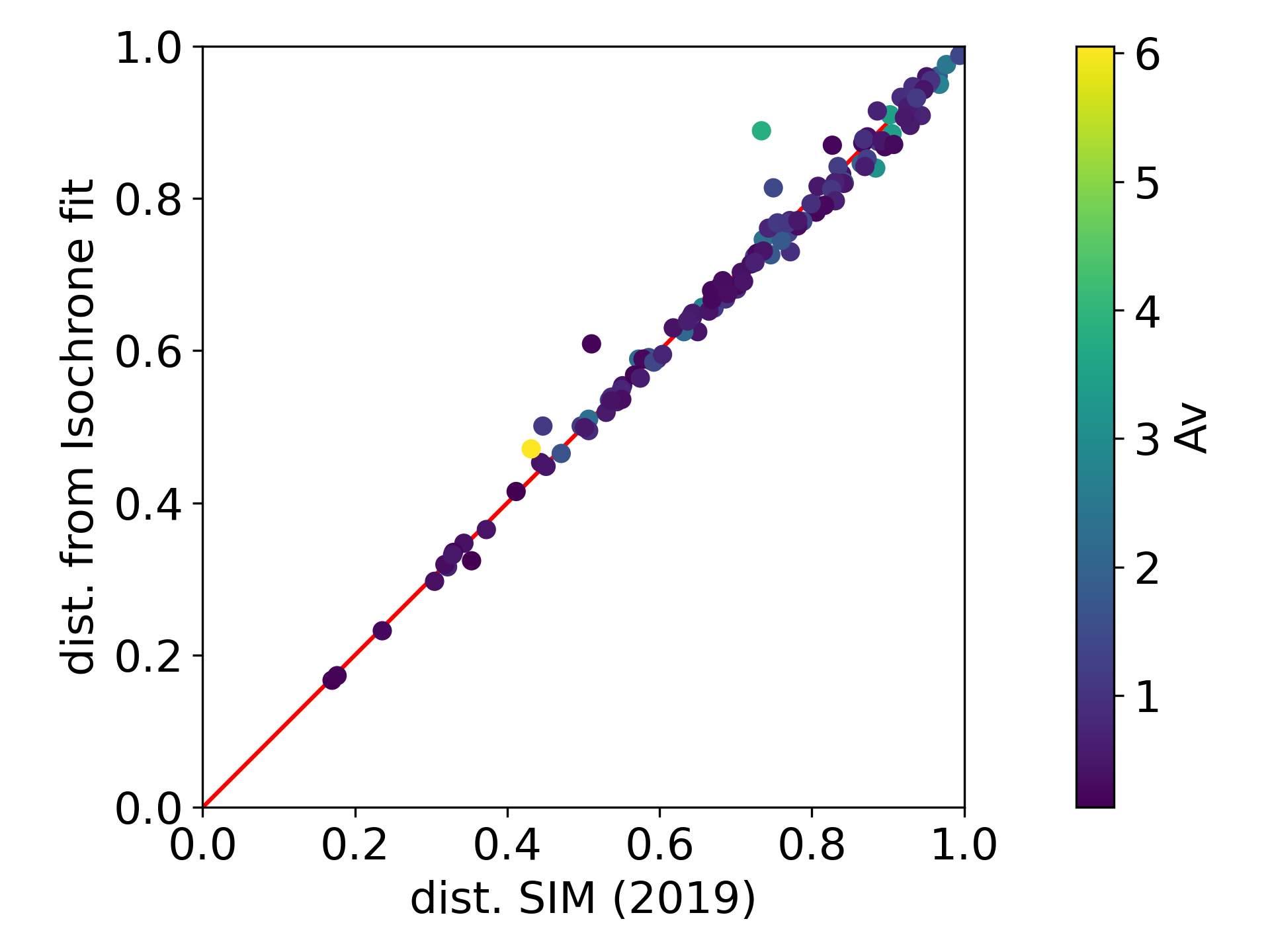}
\caption{Comparison of the parameters age, distance, $A_{V}$ and [Fe/H], for 1216, 257, 149 and 128 open cluster in common with \citet{CantateAnders}, \citet{Bossini19}, \citet{ChinesCat} and \citet{Sim2019JKAS...52..145S}, respectively.}
\label{fig:comp-lit} 
\end{figure*}

Figure \ref{fig:comp-lit} gives the comparison of the parameters for 1259, 257, 149 and 128 open clusters in common with \citet{Cantat2020a}, \citet{Bossini19}, \citet{ChinesCat} and \citet{Sim2019JKAS...52..145S}, respectively.

The comparisons should be made carefully since the mentioned studies determine open cluster parameters through isochrone fitting using different methods and strategies. Although the works cited used PADOVA isochrones \citep{Bressan2012}, 
scaled to solar metal content with $Z_{\odot} = 0.0152$, \citet{Cantat2020a}, \citet{Bossini19} and \citet{Sim2019JKAS...52..145S} opted to use a fixed metallicity while \citet{ChinesCat} considered metallicity as a free parameter in the isochrone fit. 

The study of \citet{ChinesCat} uses stars with $G$ $\le$ 17 and the other works used stars limited to $G = 18$.
Since we used the stellar membership previously published we follow the same magnitude cut-offs which imposes limitations to obtain adequate results for more distant open clusters due to the low sampling of the main sequence. For that reason the oldest known open clusters as Berkeley~15 and ESO~393~12 and many others are not present in our sample. 

The comparison of our values with those derived by \citet{Cantat2020a} shows systematic trends in $A_V$ and distance, in the sense of our $A_V$ being larger and distance being smaller than those of \citet{Cantat2020a}. Despite the large dispersion ($\sigma$ = 0.33) no trend in age is detected.
The same systematic trends in $A_V$ and distance can be seen in the comparison with the values by \citet{Bossini19}.
While all studies used the same member stars, deviations are expected due to differences in the methods, which include different extinction laws and priors in \citet{Bossini19} and \citet{Cantat2020a}.

Regarding the discrepancies in ages, we found 22 clusters with differences larger than 3$\sigma$. The objects are presented in Table \ref{table:discrepantages}. An inspection of the CMDs reveals considerable scatter and/or poor membership determination  which lead to poor isochrone fit, mainly due to lack of clear turn-off, which justifies the differences obtained between our parameters and those from \citet{Cantat2020a}.
We expect that these cases can have better results with the catalog {\it Gaia} EDR3. 

The comparison with \citet{Sim2019JKAS...52..145S} shows no expected systematic trend in distances. This is because their method estimates the distances from the mode of the parallaxes of the members corrected of zero-point offset of $-$0.029 mas from \citet{Lindegren2018}. 
The authors used the colour excess E($G_{BP}-G_{RP}$) and absorption $A_{G}$ from the Gaia DR2 catalog to estimate the interstellar extinction in their isochrone fitting. However, \citet{2018A&A...616A..17A} have shown that the extinction values published in Gaia DR2 are not very reliable and even exhibit large spreads within the same OC. These artifacts in interstellar extinction, the adoption of solar metallicity as well as some possible bias in distance may have affected the ages determined by \citet{Sim2019JKAS...52..145S}. A more detailed analysis would be needed to understand the contribution of each factor to the difference in the estimated ages, but that is beyond the scope of this study. 

Regarding the comparison with \citet{ChinesCat}, we note that there is no good agreement. The plot of the comparison of [Fe/H] shows steps of 0.25 dex used by the authors in the isochrone fit. Possibly as happened with the results of \citet{Sim2019JKAS...52..145S}, the values of the metallicity may have affected the ages determined by the authors. 
Note that since class 3 objects in \citet{ChinesCat} contains very young clusters with log(age)$\le$6.8 dex, the highest density in the young clusters shown in the plot of age comparison is expected as pointed by \citet[][their Fig. 7]{ChinesCat}.

Finally, we quantitatively evaluate the reliability of the parameters obtained from the isochrone fit by comparing them with the data. For this we use as a figure of merit the likelihood provided in equation \ref{eq:likelihood}. According to this definition the best fit is the one which maximizes the likelihood, or as in our algorithm, minimizes the objective function constructed from it. Thus, the likelihood-ratio test, or the ratio of the values of the objective function, from two distinct solutions can indicate which was the best.

\begin{figure*}
\centering
\includegraphics[scale = 0.43]{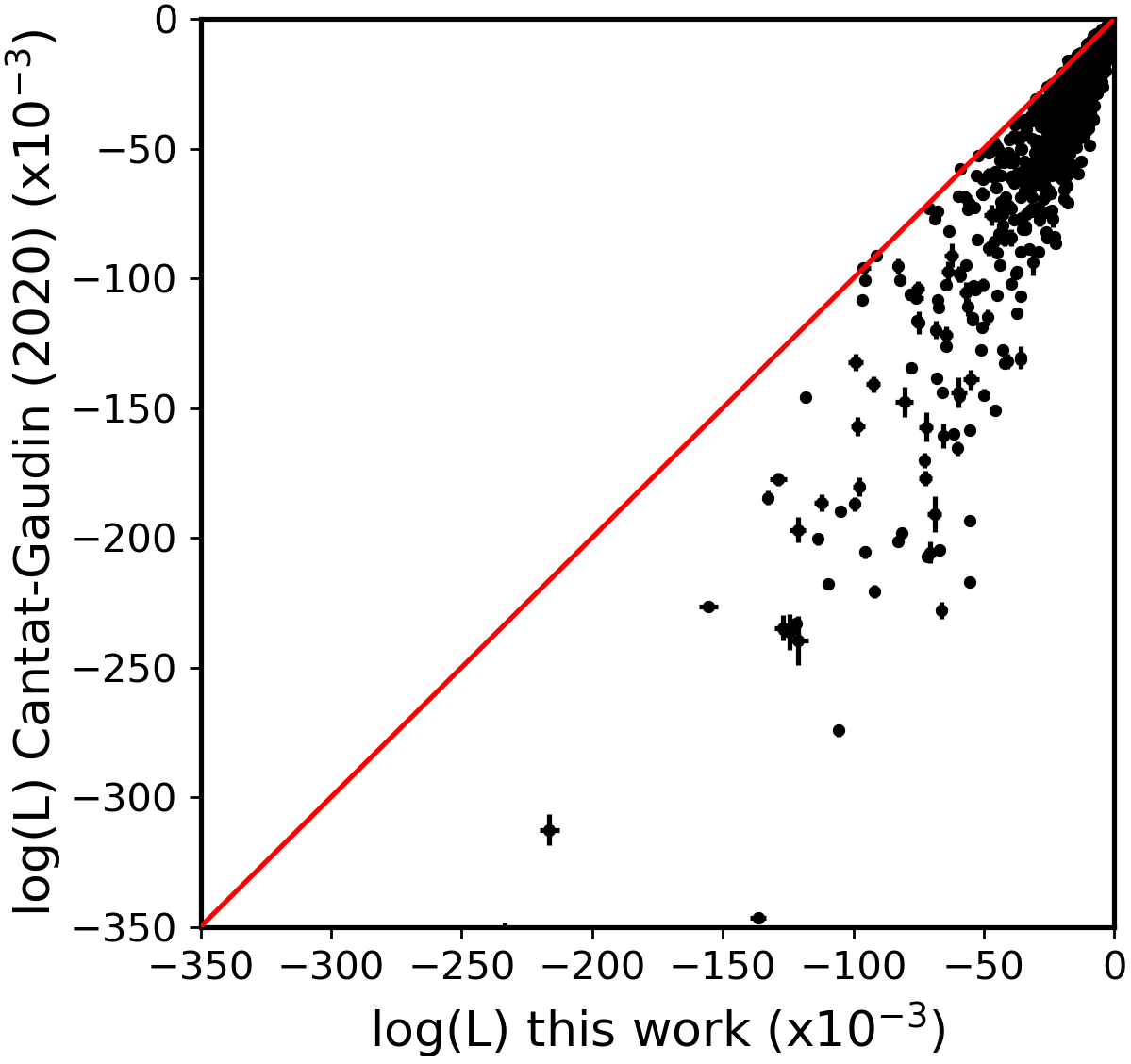}
\includegraphics[scale = 0.43]{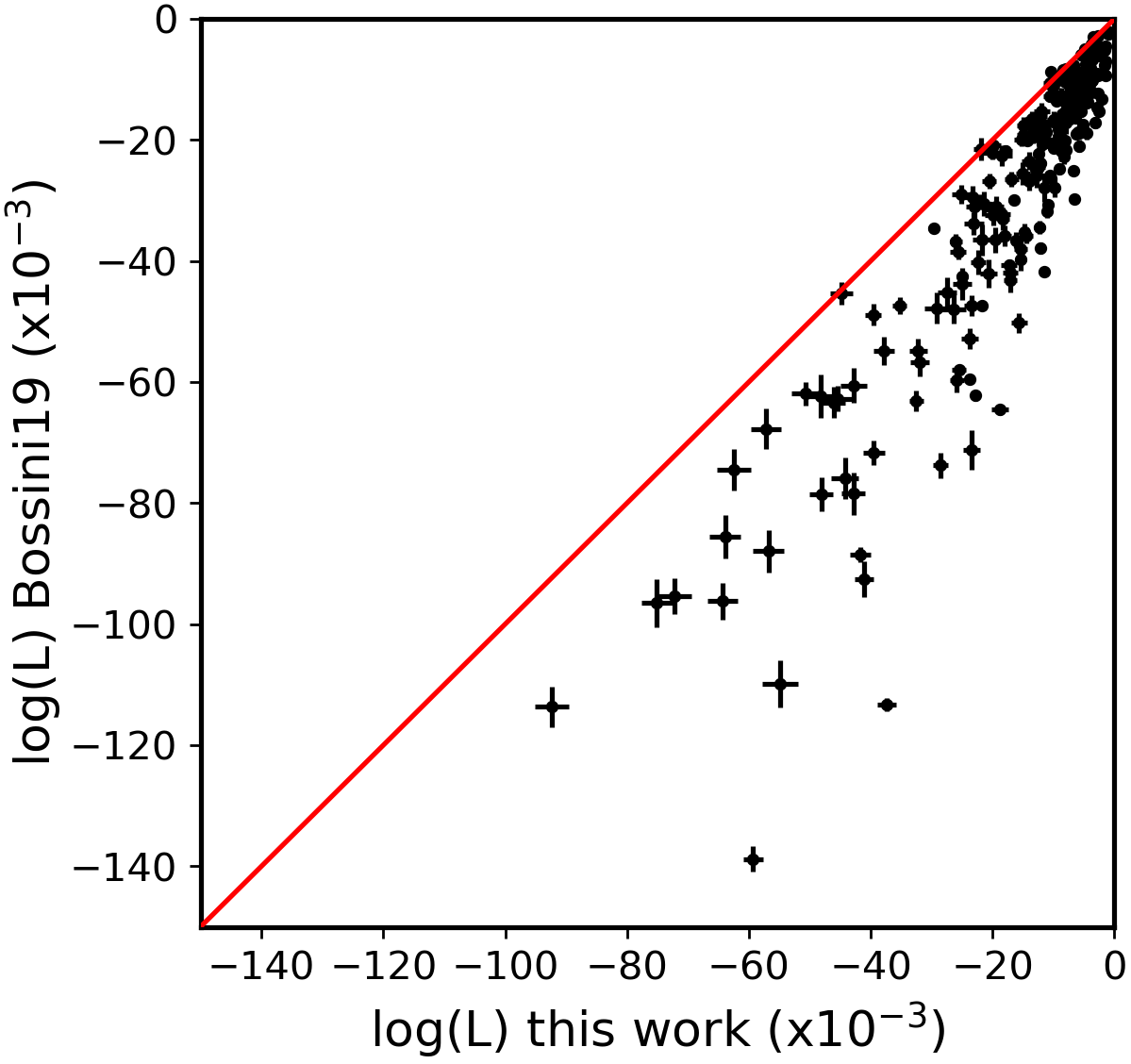}
\includegraphics[scale = 0.43]{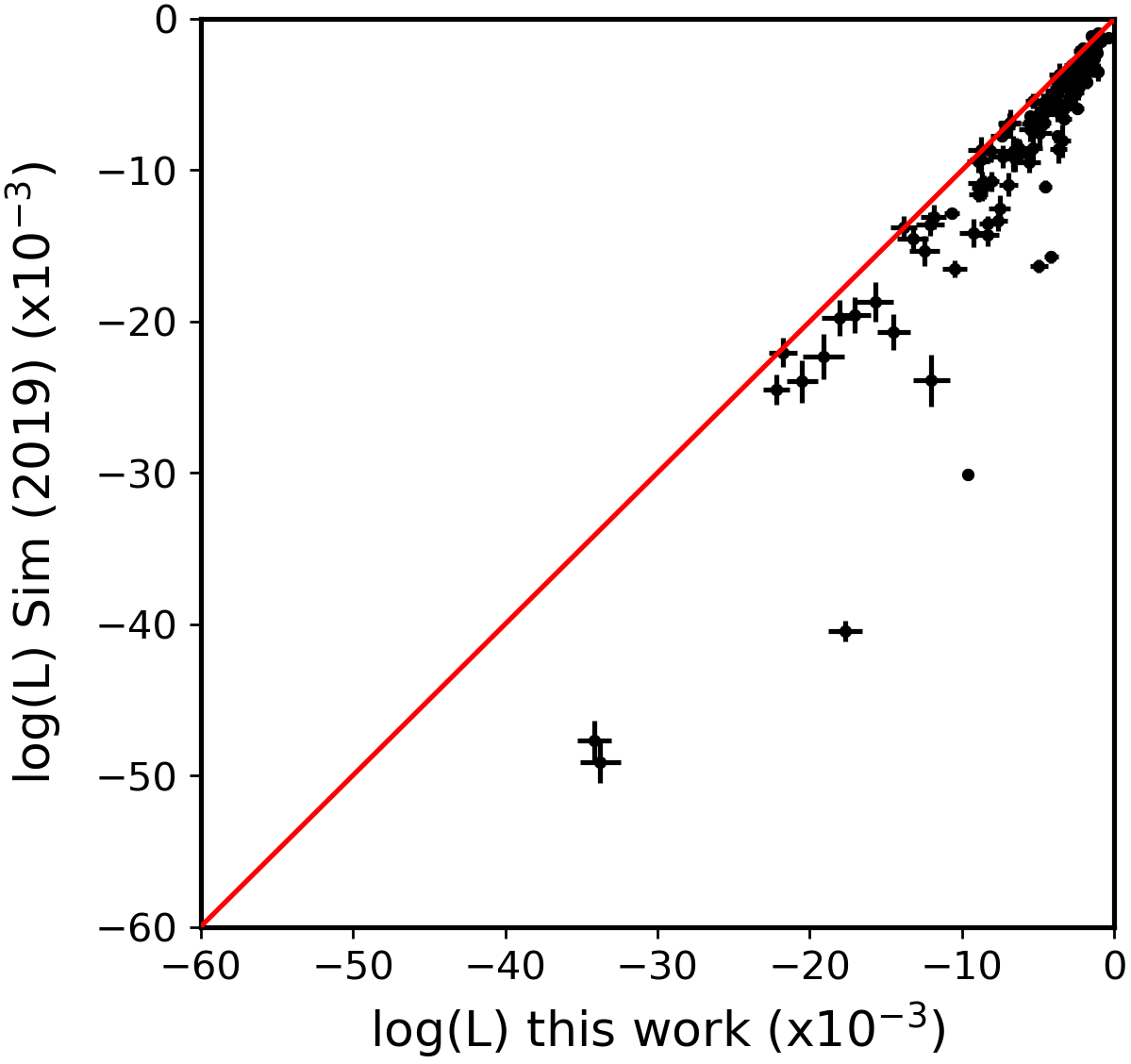} 
\caption{Comparison of the likelihood ($log(\mathcal{L})$) for open clusters in common with \citet{Cantat2020a}, \citet{Bossini19}, and \citet{Sim2019JKAS...52..145S}, respectively.}
\label{fig:comp-like} 
\end{figure*}

Since our method uses synthetic clusters obtained from model isochrones, we performed Monte-Carlo runs calculating the value of the likelihood for random samples of generated clusters with a given set of input parameters.
The process was performed 100 times and we adopted the mean of the likelihood sample as the final value and 1$\sigma$ as uncertainty.
To calculate the likelihood value from the literature parameters, we use the same data and extinction law provided by \citet{Monteiro2020}.
In Figure \ref{fig:comp-like} the logarithm of the likelihood values are presented for our sample and those of \citet{Cantat2020a}, \citet{Bossini19}, and \citet{Sim2019JKAS...52..145S}. The results show that, within our specified constraints and minimization criteria (isochrones, priors, extinction law, etc), we are obtaining the best solution.

\section{Comparison with the catalogs before {\it Gaia} DR2}

In this section we compare our results with the New Catalog of Optically Visible Open Clusters and Candidates \citep{Dias2002} (DAML) and Milky Way Star Cluster catalog \citep{Kharchenko2013} (hereafter MWSC) both published before {\it Gaia} DR2 and used in hundreds of works. A detailed discussion of the differences, advantages and disadvantages of both catalogs is presented in \citet{Monteiro2020}.

Briefly, the MWSC catalog is based in the PPMXL catalog \citep{Roeser2010}
and 2MASS catalog \citep{Skrutskie2006} to determine kinematic and photometric membership probabilities for stars in a cluster
region. The authors defined a combined stellar membership probability and determined parameters of 3006 clusters using near infra-red J,H and $K_{s}$ data.

Version 3.5 of DAML presents 2174 clusters with  parameters compiled from the literature, giving priority to observational data including the U filter, due to its importance in the determination of the $E(B-V)$ through the color-color diagram which allows more reliable determinations of distances and ages.
In the comparison with DAML catalog all results from MWSC catalog were removed.

\begin{figure*}
\centering
\includegraphics[scale = 0.36]{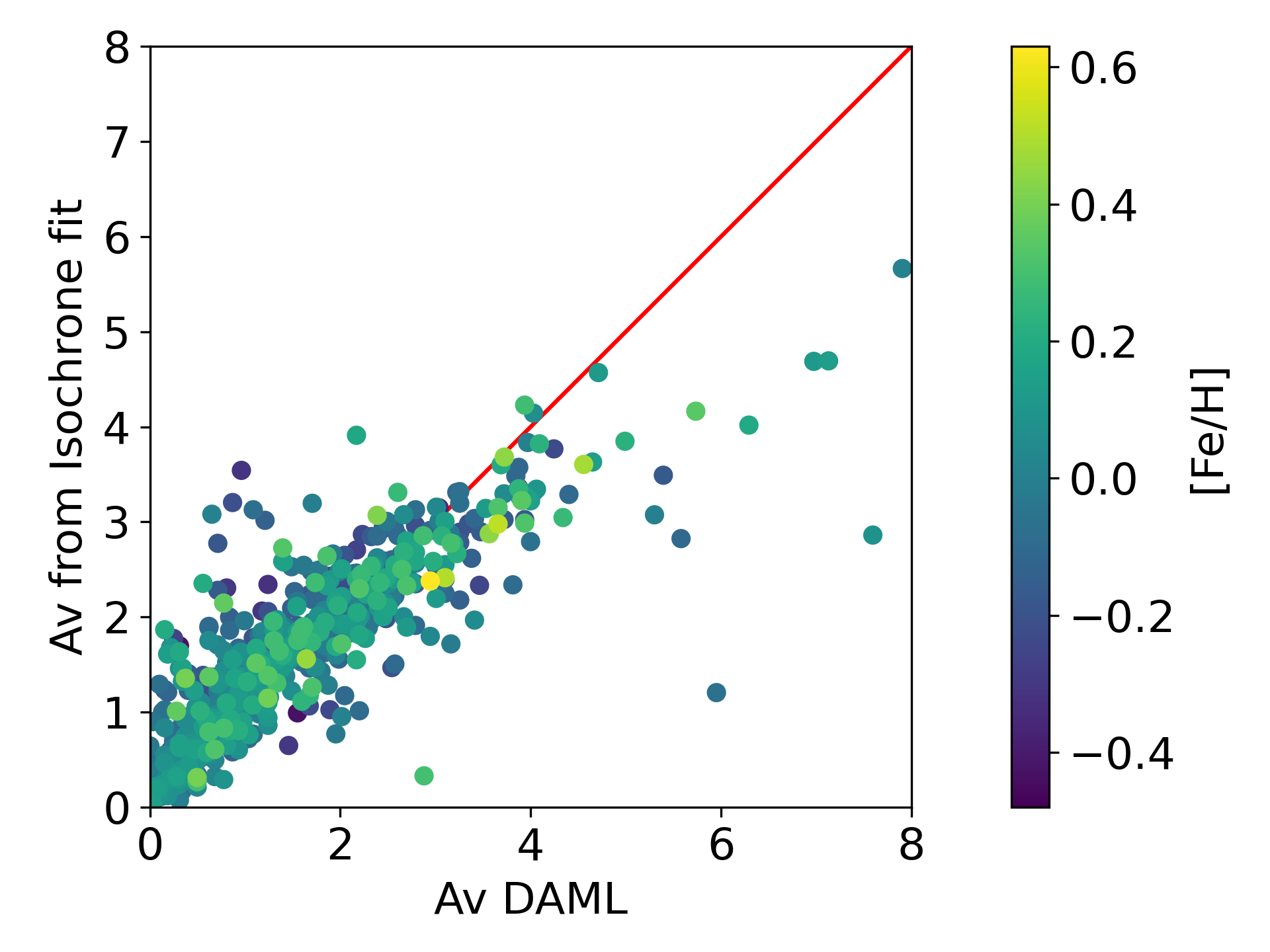}
\includegraphics[scale = 0.36]{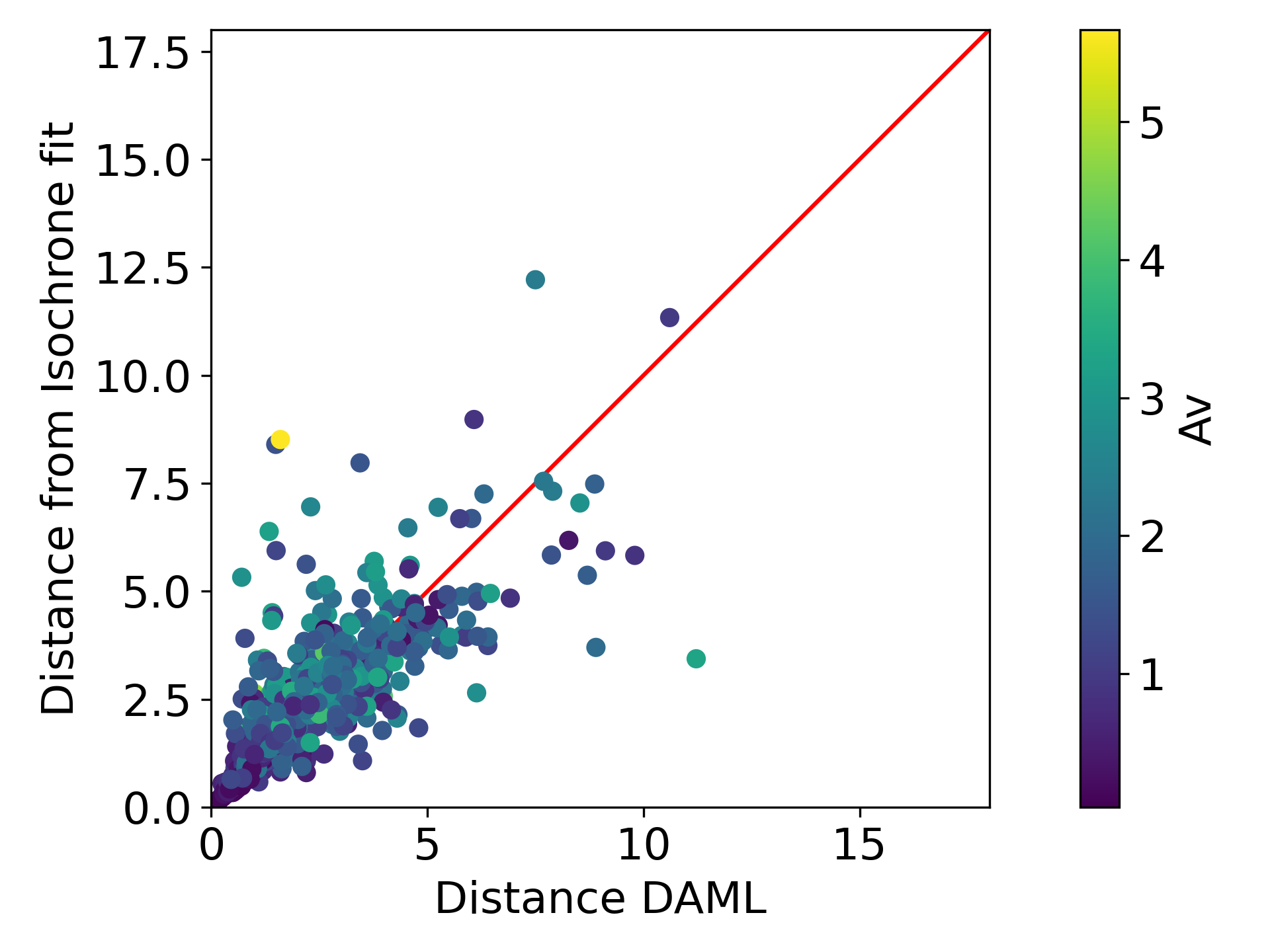}
\includegraphics[scale = 0.36]{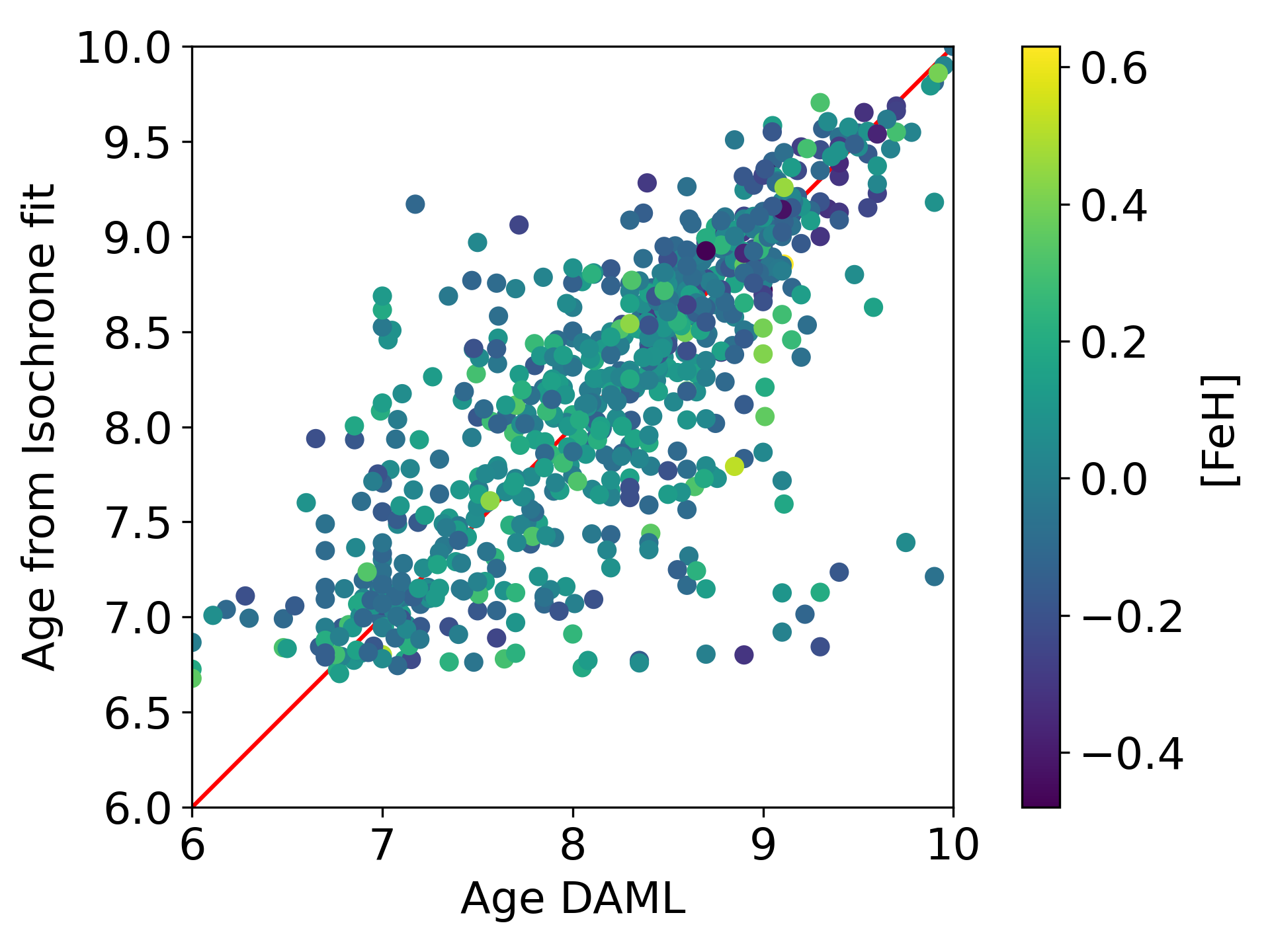} \\
\includegraphics[scale = 0.36]{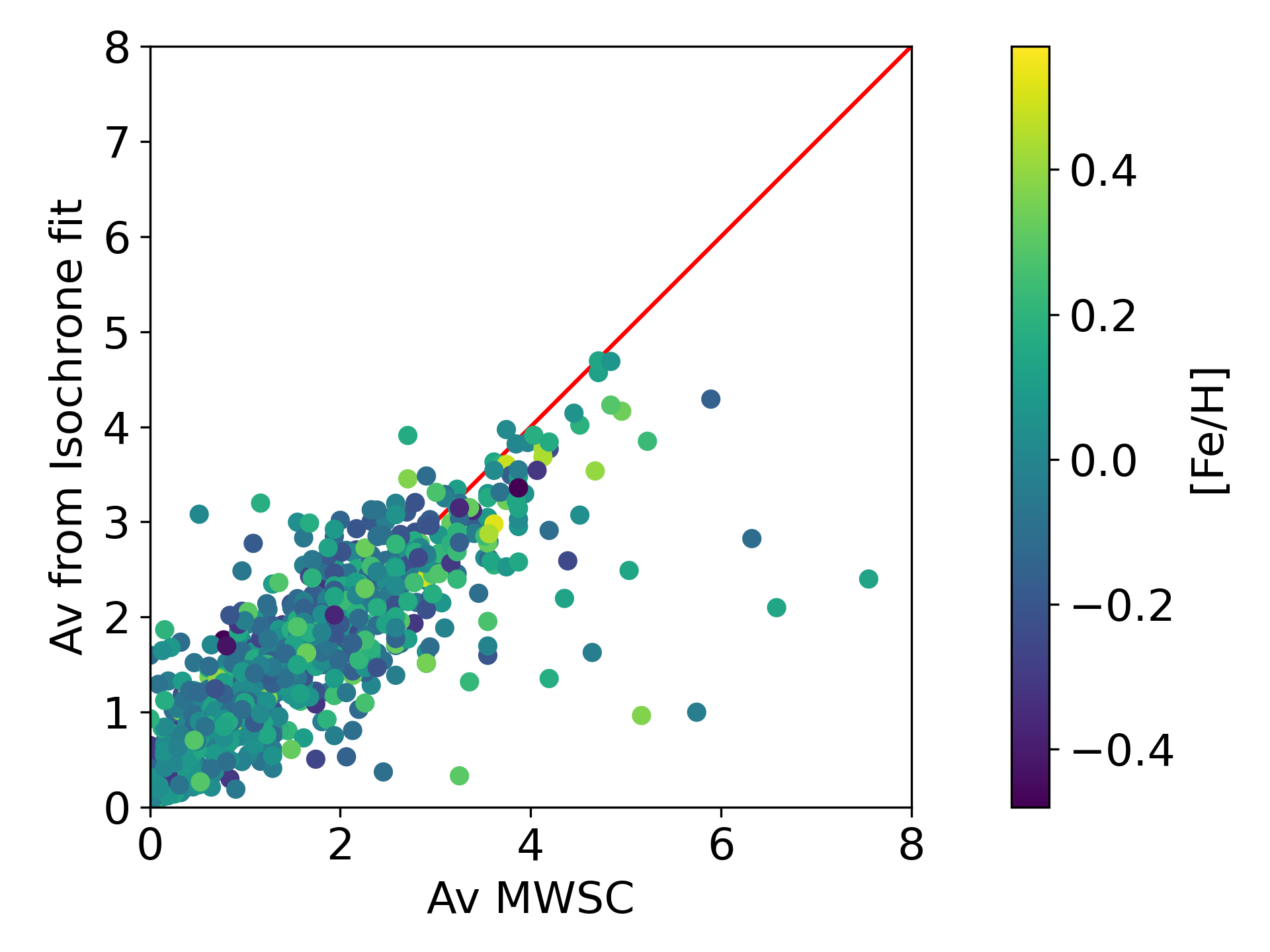}
\includegraphics[scale = 0.36]{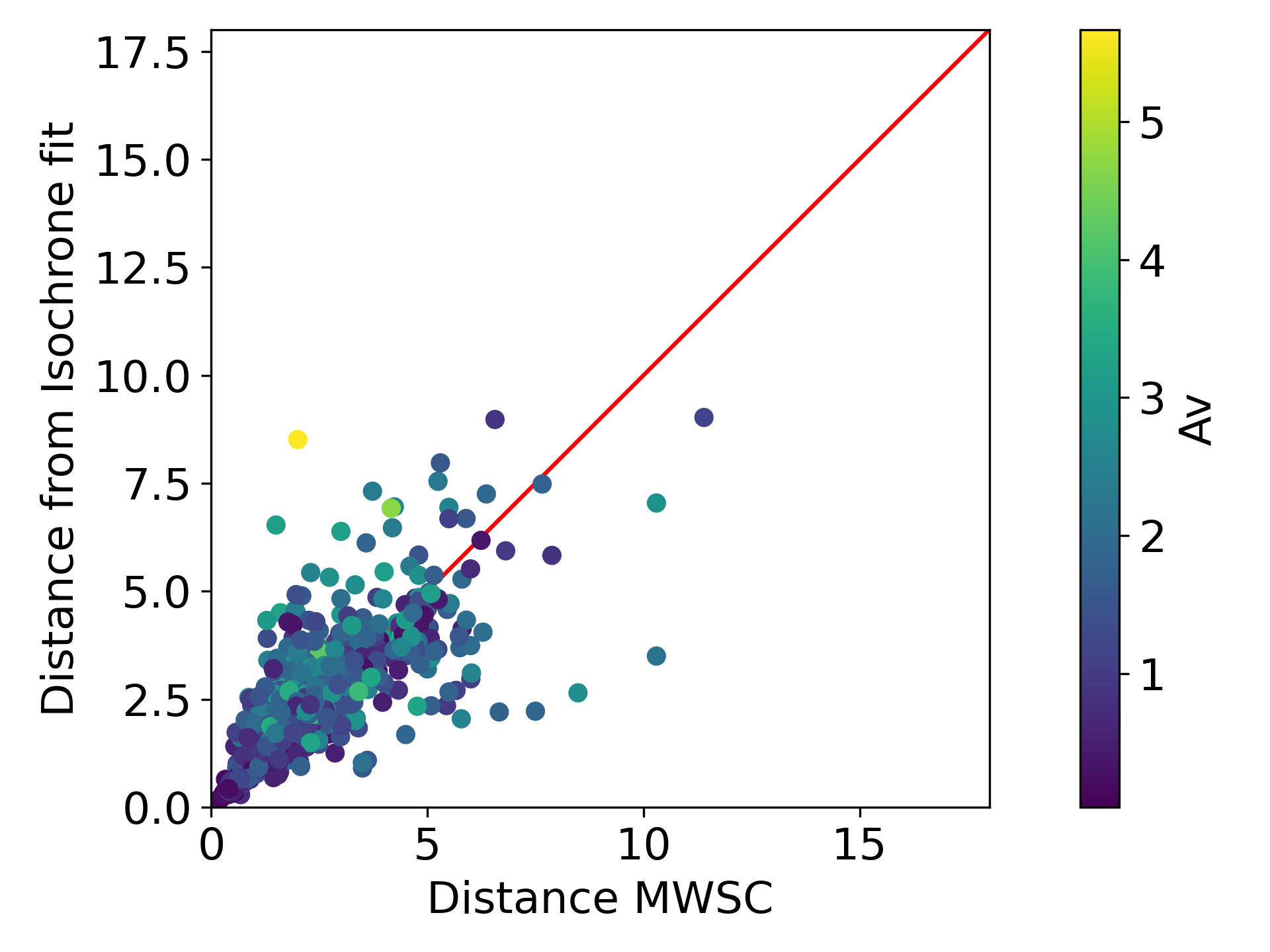}
\includegraphics[scale = 0.36]{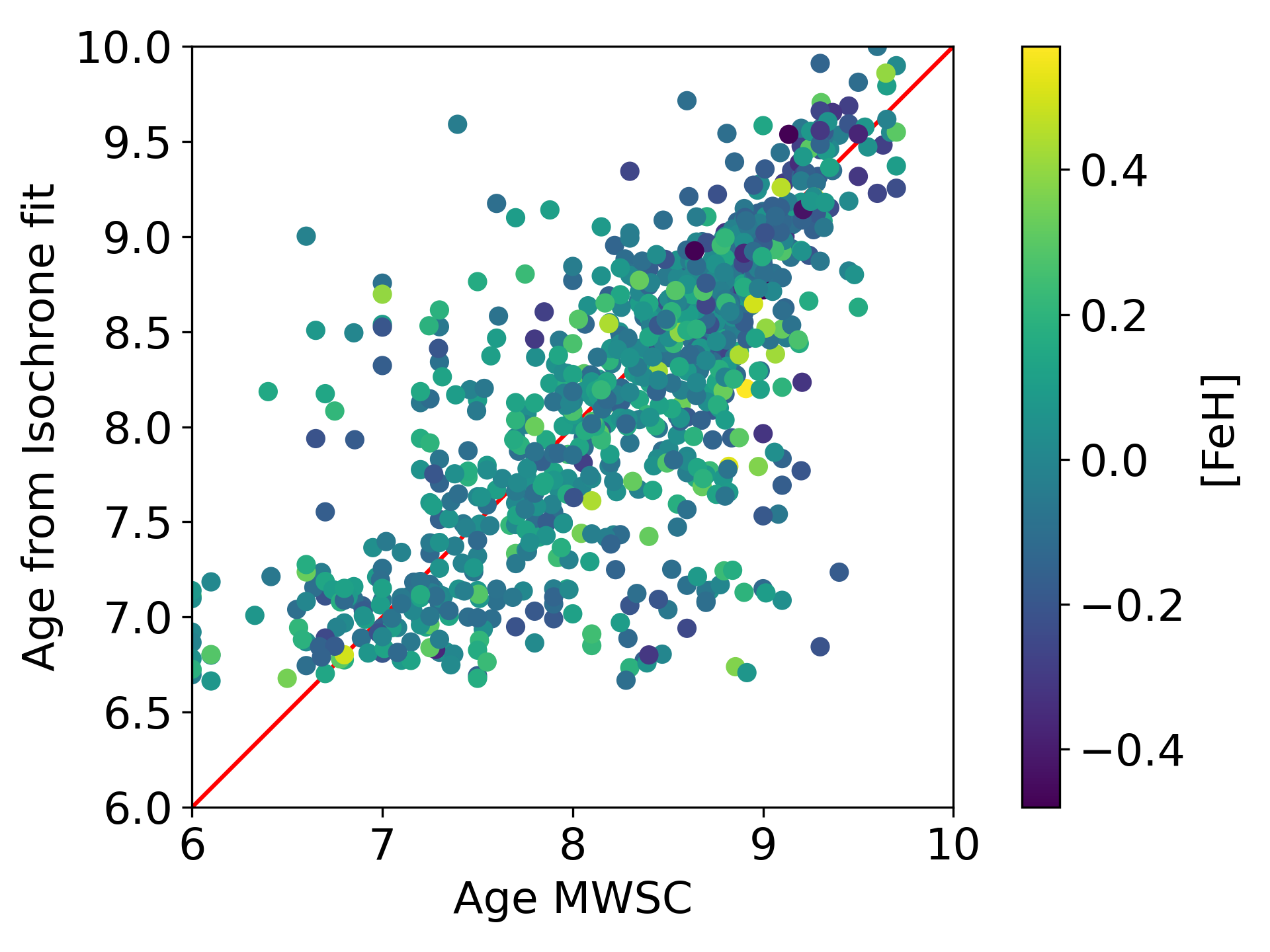}
\caption{Comparison of the parameters determined in this work with the values published by DAML and MWSC catalogs.}
\label{fig:daml02}
\end{figure*}

In Figure \ref{fig:daml02} we present the comparison of the parameters for 744 and 933 open clusters with DAML and MWSC, respectively. 
The differences obtained for each parameter are shown in Table \ref{tab:comp-lit}.  

The results of the comparison show the same characteristic scatter and trends seen in the similar comparisons by \citet{Monteiro2020}, \citet{Cantat2020a} and \citet{Bossini19}. While the Gaia based results represent overall a clear improvement over the pre-Gaia catalogs, it must be noted that distant, more reddened and older clusters are fainter and don't have Gaia based photometry and membership determinations. For those clusters the pre-Gaia catalogs remain useful and the comparisons here presented give an idea of their expected precision.

\begin{table*}
  \caption[]{Comparison of our results with those published in the literature. In the first column are give the reference. The values in the subsequent columns are in the sense our results minus literature. N is the number of common clusters.}   
\label{tab:comp-lit}
\begin{center}
\begin{tabular}{lccccccccc}
\hline 
ref                                      & distance  & $\sigma$dist  &   log(age)  & $\sigma$log(age)   & $A_V$    &  $\sigma$$A_V$  & [Fe/H]   &  $\sigma$[Fe/H] & N  \\
                                         &[pc]       &  [pc]         &[dex]         &[dex]            & [mag]    &[mag]      &[dex]     & [dex]           &\\
\hline
DAML                                   &    111    & 1046  &  -0.048 &  0.582  &  0.048 &  0.920&   0.095  &   0.243  &  744  \\  
MWSC                                   &    120    &  906  &  -0.088 &  0.571  & -0.033 &  0.751&   0.112  &   0.256  &  933    \\
Cantat$-$Gaudin                       &   -167    &  414  &  -0.057 &  0.384  &  0.289 &  0.268&          &          & 1216   \\
Bossini                                &   -138    &  262  &  -0.067 &  0.250  &  0.170 &  0.174&          &          &  257  \\
Sim                                    &     -4    &   18  &  -0.089 &  0.477  &        &       &          &          &  128   \\ 
LP                                     &           &       &  -0.557 &  0.896  &        &       &   0.090  &   0.777 &  149   \\ 
\hline
\end{tabular}
\begin{flushleft}
DAML = \citet{Dias2002}; MWSC =  \citet{Kharchenko2013}; Cantat$-$Gaudin = \citet{Cantat2020a}; Bossini = \citet{Bossini19}; Sim = \citet{Sim2019JKAS...52..145S}; LP = \citet{ChinesCat}\\
\end{flushleft}
\end{center}
\end{table*}

\section{General comments} \label{sec:comments}

The 1743 open clusters included in this work pass the empirical criteria for bona fide clusters with Gaia proposed by \citet{CantateAnders} and a visual inspection of the CMD with the fitted isochrone over-plotted.
Typical cluster detected in Gaia DR2 show small total proper motion dispersion, clear features in the CMD (e.g. main sequence, turn-off, red clump), 50$\%$ of their members are within a radius of 15 pc, parallax dispersion of the member stars adequate to the group's distance, and a minimum number of 10 member stars. 

Visual inspection of the CMDs indicates that many clusters detected with Gaia DR2 by \citet{ChinesCat} and some by \citet{Castro-Ginard2020A&A...635A..45C} are dubious or not physical objects although we recalculated the individual membership probability of the stars.

This is in agreement with the original classification by the authors. \citet{ChinesCat} considered 1747 ($72\%$) candidates as class 3 and 127 ($5\%$) as class 2, using as criterion the width of the MS, age and quality of the isochrone fitting. \citet{Castro-Ginard2020A&A...635A..45C} considered 101 ($15\%$) candidates as class B and 235 ($36\%$) as class C, considering the concentration of member stars in the astrometric parameters and the contamination in the CMD.
We estimate that objects in class A from \citet{Castro-Ginard2020A&A...635A..45C} and class 1 from \citet{ChinesCat} are likely star clusters candidates while the candidates in the other classes need confirmation.

\begin{figure}
\centering
\includegraphics[scale = 0.32]{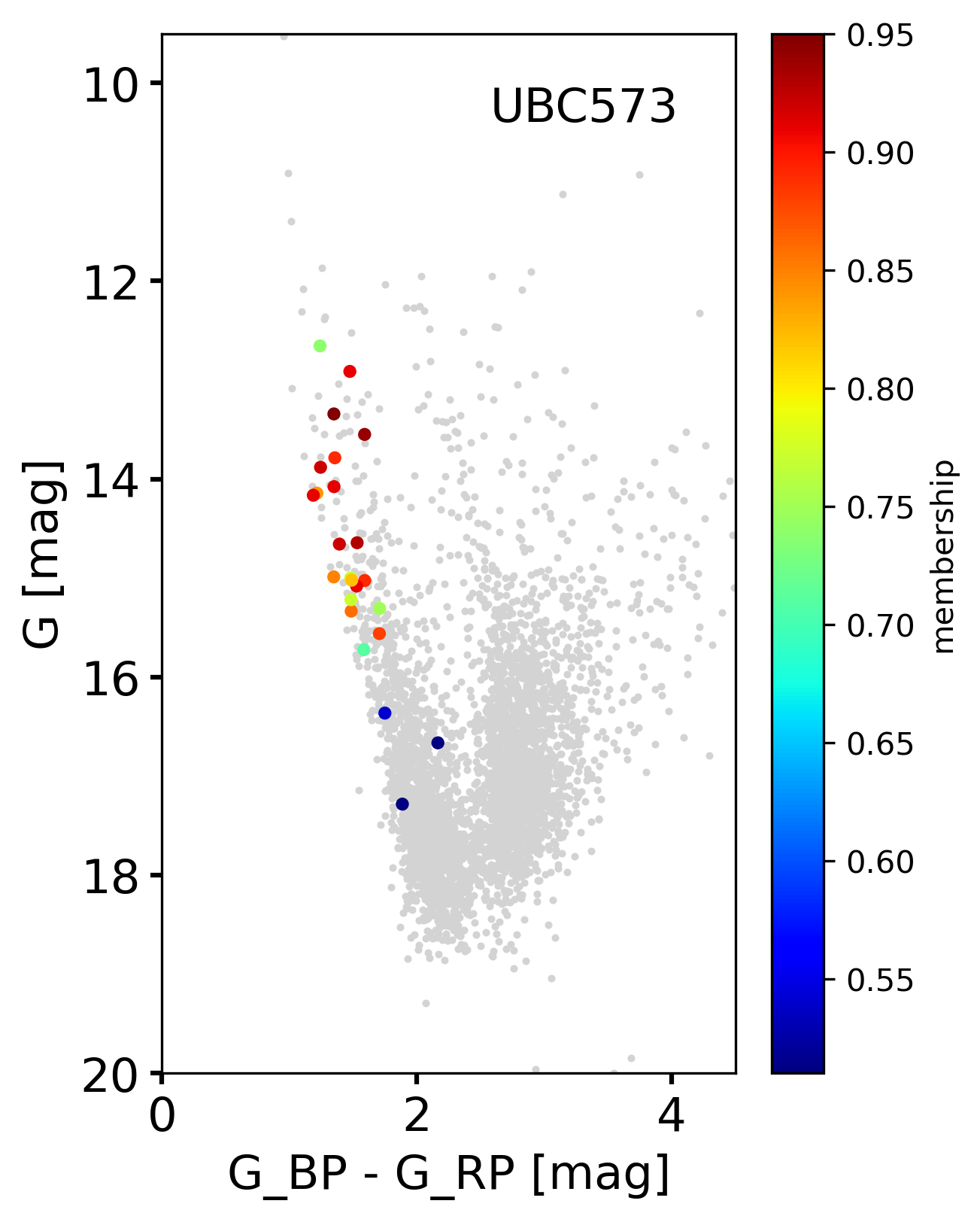}
\includegraphics[scale = 0.32]{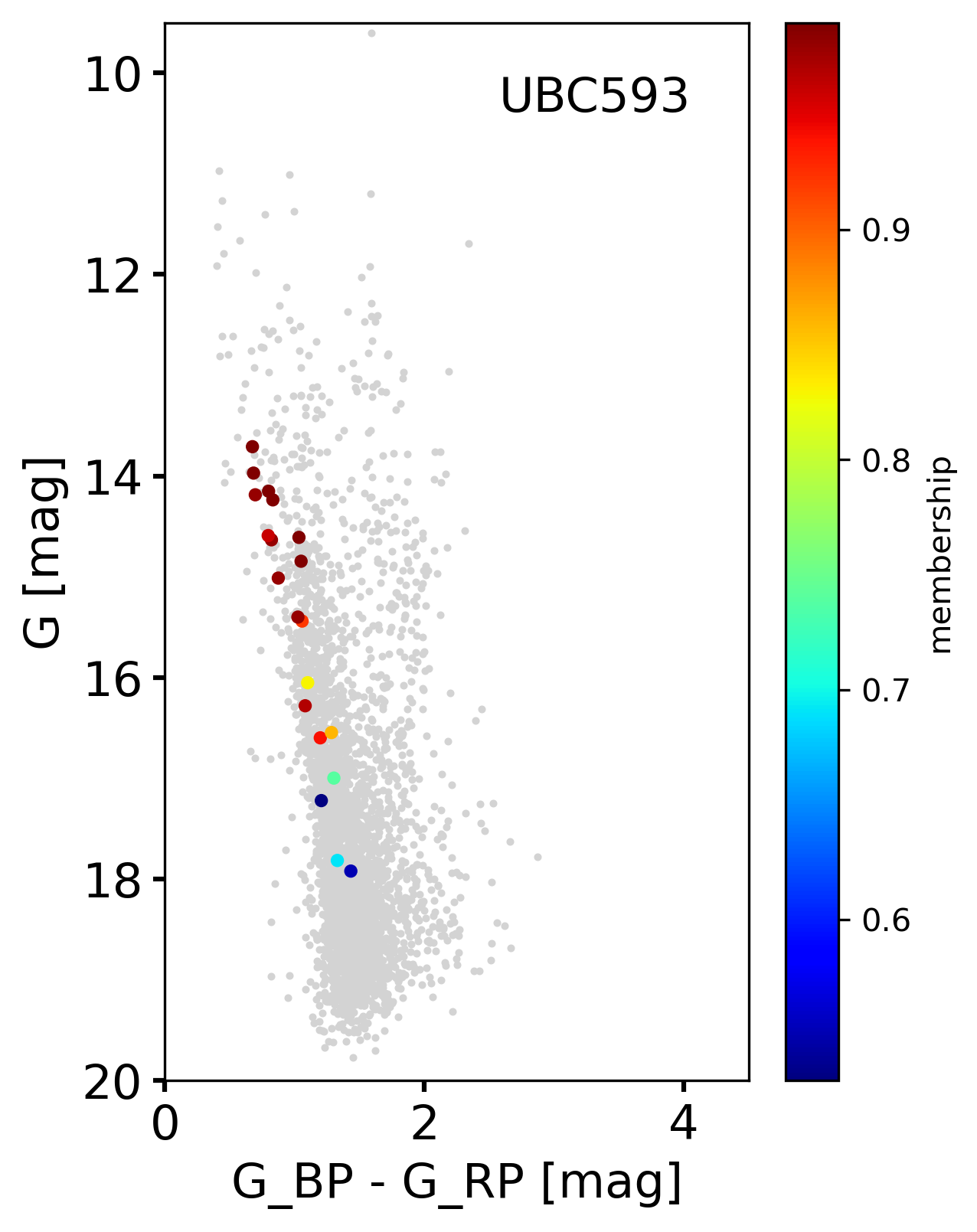}\\
\includegraphics[scale = 0.32]{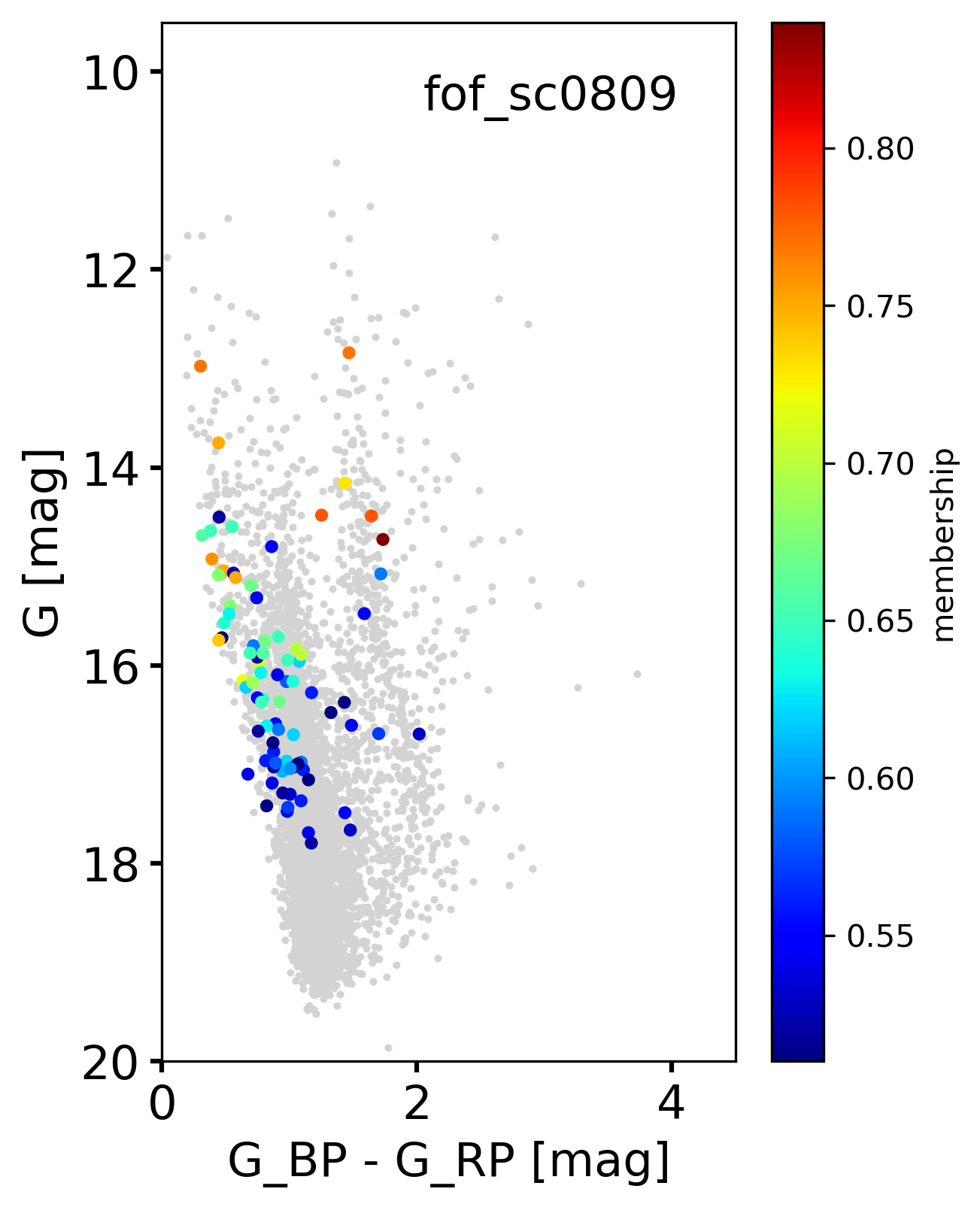}
\includegraphics[scale = 0.32]{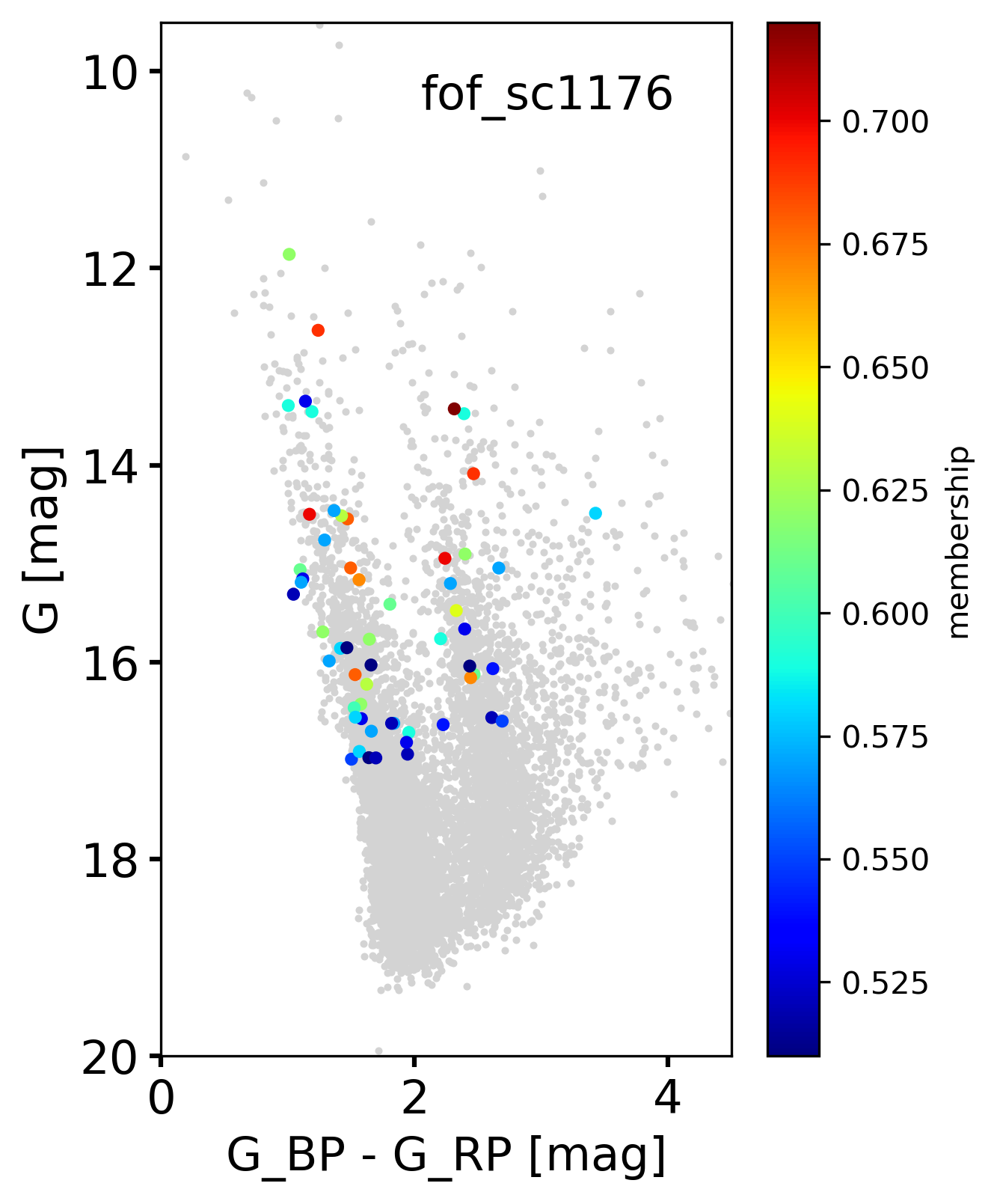}
\caption{Examples of CMDs of cases classified as dubious. 
}
\label{fig:CMDsdubious}
\end{figure}

\begin{figure}
\centering
\includegraphics[scale = 0.32]{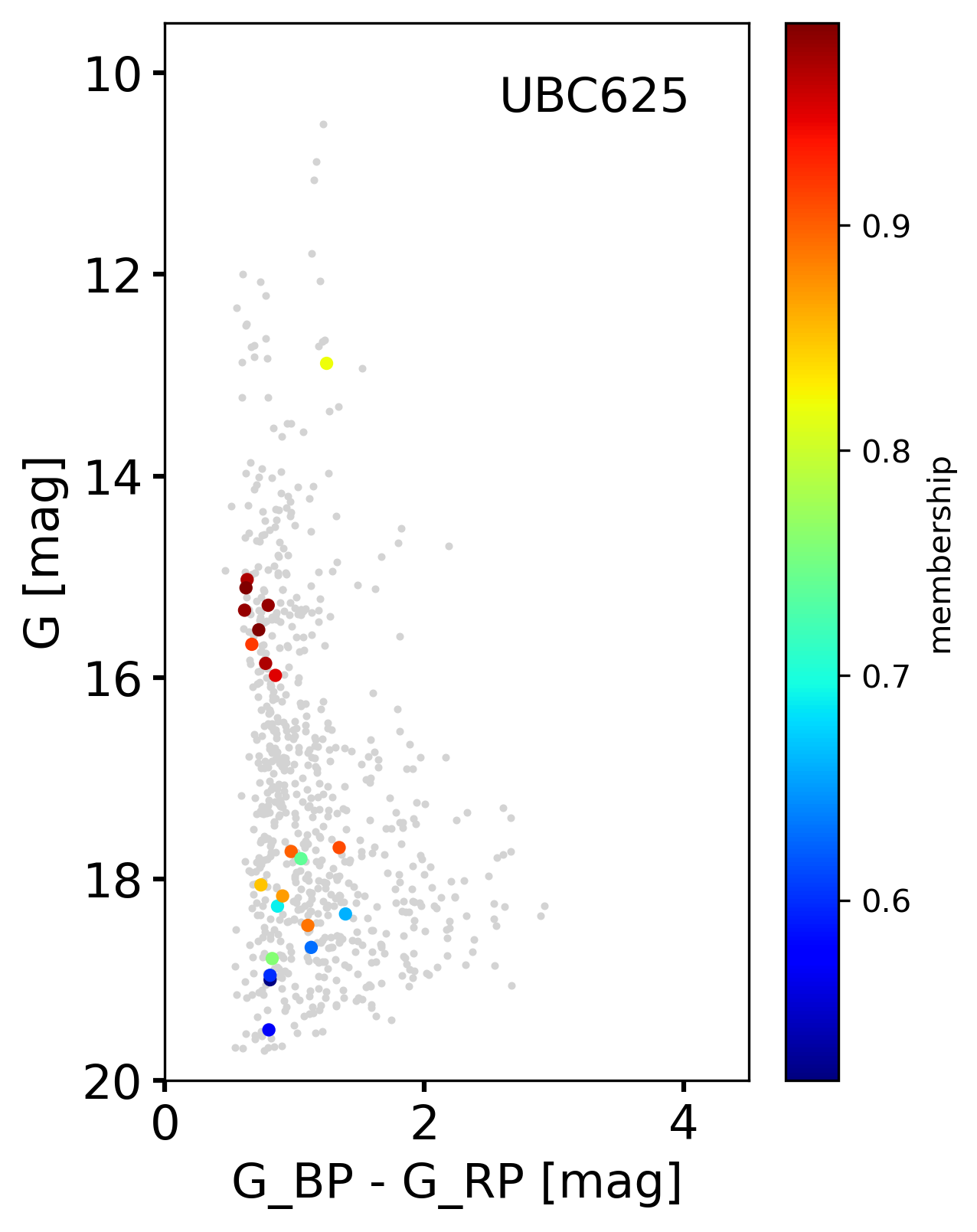}
\includegraphics[scale = 0.32]{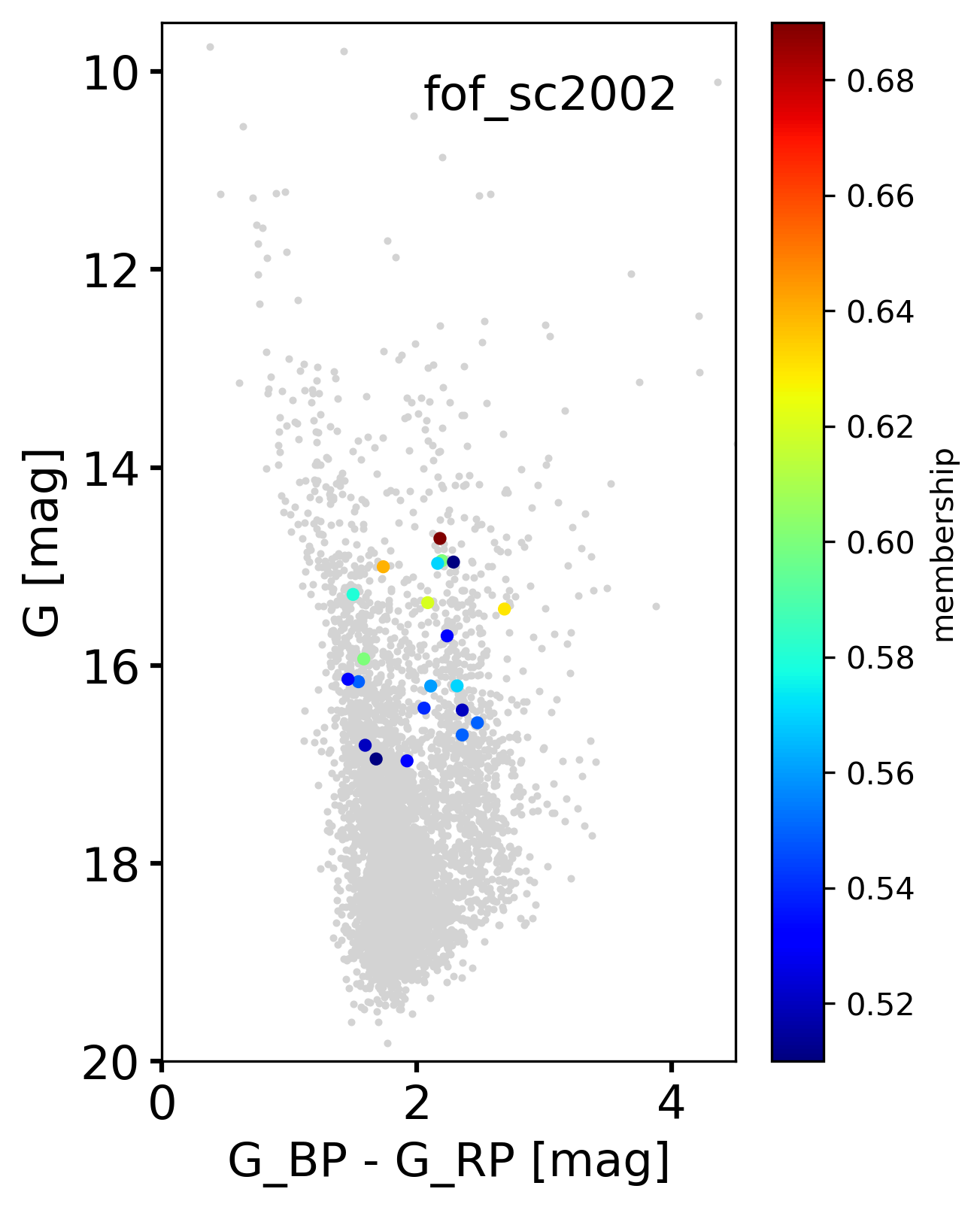}\\
\includegraphics[scale = 0.32]{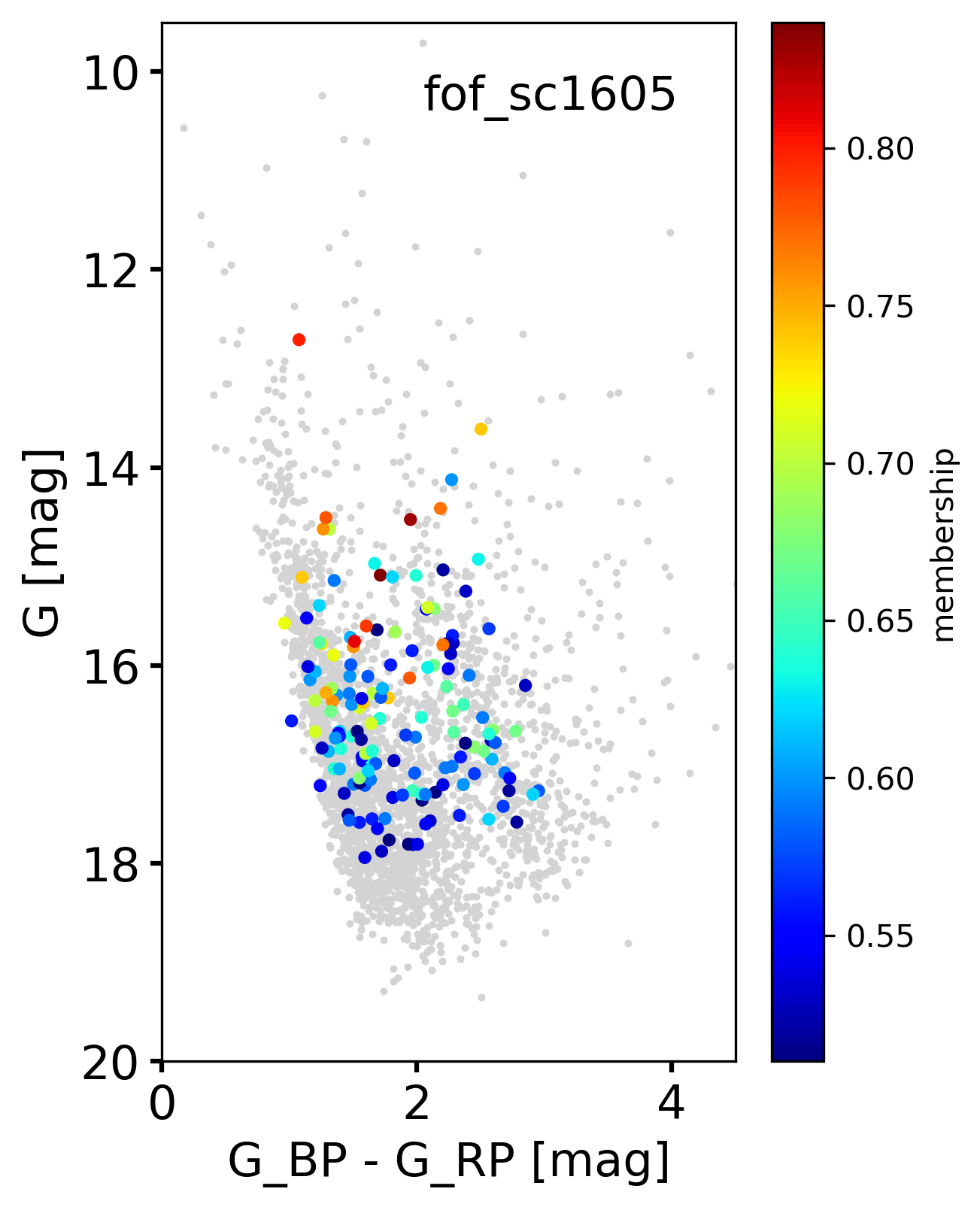}
\includegraphics[scale = 0.32]{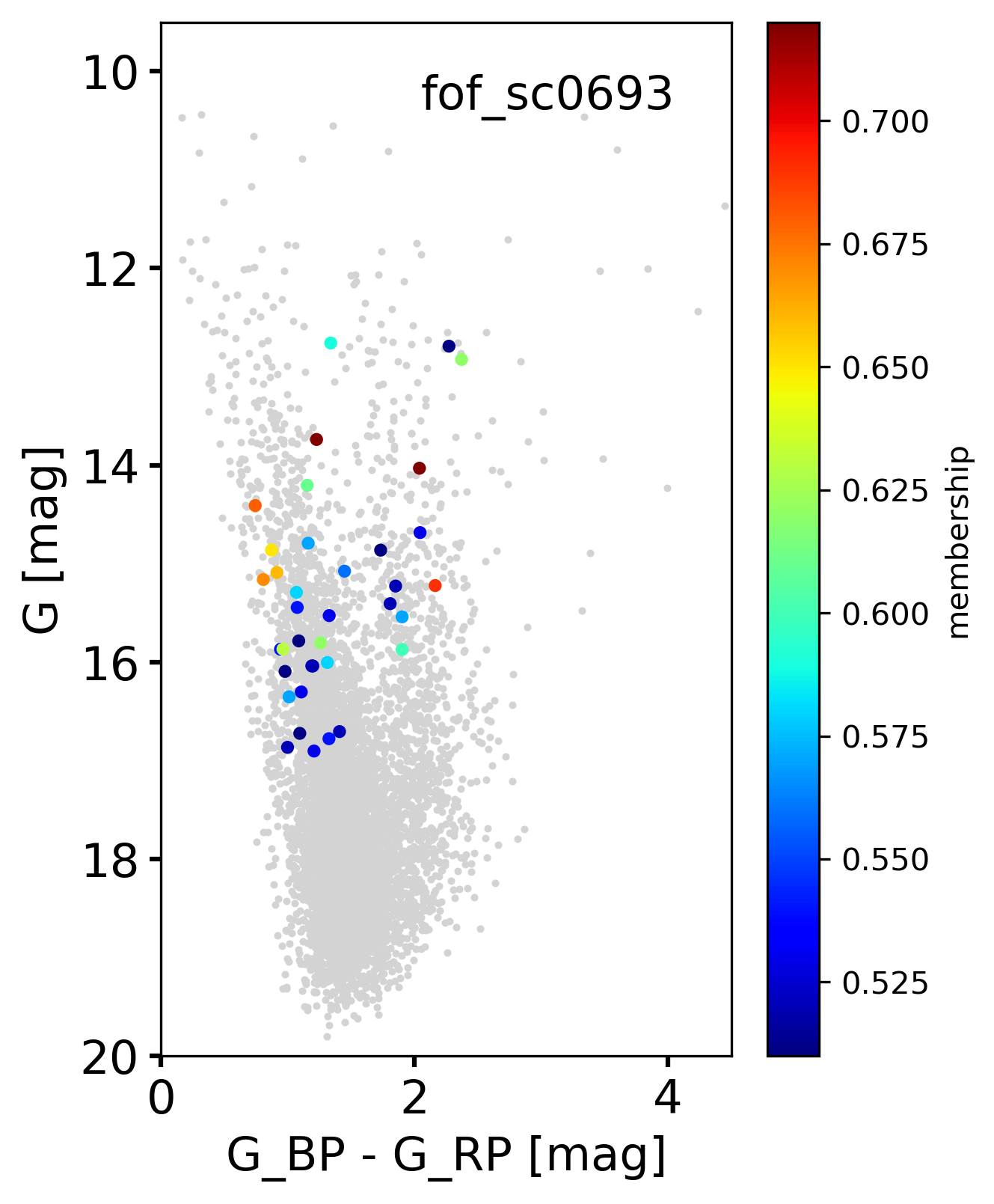}
\caption{Examples of CMDs of cases classified as not real clusters. 
}
\label{fig:CMDsnotreal}
\end{figure}

Figures \ref{fig:CMDsdubious} and \ref{fig:CMDsnotreal} show typical cases we found with dubious and non real objects for which our code was unable to obtain reliable solutions.
In short, some clusters as UBC~625 and fof~sc~2002 have few members and despite the low dispersion in proper motion, the CMDs do not show the typical clear features of a real cluster. 
Other cases as UBC~649 and fof~sc~2213 have CMDs with a gap of more than 1 mag in the possible main sequence of the cluster, that cannot be explained by the theory of stellar evolution.
A number of clusters, as fof~sc~1605, have very spread MS on the CMD without a feature of a cluster. The list of likely not real open clusters (85 not real and 82 dubious) is
given in Tables \ref{table:notclusters} and \ref{table:dubious} for further investigations.

\begin{figure}
\centering
\includegraphics[scale = 0.3]{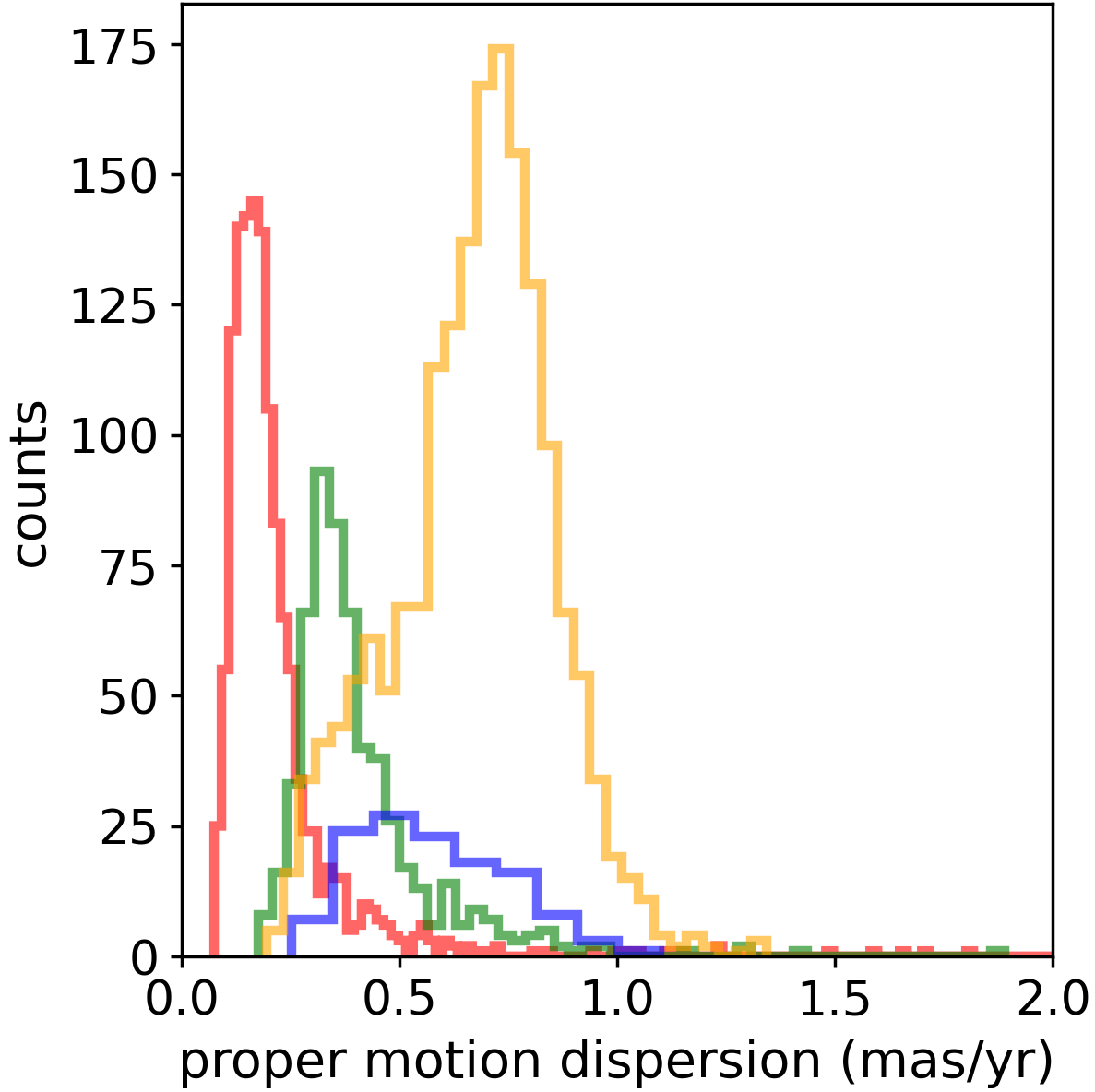}
\includegraphics[scale = 0.3]{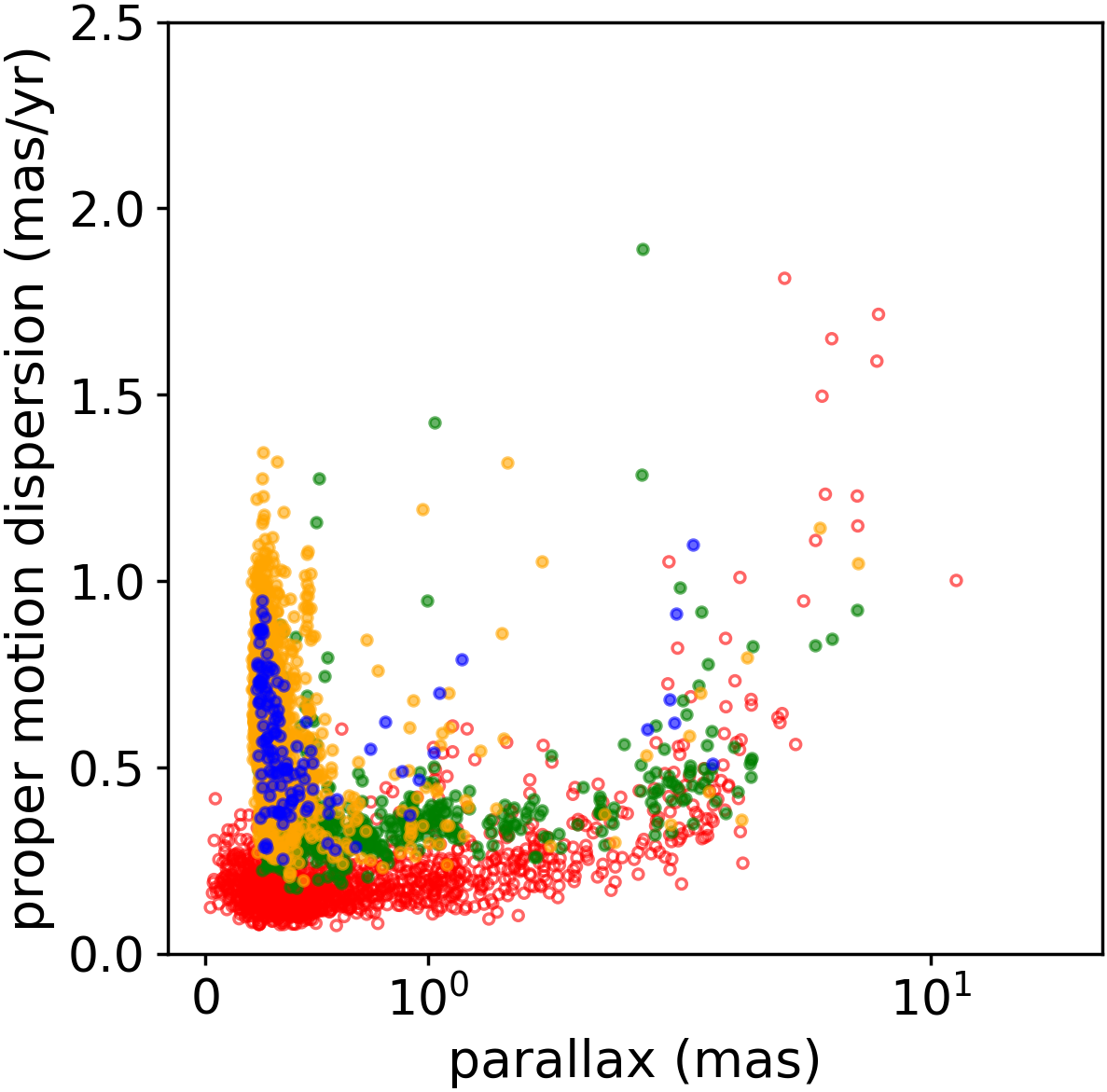}\\
\includegraphics[scale = 0.3]{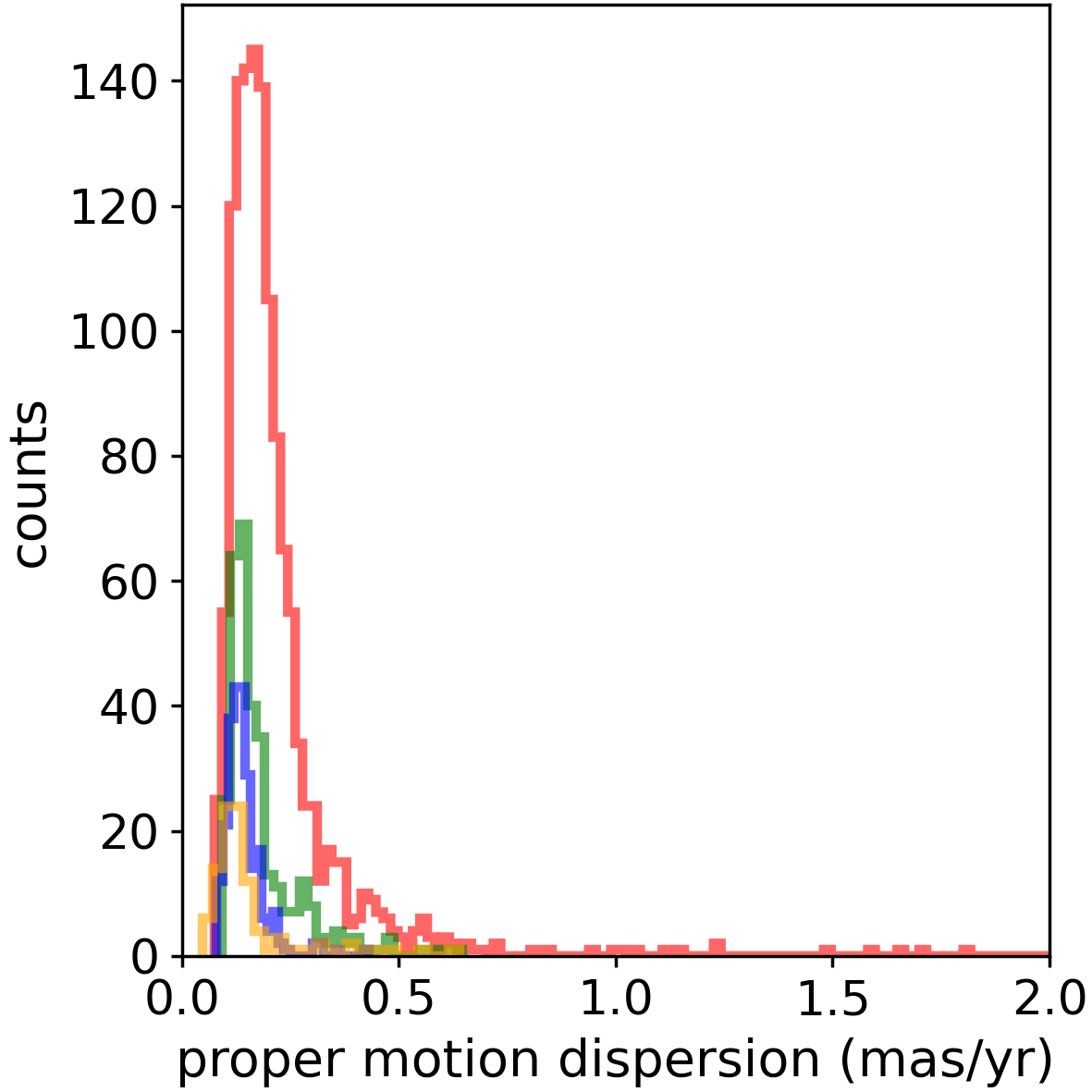}
\includegraphics[scale = 0.3]{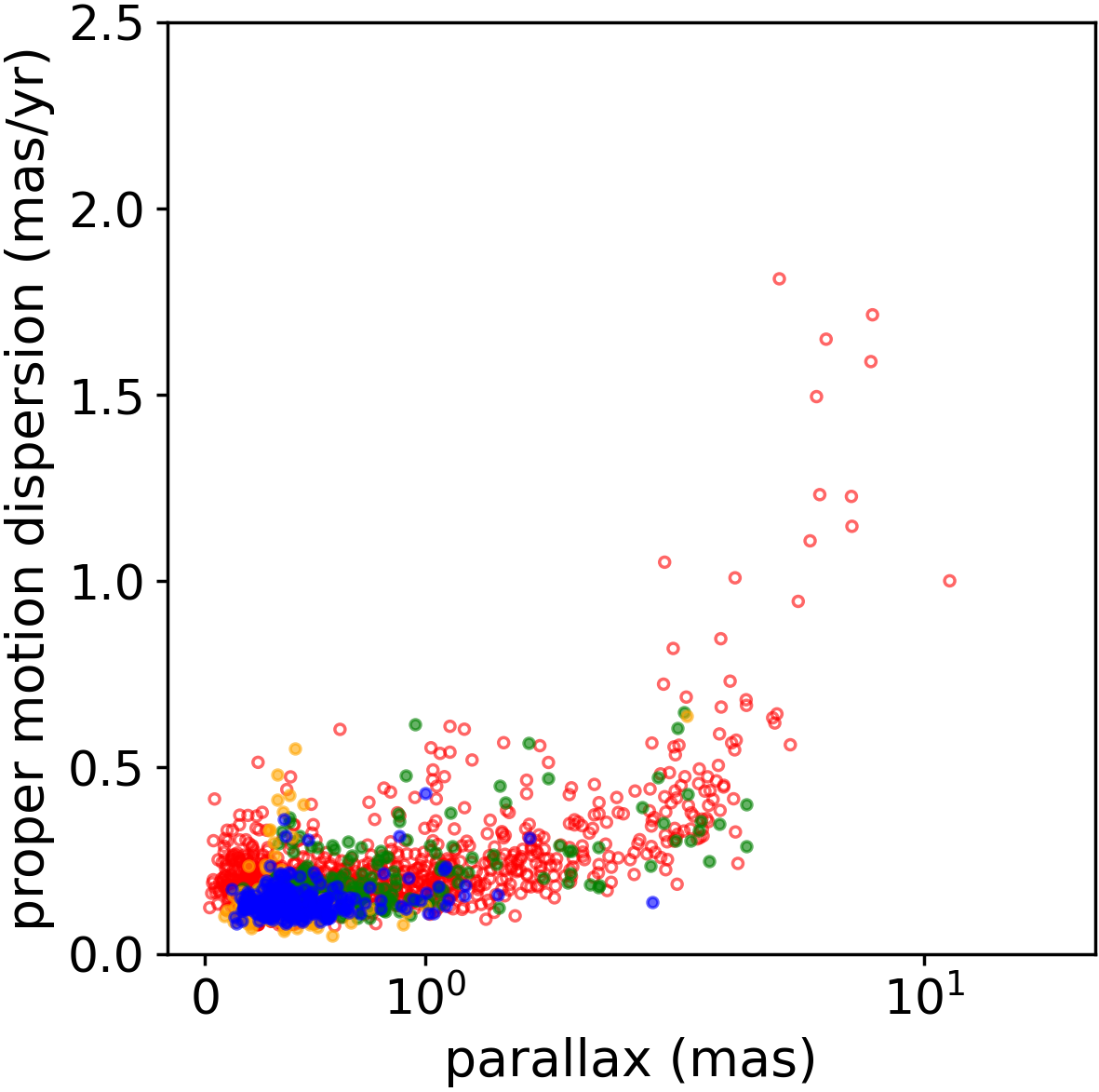}
\caption{Upper panels: Comparison of the total observed proper motion dispersion ($\sqrt{(\mu_{\alpha} \cos \delta)^2 + (\mu_{\delta})^2   }$) in mas\,yr$^{-1}$ of the open clusters from \citet{Cantat2020a} (red), \citet{ChinesCat} class 1 (green),  class 2 (blue) and class 3 (orange). The plot on the right gives the total observed proper motion dispersion as a function of the cluster mean parallax with the x-axis  on a log scale. Bottom panels: the same for open clusters from \citet{Castro-Ginard2020A&A...635A..45C} with color codes for their class A (green),  class B (blue) and class C (orange).} 
\label{fig:SPM}
\end{figure}

The large number of doubtful cases mainly from \citet{ChinesCat} can be explained as due to objects that are not real open clusters. Figure \ref{fig:SPM} presents a comparison of the total proper motion dispersion of real open clusters (from \citet{CantateAnders}) and the sample of the open clusters from \citet{ChinesCat}. The figure shows a clear separation of their class 2 (lower isochrone fit quality) and class 3  (lower isochrone fit quality and scattered CMD) samples both with many clusters with total proper motion dispersion greater than their class 1 (higher quality results) and mainly greater than the real clusters from \citet{Cantat2020a}.
On the other hand, Figure \ref{fig:SPM} shows the same analysis to UBC open clusters considering the classification from \citet{Castro-Ginard2020A&A...635A..45C}. Here we note that the  majority of cases fall within the proper motion dispersion of real open clusters from \citet{Cantat2020a}.
Although the dispersion of the proper motion is not a definitive proof, it does provide indication that a large proportion of class 2 and 3 objects from \citet{ChinesCat} are not real clusters. We understand that these cases need further confirmation. For this reason we opted to give a list of dubious cases and revisit it in a more detailed study in another work. 

\begin{figure*}
\flushleft
\includegraphics[scale = 0.35]{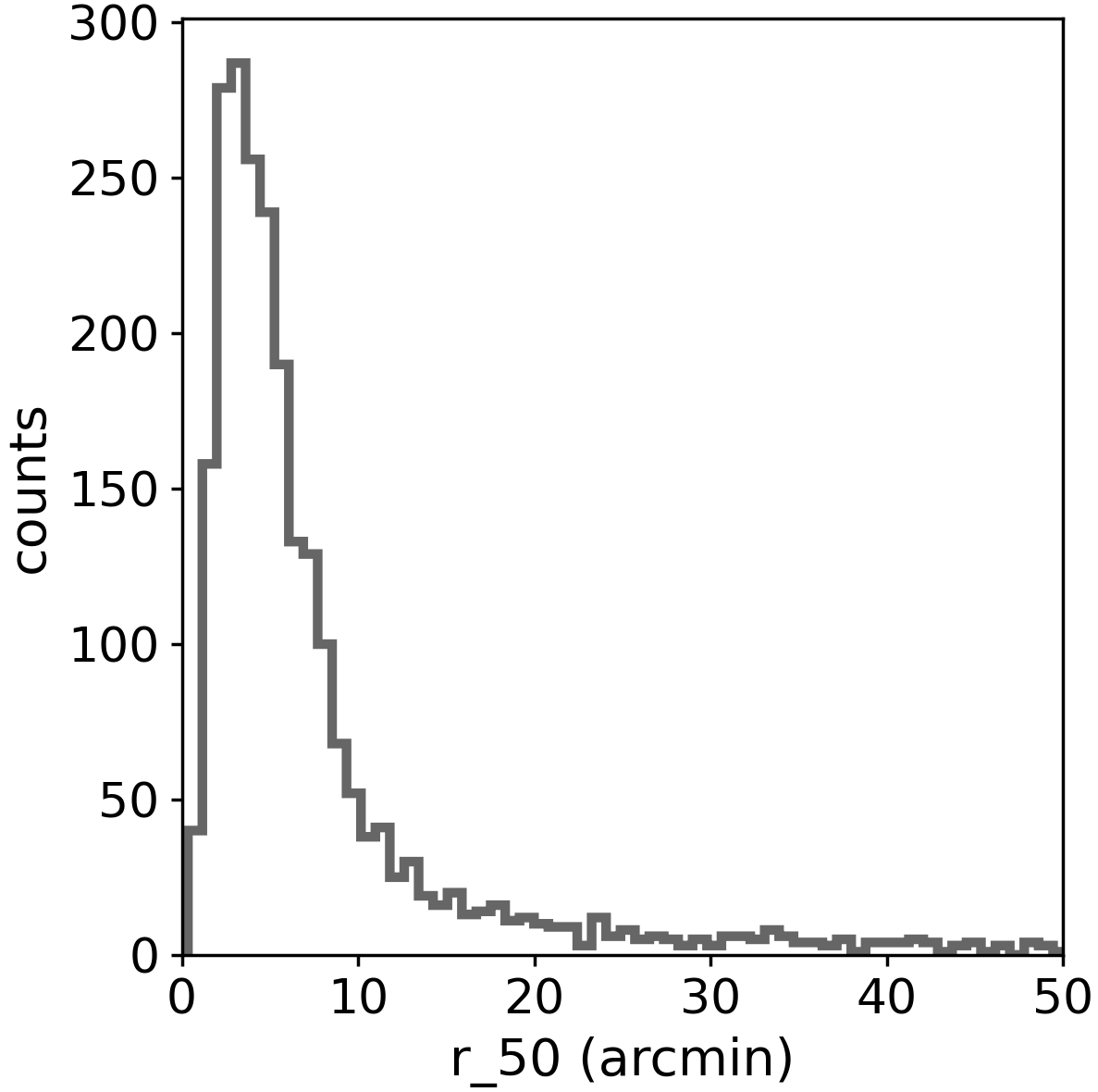}
\includegraphics[scale = 0.35]{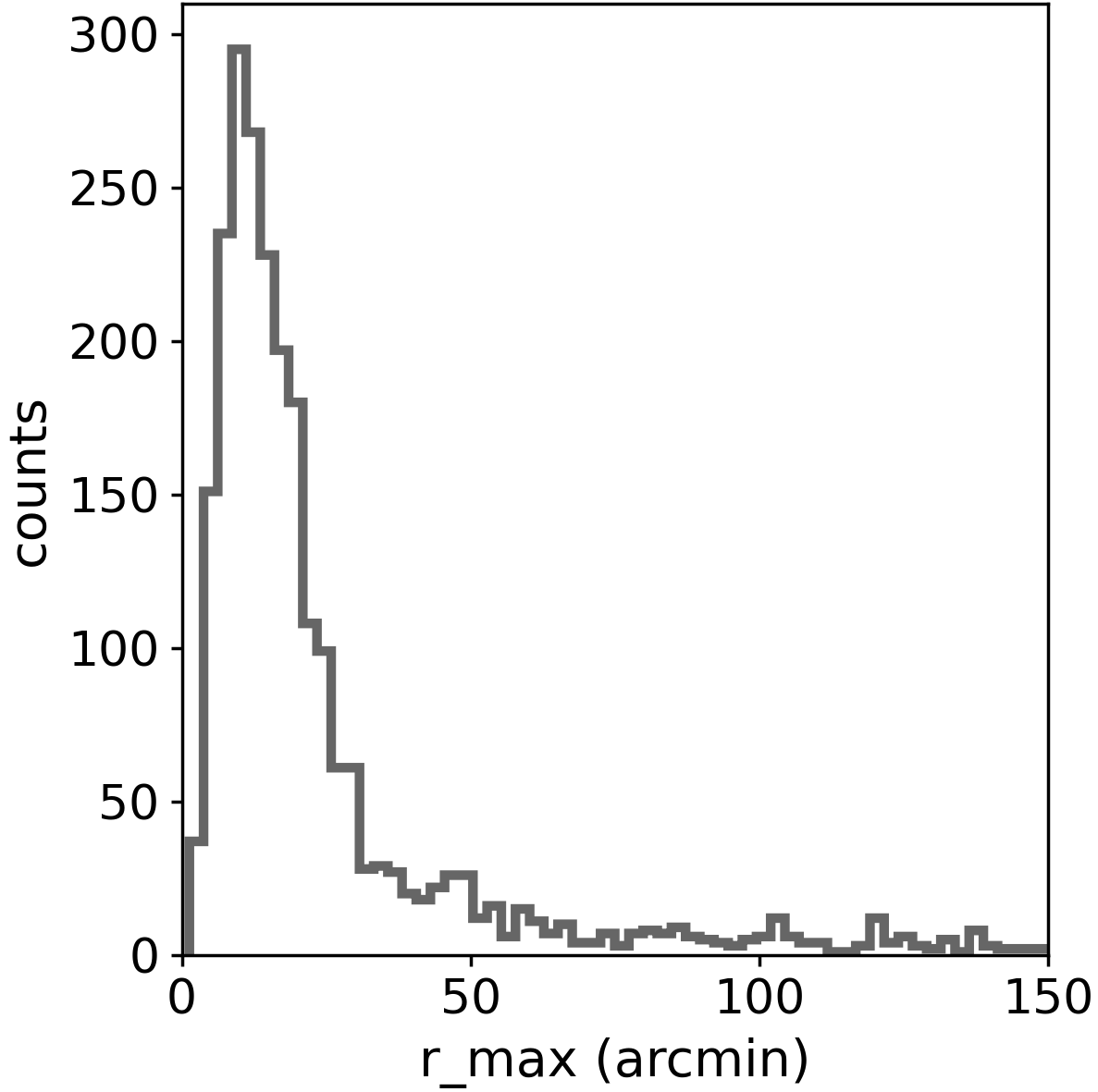}
\includegraphics[scale = 0.35]{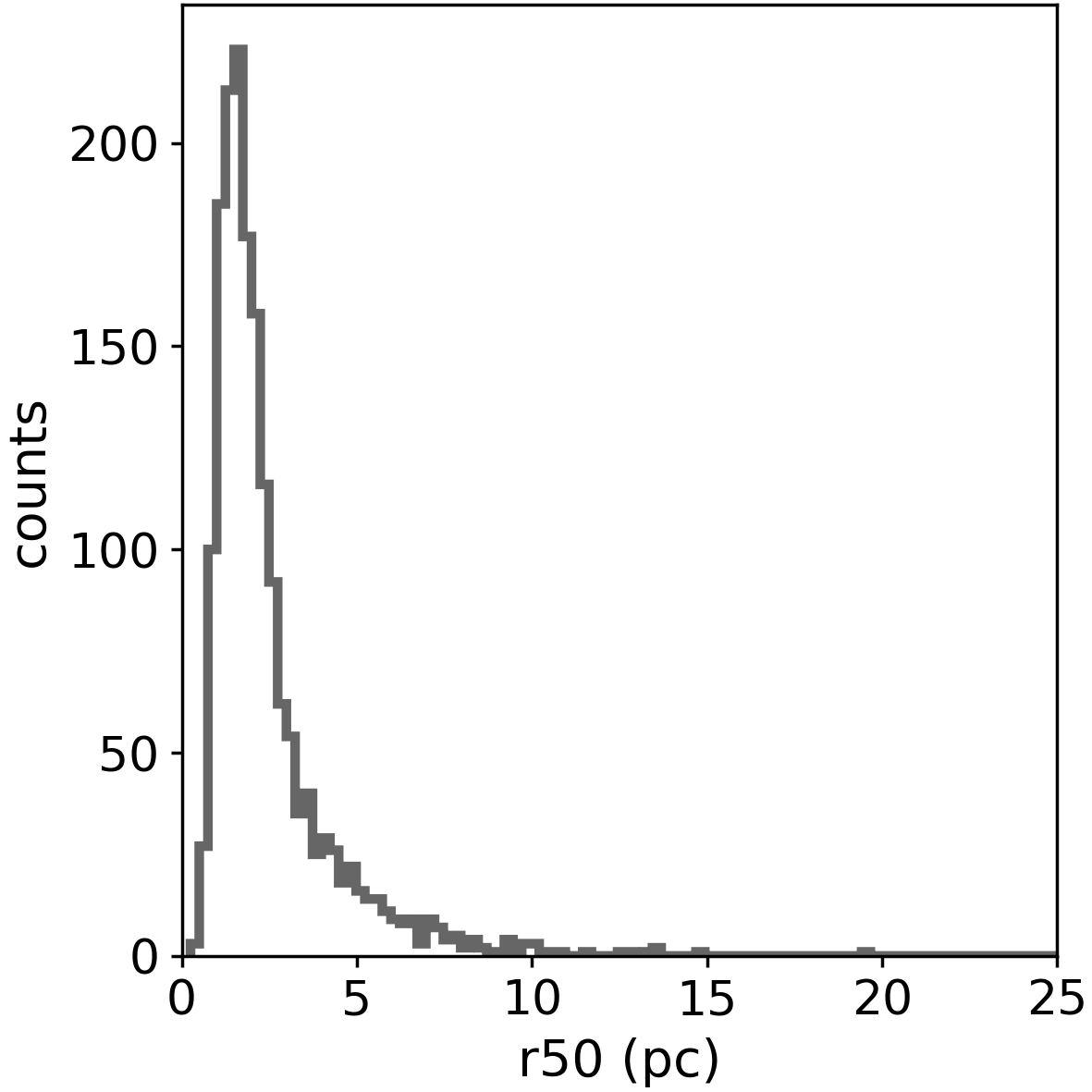}
\includegraphics[scale = 0.35]{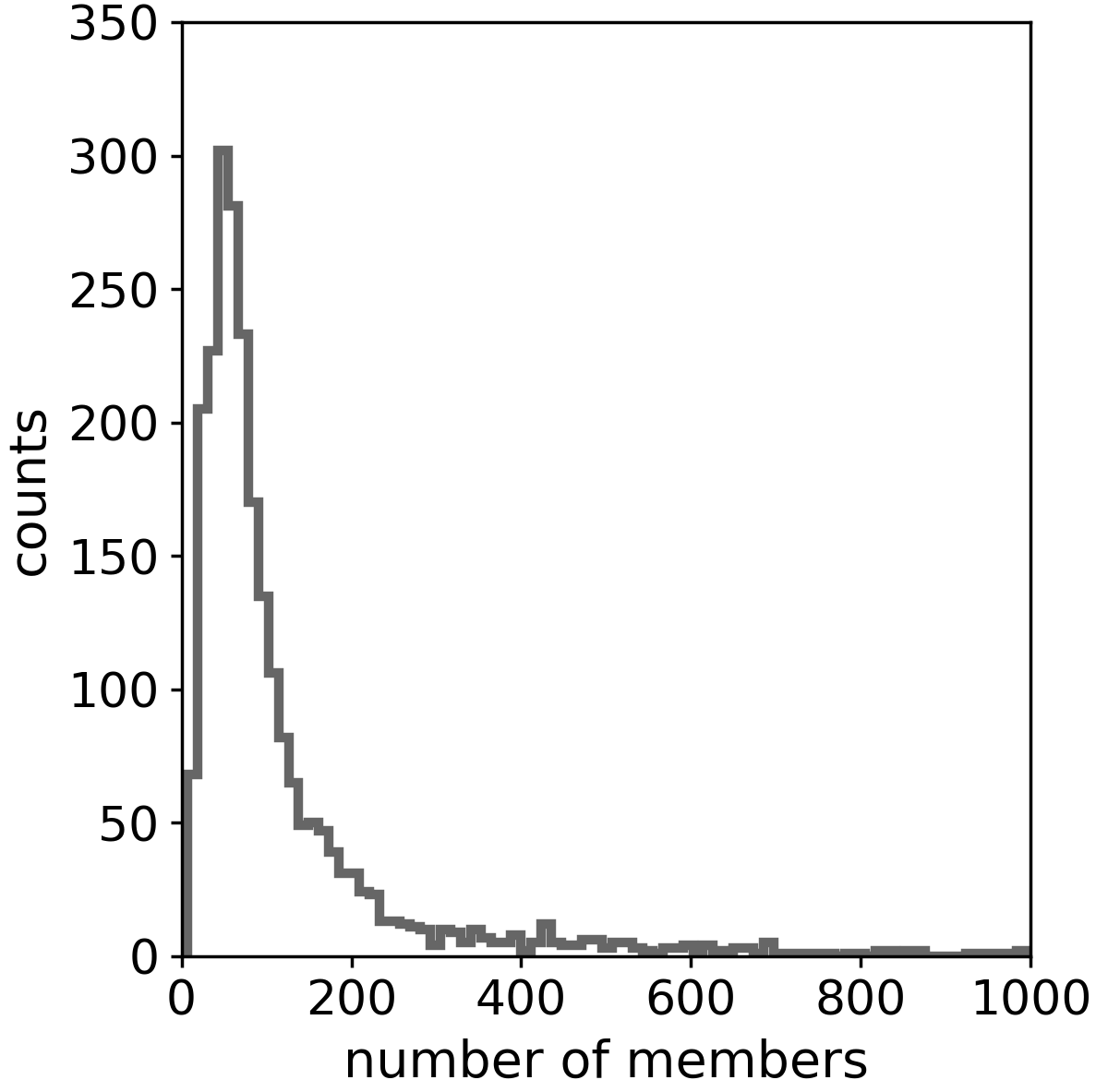}\\
\includegraphics[scale = 0.35]{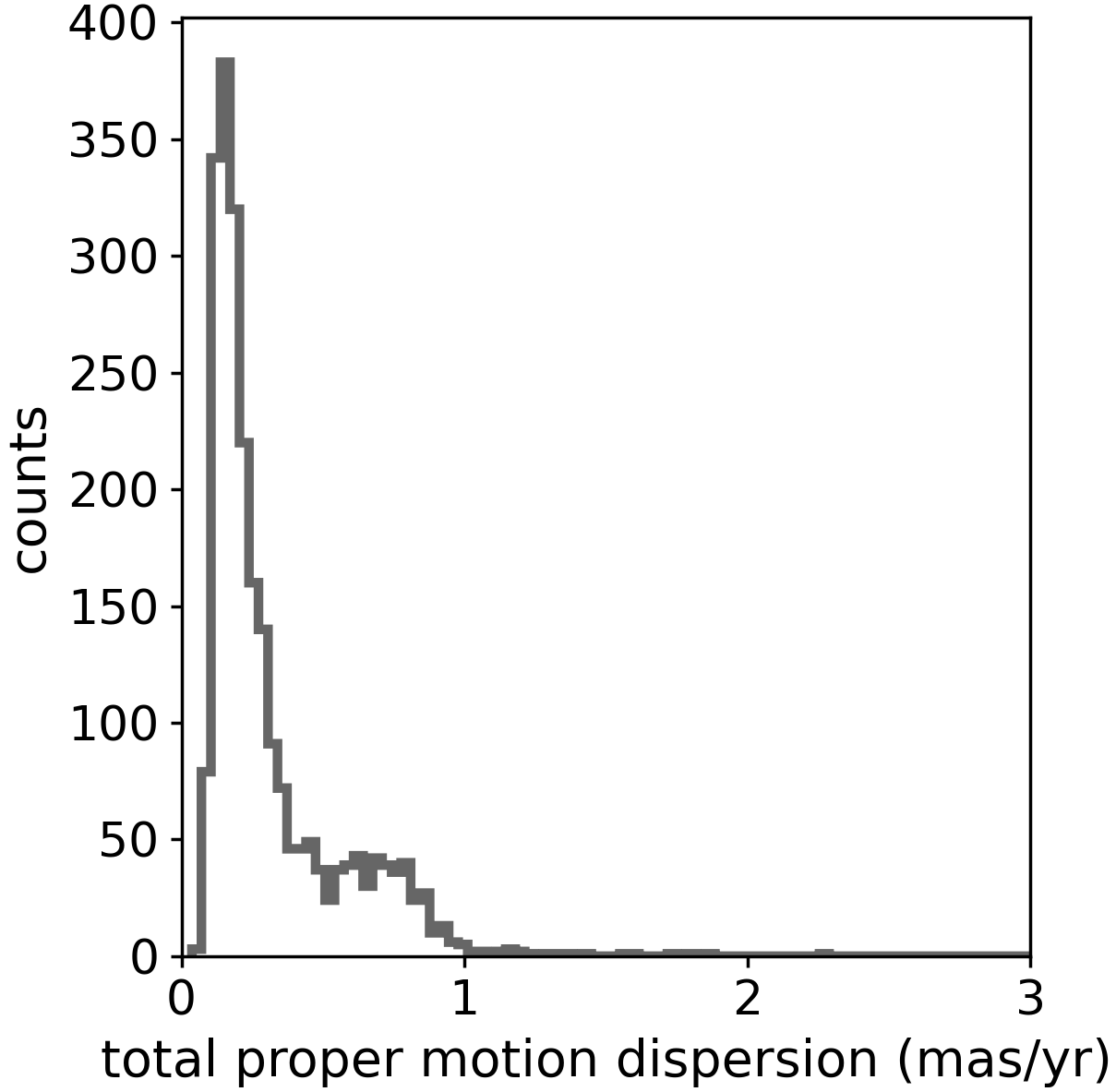}
\includegraphics[scale = 0.35]{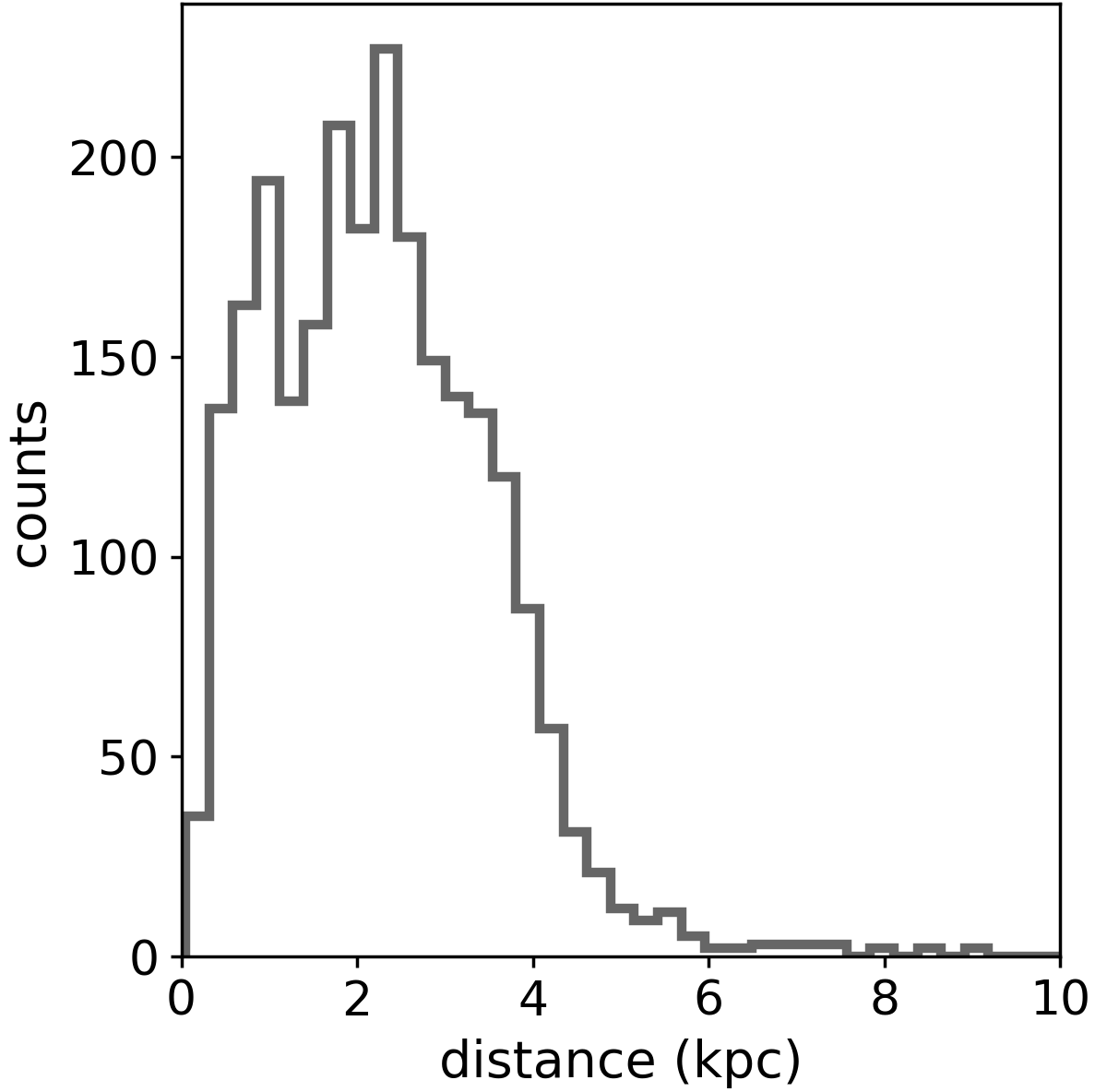}
\includegraphics[scale = 0.35]{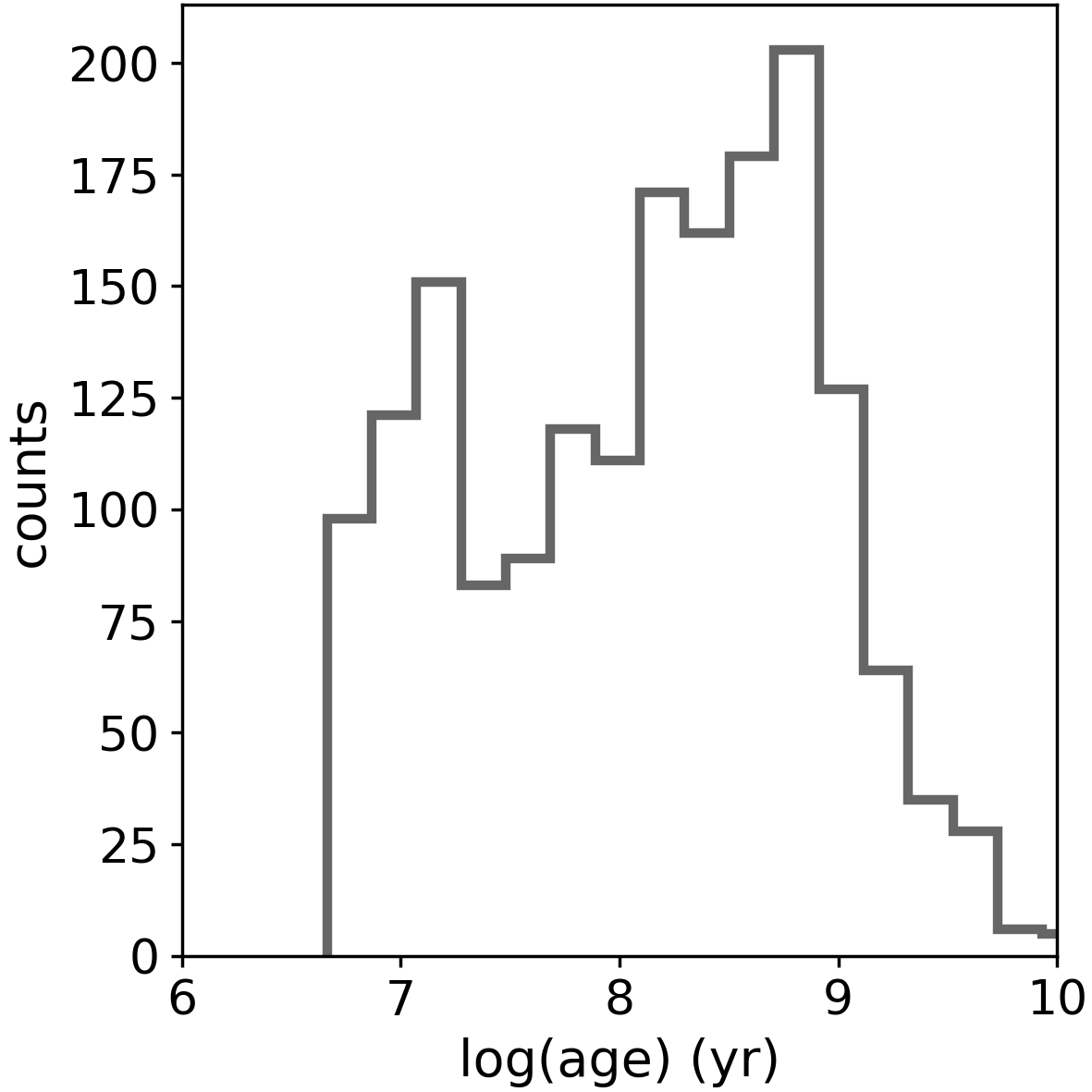}
\includegraphics[scale = 0.35]{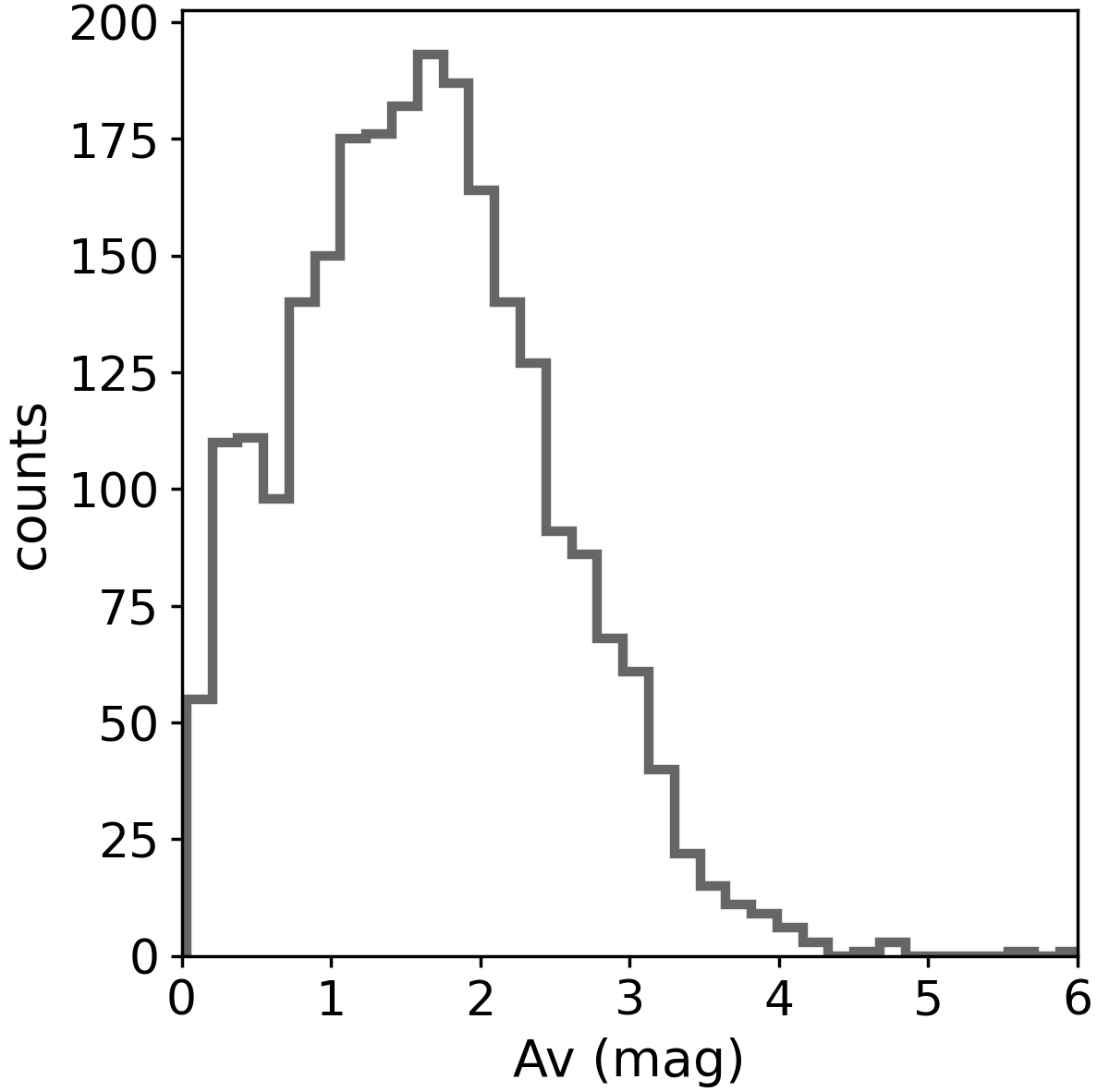}
\caption{Distribution of the parameters of the 1743 open clusters presented in this work. $r_{50}$ is the apparent radius containing $50\%$ of the cluster's members and the maximum radius $r_{max}$ is the greatest angular distance between the member stars and the central coordinates of each cluster. The total proper motion dispersion is given by ($\sqrt{(\mu_{\alpha} \cos \delta)^2 + (\mu_{\delta})^2  }$).} 
\label{fig:statistics}
\end{figure*}

\begin{figure}
\centering
\includegraphics[scale = 0.5]{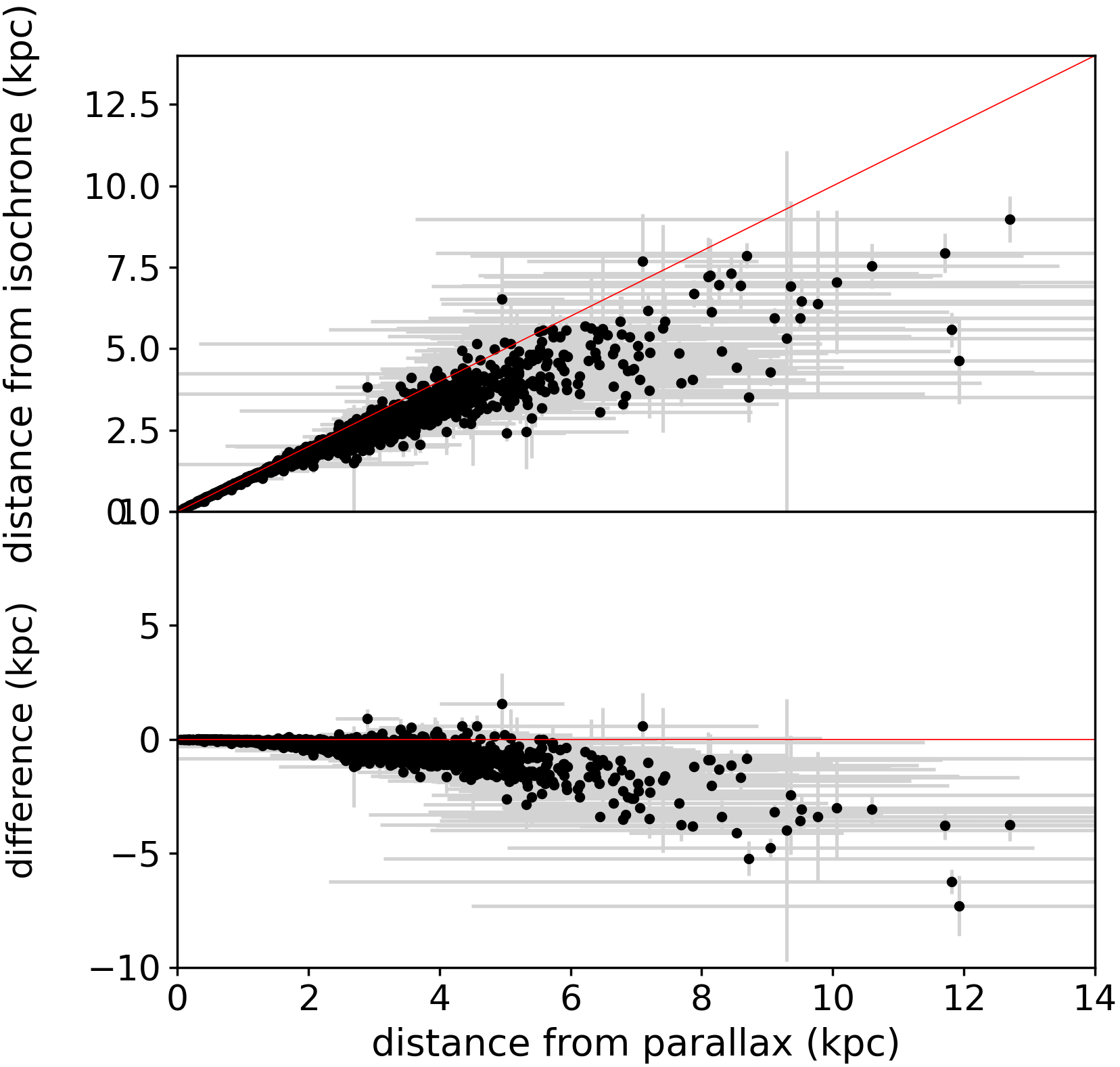}
\caption{Comparison of the distances determined from isochrone fits and from parallaxes. The bottom plot shows the difference in distance in the sense distance from isochrones minus from parallaxes as a function of the distance from parallaxes.}
\label{fig:offset}
\end{figure} 

\begin{figure}
\centering
\includegraphics[scale = 0.7]{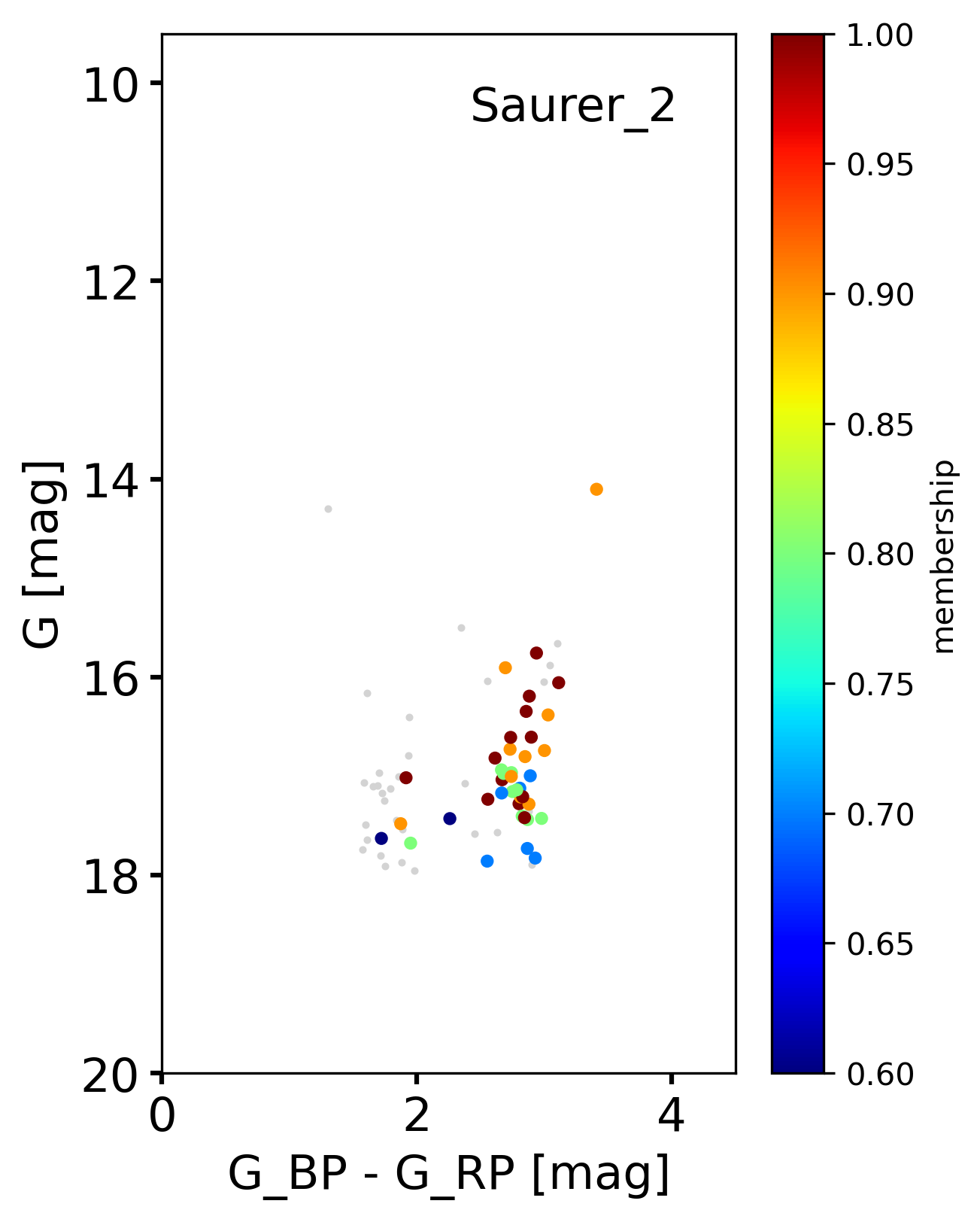}
\caption{Saurer 2 is a typical old and distant cluster for which the data do not define a main sequence and our code fails in the estimation of the parameters.}
\label{fig:Saurer2}
\end{figure}

Finally, we point out that older and more distant open clusters, such as Saurer~2 presented in Figure \ref{fig:Saurer2}, do not define clear main sequences with published members and our code results in poor parameter estimates. Nevertheless, the reality of Saurer~2, which was not discovered based on {\it Gaia} data, is not in question. This is a case of real cluster that is too faint to have enough {\it Gaia}DR2 based membership determinations to define a clear CMD sequence. These cases were eliminated from the final list in the visual inspection phase.

\section{Statistics}

Although determinations of distance, extinction, number of members and proper motion dispersion may change using deeper photometric surveys or improved astrometric precision, the statistical description of the results from this work can provide a useful reference, in addition to the bona-fide criteria from \citet{CantateAnders}, for assessing if an object is or not a real OC.

Figure \ref{fig:statistics} shows the distribution of the parameters of the 1743 open clusters investigated in this study.
It is noted that $70\%$ of the clusters have up to 50 members, $90\%$ have total proper motion dispersion less than 0.5 mas\,yr$^{-1}$ (regardless of the distance ) and $99\%$ have a radius smaller than 15 pc (considering $r_{50}$). 

The clusters have distances from the Sun that vary from 47 pc (Melotte~25 = Hyades) to 8978 pc (Tombaugh~2) 
with $21\%$ of the sample located within 1 kpc from the Sun. The maximum $A_V$ determined in this study was about 6 mag for cluster UPK~402. The range in log(age) is 6.7 dex (FSR~1352) to 10.0 dex (Berkeley~17).

Finally, we also investigated the relations between cluster parameters and distance to the Galactic Center, but no significant correlation was found.

\section{Comparison of distances from parallaxes and isochrone fiting} \label{sec:distcomp}

Studies comparing the distances obtained using {\it Gaia} DR2 parallaxes with those from other methods show the existence of a systematic offset. The value of the offset ranges from $-$0.082 mas to $-$0.029 mas, depending on the objects and the method used, as shown by \citet{Arenou2018}.

\citet{Stassun2018ApJ...862...61S} estimated a systematic offset of ($-$0.082$\pm$0.033) mas using a sample of 158 eclipsing binary. \citet{Riess2018ApJ...861..126R} determined a global offset of $($-$0.046\pm0.013)$ mas using 50 Cepheids.
\citet{Zinn2019ApJ...878..136Z} finds $($-$0.0528\pm0.0024)$ mas using more than 3000 stars with asteroseismically determined. \citet{Kounkel2018AJ....156...84K} used 55 young stars with Very Large Baseline Array parallaxes and finds an offset of $($-$0.074\pm0.034)$ mas. 
A global zero-point of $-$0.029 mas was found by \citet{Lindegren2018} using the high-precision sample of about 493 thousands quasars. 
\citet{Arenou2018} presented a residual zero-point in parallaxes of $($-$0.067\pm0.120)$ mas for MWSC and $($-$0.064\pm0.170)$ mas for DAML, comparing the stars in about 200 clusters. Finally, \citet{Perren2020A&A...637A..95P} performed a combined analysis of UBVI photometry and Gaia DR2 data with the ASteCA code for inference of cluster parameters and report an offset of $-$0.026 mas using a sample of 10 clusters.

Figure \ref{fig:offset} presents a comparison of the distances determined from isochrone fits and from parallaxes for the 1743 open clusters in this study. The distances obtained from parallaxes were determined with a maximum likelihood estimation assuming a normal distribution for individual member stars and taking into account individual parallax uncertainties. The errors were estimated by considering a symmetric distribution so that $\sigma = r_{95} - r_{5}/(2\times1.645)$, which is equivalent to a 1$\sigma$ Gaussian uncertainty, where $r_{5}$ and $r_{95}$ are the 5th and 95th percentile confidence intervals. To estimate the individual cluster offsets we maximize the following maximum likelihood formulation:

\begin{equation}
\begin{aligned}
	\mathcal{L} \propto & \prod_{i=1} P(x|d_{iso},\varpi_i,\sigma_{\varpi_i})  = \\
	& \prod_{i=1} \frac{1}{ \sqrt{2 \pi \sigma_{\varpi_i}^2} }  \exp \left( - \frac{  [ (\varpi_i+x) - \frac{1}{d_{iso}}] ^2  }{ 2 \sigma_{\varpi_i}^2 } \right)
\end{aligned}
\end{equation} 

\noindent where $P(x|d_{iso},\varpi_i,\sigma_{\varpi_i}) $  is the probability of obtaining an offset $x$ given the cluster $d_{iso}$, the distance from the isochrone fit (in kpc), the value of $\varpi_i$ (in mas) for the parallax of cluster member star $i$ and its uncertainty $\sigma_{\varpi_i}$. An important assumption here is that the size of the cluster is smaller than its distance and so the procedure will not be ideal for large nearby objects. We have also neglected the correlations between parallax measurements.

With the procedure outlined above we found a global offset of $(-0.05\pm0.04)$mas and $(-0.05\pm0.03)$mas using the mean and median of the sample of differences, respectively. Our analysis follows the suggestion of \citet{Zinn2019ApJ...878..136Z} that the offset may increase with the distances, as can be seen in Figure \ref{fig:offset}.

\section{Conclusions} 

We presented an update of parameters for 1743 Galactic open clusters from isochrone fits to {\it Gaia} DR2 data, using the improved extinction polynomial and metallicity gradient prior as presented in \citet{Monteiro2020}, and in addition a weak prior in interstellar extinction.

The isochrone fitting code, described in \citet{Monteiro2020, Monteiro2017} uses a
cross-entropy global optimization procedure to fit isochrones to $G_{BP}$ and $G_{RP}$ magnitudes from {\it Gaia} DR2 to determine the distance, age, $A_V$ and [Fe/H] of the clusters. The membership probability of stars and their nominal photometric errors are taken into account. The parameters are given with associated uncertainties, with mean values for the sample are about 170 pc in distance, 0.24 dex in log(age) and 0.12 mag in $A_v$. 

In this study, we used stellar membership probabilities published in the literature.  However, for the clusters from \citet{ChinesCat} and \citet{Castro-Ginard2020A&A...635A..45C} the values were recalculated applying a maximum likelihood approach which allowed the construction of clearer CMDs and better weights for the isochrone fitting code. 

The mean radial velocities of the clusters were determined from {\it Gaia} DR2 radial velocities of member stars. In total, radial velocity were calculated for 831 clusters of which 198 had no previous published estimates.
 
The parameters of the sample indicate that the clusters detected in {\it Gaia} DR2 50$\%$ of their members are within a radius of 15 pc and total proper motion dispersion smaller than $0.5 - 0.8$ $mas~yr^{-1}$.  

Analysis of the differences in distances determined from isochrone fitting and from parallax indicates an offset in the {\it Gaia} DR2 parallax of $(-0.05\pm0.04)$mas and $(-0.05\pm0.03)$mas using the mean and median of the differences, respectively. 

In addition to the cluster parameters, all membership probabilities and individual stellar data are also made available, which can be useful for other research, as well as for the selection of targets for spectroscopy.

This paper is a follow-up to \citet{Monteiro2020}. Both papers are part of an ongoing project to bring DAML into the Gaia era.

\section*{Data Availability Statement}
The data underlying this article, that support the plots
and other findings, are available in the article and in its online supplementary material.  

This work has made use of data from the European Space Agency(ESA) Gaia (http://www.cosmos.esa.int/gaia) mission, processedby the Gaia Data Processing and Analysis Consortium (DPAC,http://www.cosmos.esa.int/web/gaia/dpac/consortium).

We also employed catalogs from CDS/Simbad (Strasbourg)
and Digitized Sky Survey images from the Space Telescope Science
Institute (US Government grant NAG W-2166)

\section*{Acknowledgements}
We thank the referee for the valuable suggestions that improved the quality of the paper.
W.S.Dias acknowledges CNPq (fellowship 310765/2020-0) and the S\~ao Paulo State Agency
FAPESP (fellowship 2013/01115-6). H. Monteiro would like to thank
FAPEMIG grants APQ-02030-10 and CEX-PPM-00235-12. AM acknowledges the support from the Portuguese FCT Strategic Programme UID/FIS/00099/2019 for CENTRA.
This research was performed using the facilities of the Laborat\'orio de Astrof\'isica Computacional da Universidade Federal de Itajub\'a (LAC-UNIFEI).
This work has made use of data from the European Space Agency (ESA) mission Gaia (http://www.cosmos.esa.int/gaia), processed
by the Gaia Data Processing and Analysis Consortium (DPAC,
http://www.cosmos.esa.int/web/gaia/dpac/consortium). We employed catalogues from CDS/Simbad (Strasbourg)
and Digitized Sky Survey images from the Space Telescope Science
Institute (US Government grant NAG W-2166).




\bibliographystyle{mnras}
\bibliography{refscat} 

\clearpage

\appendix

\section{List of clusters with discrepant ages}

\begin{table*}[!hb]
\caption{List of 22 objects with differences out of 3$\sigma$ in ages with   \citet{Cantat2020a}.}
\begin{tabular}{lll}
\hline
Alessi~18            &  SAI~25    &  UBC~594    \\    
Barkhatova~1         &  UBC~276   &  UBC~668     \\    
Berkeley~1           &  UBC~322   &               \\    
Berkeley~79          &  UBC~428   &               \\    
COIN-Gaia~41         &  UBC~432   &               \\    
FSR~1363             &  UBC~473   &              \\    
Juchert~20           &  UBC~474   &              \\    
LP~1218              &  UBC~491   &              \\    
NGC~1977             &  UBC~521   &              \\    
NGC~6664             &  UBC~548   &              \\ 
\hline
\end{tabular}%
\label{table:discrepantages}
\end{table*}

\section{List of likely not real clusters}

\hfill \\
\hfill \\
\hfill \\
\hfill \\

\begin{table*}[!hb]
\caption{List of objects that were found to likely not to be real open clusters. Clusters with the \emph{fof} and \emph{UBC} prefixes are from \citet{ChinesCat} and  \citet{Castro-Ginard2020A&A...635A..45C}, respectively.}
\begin{tabular}{lllll}
\hline
\hline
fof sc0160                        & fof sc0880            & fof sc1467        &  fof sc1676           & fof sc2175            \\   
fof sc0259                        & fof sc0888            & fof sc1477        &  fof sc1688           & fof sc2213            \\   
fof sc0317                        & fof sc0889            & fof sc1481        &  fof sc1701           & UBC 10b                \\  
fof sc0401                        & fof sc0895            & fof sc1484        &  fof sc1702           & UBC 625                \\  
fof sc0456                        & fof sc0988            & fof sc1485        &  fof sc1703           & UBC 649                \\  
fof sc0471                        & fof sc1024            & fof sc1488        &  fof sc1712           &                       \\   
fof sc0611                        & fof sc1042            & fof sc1514        &  fof sc1722           &                       \\   
fof sc0659                        & fof sc1052            & fof sc1515        &  fof sc1723           &                       \\   
fof sc0660                        & fof sc1144            & fof sc1537        &  fof sc1737           &                       \\   
fof sc0674                        & fof sc1148            & fof sc1591        &  fof sc1752           &                       \\   
fof sc0692                        & fof sc1152            & fof sc1605        &  fof sc1782           &                       \\   
fof sc0693                        & fof sc1154            & fof sc1615        &  fof sc1811           &                       \\   
fof sc0730                        & fof sc1159            & fof sc1627        &  fof sc2002           &                       \\   
fof sc0732                        & fof sc1210            & fof sc1638        &  fof sc2004           &                       \\   
fof sc0736                        & fof sc1243            & fof sc1649        &  fof sc2097           &                       \\   
fof sc0814                        & fof sc1245            & fof sc1664        &  fof sc2112           &                       \\   
fof sc0834                        & fof sc1262            & fof sc1665        &  fof sc2158           &                       \\   
fof sc0837                        & fof sc1305            & fof sc1667        &  fof sc2167           &                       \\   
fof sc0841                        & fof sc1307            & fof sc1670        &  fof sc2169           &                       \\   
fof sc0871                        & fof sc1308            & fof sc1671        &  fof sc2173           &                       \\   
\hline
\end{tabular}%
\label{table:notclusters}
\end{table*}

\begin{table*}[!hb]
\caption{List of objetcs that were found dubious open clusters. Clusters with the \emph{fof} and \emph{UBC} prefixes are from \citet{ChinesCat} and  \citet{Castro-Ginard2020A&A...635A..45C}, respectively.}
\begin{tabular}{lllll}
\hline
\hline
fof sc0253             & fof sc0949           & fof sc1547             & fof sc1951              &  UBC 644  \\    
fof sc0396             & fof sc0996           & fof sc1548             & fof sc1988              &          \\    
fof sc0522             & fof sc1026           & fof sc1566             & fof sc1999              &          \\    
fof sc0559             & fof sc1083           & fof sc1574             & fof sc2032              &          \\    
fof sc0661             & fof sc1084           & fof sc1602             & fof sc2035              &          \\    
fof sc0696             & fof sc1169           & fof sc1637             & fof sc2089              &          \\    
fof sc0728             & fof sc1176           & fof sc1642             & fof sc2155              &          \\    
fof sc0743             & fof sc1192           & fof sc1668             & fof sc2159              &          \\    
fof sc0751             & fof sc1194           & fof sc1689             & fof sc2160              &          \\    
fof sc0809             & fof sc1195           & fof sc1711             & fof sc2171              &          \\    
fof sc0810             & fof sc1202           & fof sc1714             & UBC 325                  &           \\    
fof sc0838             & fof sc1203           & fof sc1715             & UBC 359                  &           \\    
fof sc0839             & fof sc1208           & fof sc1716             & UBC 416                  &           \\    
fof sc0883             & fof sc1225           & fof sc1727             & UBC 505                  &           \\    
fof sc0903             & fof sc1239           & fof sc1765             & UBC 573                  &           \\    
fof sc0905             & fof sc1385           & fof sc1766             & UBC 575                  &           \\    
fof sc0909             & fof sc1387           & fof sc1774             & UBC 577                  &           \\    
fof sc0929             & fof sc1428           & fof sc1791             & UBC 579                  &           \\    
fof sc0946             & fof sc1452           & fof sc1792             & UBC 592                  &           \\    
fof sc0948             & fof sc1461           & fof sc1808             & UBC 593                  &           \\    
\hline
\end{tabular}%
\label{table:dubious}
\end{table*}


\bsp	
\label{lastpage}
\end{document}